\documentclass[11pt]{article}
\usepackage{amsmath, amssymb, graphics}

\usepackage{float}

\usepackage[nosort]{cite}
\newcommand{\mathsym}[1]{{}}
\usepackage{verbatim}
\usepackage{amsfonts,euscript,amssymb,stmaryrd,braket}
\usepackage{graphics,tikz}
\usetikzlibrary{shapes.misc,arrows,decorations.markings,patterns,calc,shapes.geometric}
\usepackage{caption}
\usepackage{subcaption}
\usepackage{slashed}
\definecolor{hyperref}{RGB}{026,028,185}
\usepackage[bookmarks=true,colorlinks=true,linkcolor=hyperref,citecolor=hyperref,urlcolor=hyperref,bookmarksnumbered]{hyperref}
 \usepackage{booktabs}
 \usepackage{multirow}
\usepackage{tabularx}
\usepackage{longtable}
\usepackage{bbold,bbding}
\usepackage{amsthm} 
\usepackage{esint}
\newcommand{\bal}{\begin{equation}\begin{aligned}}
\newcommand{\eal}{\end{aligned} \end{equation}}
\def\id{\protect{{1 \kern-.28em {\rm l}}}}

\makeatletter
\renewcommand\section{\@startsection {section}{1}{\z@}%
                                   {-3.5ex \@plus -1ex \@minus -.2ex}%
                                   {2.3ex \@plus.2ex}%
                                   {\normalfont\large\bfseries}}
\renewcommand\subsection{\@startsection{subsection}{2}{\z@}%
                                   {-3.25ex\@plus -1ex \@minus -.2ex}%
                                   {1.5ex \@plus .2ex}%
                                   {\normalfont\normalsize\bfseries}}

\makeatother

\numberwithin{equation}{section}

\usepackage[utf8]{inputenc}
\usepackage{epsfig}
\usepackage{graphicx}
\usepackage[multiple]{footmisc}
\usepackage{amssymb,amsmath}
\usepackage{fancybox,framed,tikz}
\usetikzlibrary{decorations.pathmorphing,patterns}
\usepackage{dsfont}
\usepackage{mathtools}
\usepackage{braket}
\usepackage{slashed}
\usepackage{rotating}
\usepackage{bbold,amsfonts}
\usepackage{multirow}
\usepackage{verbatim}
\usepackage{upgreek}

\tikzset{cross/.style={cross out, draw=black, minimum size=2*(#1-\pgflinewidth), inner sep=0pt, outer sep=0pt},
cross/.default={1pt}}

\newcommand{\be}{\begin{equation}}
\newcommand{\ee}{\end{equation}}

\newcommand{\zb}{\bar{z}}

\newcommand{\Tr}{\textup{Tr}}

\usepackage{color}

\definecolor{mypink1}{rgb}{0.958, 0.188, 0.478}

\newcommand{\ba}{\begin{eqnarray}}
\newcommand{\ea}{\end{eqnarray}}

\def\Disc{\text{Disc}}

\hypersetup{}

\tikzset{Witten diagram/.style={execute at begin picture={%
\draw[blue ,fill=blue!05] circle[radius=\pgfkeysvalueof{/tikz/Witten/radius}];
\path node (X){\phantom{X}};
},baseline={(X.base)}},vertex/.style={circle,fill,inner sep=1.414pt,node
contents={}},
Witten/.cd,radius/.initial=1.414cm}

\newcommand\veps{{\varepsilon}}

\DeclareMathOperator{\tr}{Tr}
\DeclareMathOperator{\Pexp}{\mathcal{P}exp}
\DeclareMathOperator{\sgn}{sgn}

\def\Dm{{\mathcal{D}}}

\def\Jm{{\mathcal{J}}}

\def\Nm{{\mathcal{N}}}
\def\Om{{\mathcal{O}}}

\def\Vm{{\mathcal{V}}}

\tikzset{
  defect/.style={
    thick,
    line width=0.12em,
  },
  prop/.style={},
  valign/.style={
    baseline={([yshift=-.55ex]current bounding box.center)}
  },
  cross/.style={
    cross out,
    draw=black,
    minimum size=.5em,
    inner sep=0pt,
    outer sep=0pt
  },
  cross/.default={1pt} 
}

\tikzset{
  vtx/.style={
    circle,
    draw=blue,
    fill=blue,
    inner sep=1pt
  },
  circl/.style={
    circle,
    draw=black,
    fill=black,
    inner sep=1pt
  },
  wcirc/.style={
    circle,
    draw=white,
    fill=white,
    inner sep=2pt
  },
  bcirc/.style={
    circle,
    draw=black,
    fill=black,
    inner sep=1pt
  },
  dcirc/.style={
    circle,
    draw=blue,
    fill=blue,
    inner sep=1pt
  },
  rcirc/.style={
    circle,
    draw=red,
    fill=red,
    inner sep=1pt
  },
  phi/.style={
    dashed
  },
  sigma/.style={
    thick,
    dashed
  },
  vl1/.style={
    thick,
    blue
  },
  vl2/.style={
    thick,
    dashed,
    blue
  },
  zprop/.style={
    thick
  },
  valign/.style={
    baseline={([yshift=-.55ex]current bounding box.center)}
  },
  cross/.style={cross out, draw=black, minimum size=.4em, inner sep=0pt, outer sep=0pt},
cross/.default={1pt}
}

\begin{document}
\renewcommand{\thefootnote}{\arabic{footnote}}

\overfullrule=0pt
\parskip=2pt
\parindent=12pt
\headheight=0in \headsep=0in \topmargin=0in \oddsidemargin=0in

\begin{center}
\vspace{1.2cm}
{\Large\bf \mathversion{bold}
{Analytic bootstrap for magnetic impurities}\\
}
 
\author{ABC\thanks{XYZ} \and DEF\thanks{UVW} \and GHI\thanks{XYZ}}
 \vspace{0.8cm} {
 Lorenzo~Bianchi$^{a,b}$ \footnote{\tt lorenzo.bianchi@unito.it}, Davide Bonomi$^{a,c}$ \footnote{{\tt davide.bonomi@city.ac.uk}}, Elia de Sabbata$^{a,b}$ \footnote{{\tt elia.desabbata@unito.it}}, Aleix Gimenez-Grau$^{d}$ \footnote{\tt gimenez@ihes.fr}
 }
 
 \vskip  0.5cm

\small
{\em
$^{a}$  
Dipartimento di Fisica, Universit\`a di Torino and INFN - Sezione di Torino\\ Via P. Giuria 1, 10125 Torino, Italy\\    

$^{b}$  
I.N.F.N. - sezione di Torino,\\
Via P. Giuria 1, I-10125 Torino, Italy\\    

$^{c}$   Department of Mathematics, City, University of London,\\
Northampton Square, EC1V 0HB London, United Kingdom\\

$^{d}$ Institut des Hautes \'Etudes Scientifiques, 91440 Bures-sur-Yvette, France
\vskip 0.02cm

}
\normalsize

\end{center}

\vspace{0.3cm}
\begin{abstract} 
We study the O(3) critical model and the free theory of a scalar triplet in the presence of a magnetic impurity. We use analytic bootstrap techniques to extract results in the $\veps$-expansion. First, we extend by one order in perturbation theory the computation of the beta function for the defect coupling in the free theory. Then, we analyze in detail the low-lying spectrum of defect operators, focusing on their perturbative realization when the defect is constructed as a path-ordered exponential. After this, we consider two different bulk two-point functions and we compute them using the defect dispersion relation. For a free bulk theory, we are able to fix the form of the correlator at all orders in $\veps$. In particular, taking $\veps\to1$, we can show that in $d=3$ one does not have a consistent and non-trivial defect CFT. For an interacting bulk, we compute the correlator up to second order in $\veps$. Expanding these results in the bulk and defect block expansions, we are able to extract an infinite set of defect CFT data. We discuss low-spin ambiguities that affect every result computed through the dispersion relation and we use a combination of consistency conditions and explicit diagrammatic calculations to fix this ambiguity.
\end{abstract} 

\newpage

\tableofcontents
 \newpage  
\section{Introduction and discussion}

One of the fields where the conformal bootstrap has delivered the most astonishing results is the study of statistical models at criticality. After the celebrated prediction for the critical exponents of the three-dimensional Ising models \cite{El-Showk:2012cjh,El-Showk:2014dwa}, several other remarkable results have been obtained (see \cite{Poland:2018epd,Rychkov:2023wsd} for recent reviews).
This proves the great potential of the idea that conformal field theories (CFTs) can be explored to high precision leveraging only symmetries and internal consistency. Parallel to these numerical developments, a set of important analytical tools have been introduced, which proved very useful when the theory admits a perturbative expansion in some small parameter (not necessarily the coupling) and some control over the perturbative spectrum \cite{Alday:2016njk,Caron-Huot:2017vep,Carmi:2019cub}. In particular, among other applications, these methods were used to reproduce and extend results in the $\veps$-expansion for the $O(N)$ critical model \cite{Alday:2017zzv,Henriksson:2018myn} (see \cite{Henriksson:2022rnm} for an extensive review).

The spectrum of a CFT contains both local and extended operators. The development of bootstrap techniques for the latter have recently emerged as an independent research line, the defect bootstrap program. In a defect setup, there are essentially two observables which are suitable to a bootstrap analysis: the defect four-point function and the bulk two-point function. For general defects, these quantities depend on two kinematical cross ratios and, in this respect, they constitute the direct analogue of four-point correlators in homogeneous CFTs. While the defect four-point function is accessible also to the numerical methods of \cite{Rattazzi:2008pe}, the bulk two-point function generically lacks the positivity properties that are necessary for this approach. Nevertheless analytical methods based on the Lorentzian inversion formulae \cite{Lemos:2017vnx,Liendo:2019jpu} and dispersion relations \cite{Bianchi:2022ppi,Barrat:2022psm} proved very efficient in predicting defect CFT data in holographic setups \cite{Barrat:2021yvp,Gimenez-Grau:2023fcy} and the $\veps$--expansion \cite{Gimenez-Grau:2021wiv,Bianchi:2022sbz,Gimenez-Grau:2022ebb}.

Motivated by the success of the conformal bootstrap for homogeneous critical models, it is natural to ask which predictions we can make for statistical models in the presence of defects. Focusing on the $O(N)$ critical model, there is a variety of extended excitations one can introduce, including line defects \cite{Vojta_2000,sachdev_quantum_1999,Sachdev_2001,Sachdev_2003,Liu_2021,Cuomo:2021kfm,Cuomo:2022xgw,Rodriguez-Gomez:2022gbz,Gimenez-Grau:2022czc,Bianchi:2022sbz,Gimenez-Grau:2022ebb,Rodriguez-Gomez:2022xwm,Rodriguez-Gomez:2022gif,Nishioka:2022qmj,Pannell:2023pwz}, boundaries \cite{Liendo:2012hy,Bissi:2018mcq,ParisenToldin:2020gpb,Metlitski:2020cqy,Toldin:2021kun,Padayasi:2021sik,Nishioka:2022odm} and surface operators \cite{Deng:2005dh,Krishnan:2023cff,Trepanier:2023tvb,Raviv-Moshe:2023yvq,Giombi:2023dqs}. In this paper we focus on the former and, more specifically, on a class of line defects commonly denoted as magnetic impurities (analytic bootstrap studies of another line defect, the localized magnetic field, are available in \cite{Bianchi:2022sbz,Gimenez-Grau:2022ebb}). Magnetic impurities were originally introduced in \cite{Vojta_2000,sachdev_quantum_1999} with the aim of modeling a doped two-dimensional anti-ferromagnet at the quantum critical point. More recently, there has been a renewed interest for these defects. In \cite{Liu_2021}, they were analyzed at large $N$ with the long-term goal of understanding the interplay between symmetry protected topological phases and quantum criticality, whereas \cite{Cuomo:2022xgw} found a semiclassical description for this defect in the limit of large spin. In \cite{Beccaria:2022bcr}, it was also noted how these defects emerge in a specific scaling limit of superconformal Wilson lines in $\mathcal{N}=4$ SYM theory.

The universality class of the $O(N)$ critical model is well described by a theory of $N$ massless scalar fields with a quartic interaction tuned at the Wilson-Fisher fixed point. In this context, magnetic impurities admit several different microscopic realizations, such as a path-ordered exponential or a one-dimensional fermionic theory coupled to the scalar fields in the bulk. A natural question is whether these microscopic realizations would lead to a non-trivial conformal defect also when the bulk is free, \emph{i.e.} at the Gaussian fixed point with the quartic coupling tuned to zero.\footnote{The more general question of whether a non-trivial conformal defect can exist in a free theory based on general consistency conditions has been investigated in \cite{Lauria:2020emq}. Strictly speaking that analysis is valid for a single scalar field in integer number of dimensions. We expect their results to apply also for $N$ free scalars with a defect that preserves the $O(N)$ symmetry. However, the reason that we find a non-trivial defect is because we work in non-integer number of dimensions, a situation not addressed in \cite{Lauria:2020emq}.} This question has already been raised in \cite{Cuomo:2022xgw}, where the combination of large spin and $\veps$-expansion techniques suggested that a non-trivial fixed point exists for some regime of the parameters. Even though the former analysis hints that this region does not include $\veps=1$, \emph{i.e.} the three-dimensional bulk, this point is outside the radius of convergence of the large spin and $\veps$ expansions. One way to tackle this question could be working directly in $d=3$ and using numerical techniques such as Monte Carlo simulations, numerical bootstrap, or the recently proposed fuzzy-sphere regularization \cite{Zhu:2022gjc,Hu:2023ghk}. Instead, in this work we show that analytic bootstrap techniques can be used to prove that at $\veps=1$ and for any finite value of the spin, no non-trivial conformal defect exists when the bulk is free.

A first step towards a numerical bootstrap analysis of line defects in the $O(N)$ model has been taken in \cite{Gimenez-Grau:2022czc}. As we mentioned, the natural observable for the numerical methods is the defect four-point function. The result are exclusion plots in the space of dCFT data. Since the numerical bootstrap philosophy is purely based on imposing symmetries and consistency conditions, one of the obstacles when dealing with purely defect observable is to impose the fact that we are dealing with a defect correlator and not with a one-dimensional homogeneous CFT. Furthermore, to isolate a specific conformal defect (as it has been done, for instance, for the Ising model in the homogeneous case) one needs to input additional assumption on the spectrum of light operators. For this reason, in this paper we carry out a careful analysis and classification of the low-dimensional defect operators that characterize the magnetic impurity.
  
The most interesting case for condensed matter applications is certainly when the bulk is interacting. In that case, the beta function of the defect coupling will depend both on the defect and on the bulk coupling. The idea is to tune the bulk coupling to the critical value, and then find a defect fixed point by solving for zeroes of the defect beta function. Contrary to the localized magnetic field case, where the resulting defect coupling is not small for small values of $\veps$, in this case the critical defect coupling is proportional to $\veps$. This will allow us to compute dCFT data up to order $\veps^2$. Let us summarize our findings.

\subsection*{Summary of the results}
The goal of this work is to investigate the details of the low-lying defect spectrum and apply analytic bootstrap techniques to the bulk two-point function in the presence of magnetic impurities. Among the various explicit realizations of the magnetic impurity, we choose to work with the path-ordered exponential (see \eqref{defect} for the precise definition). This specific realization allows to preserve the full global symmetry algebra only for $N=3$, therefore this is the case we are going to consider in this paper. The extension to general $N$, through different realizations of the defect (see appendix \ref{app:representations}), is certainly an interesting future direction.

We first consider the case where the bulk theory is free and we compute the beta function for the defect coupling $\zeta$ up to order $\zeta^7$, finding for the first time an explicit dependence on the spin $j$
\begin{align}
 \beta_{\zeta}=\mu \frac{\partial \zeta}{\partial \mu}
 =
 - \frac{\veps}{2} \, \zeta
 + \zeta^3
 - \zeta^5
 + \left( 2 - \pi^2 \left(j(j+1)-\frac{1}{3}\right)\! \right) \zeta^7
 + \ldots
\end{align}
This result extends the large spin result of \cite{Cuomo:2022xgw} and the large $N$ result of \cite{Beccaria:2022bcr} for the ladder approximation of Wilson lines in $\mathcal{N}=4$ SYM. In that case, taking a specific limit of the supersymmetric Wilson line in $\mathcal{N}=4$ SYM the bulk theory essentially reduces to a free scalar theory with fields in the adjoint representation of $\mathfrak{su}(N)$ and the Wilson line is identical to the defect we consider here. The authors of \cite{Beccaria:2022bcr} considered this setup in the planar limit, but the construction works at any value of $N$ and our result is the non-planar extension for the specific case of $\mathfrak{su}(2)$.

Then we move to the analysis of some important operators in the defect spectrum and we focus in particular on their explicit realization in the path ordered exponential picture. It is a known fact that magnetic impurities contain a particular operator $\hat S^a$, called spin operator, which transforms in the adjoint representation of $\mathfrak{su}(2)$ and whose dimension, for the case of a free bulk, is fixed to $\hat{\Delta}_{\hat S}=\veps/2$. For an interacting bulk, the dimension receives corrections starting at $O(\veps^2)$. It is interesting to understand how this operator is realized inside the path ordered exponential. Its dimension vanishes for $\veps \to 0$ therefore it cannot be constructed using the bulk fields and their derivatives. It turns out that the spin operator is realized by inserting a generator $T^a$ inside the path ordered exponential (see \eqref{correlatorS} for the exact prescription). Although $T^a$ is a constant matrix, its insertion inside the trace produces a non-trivial dependence on the insertion point thanks to the path ordering prescription. We show how to perform perturbative computations using this prescription, confirming the expectations for its scaling dimension and computing its two-point function, whose value is physical because the normalization of the operator is fixed by a Ward identity.

Other interesting operators in the defect spectrum include the displacement operator, associated to the explicit breaking of translation invariance, and the operator $\hat{\Phi}=\phi^a T_a$, whose dimension is completely determined by the beta function of the defect coupling (see \eqref{DeltaPhi}). In Table \ref{table:1} and \ref{table:2} we provide a list of the lowest twist spectrum of the defect theory, specifying their quantum numbers and their scaling dimensions in perturbation theory.

In Section \ref{sec:bootstrap}, we consider the bulk two-point functions of $\phi^a$ and $\phi^2$ and we compute them using analytic bootstrap methods. Interestingly, for the case of a free bulk, the form of the correlator $\braket{\phi^a \phi ^b}$ is completely fixed at all orders in $\veps$ and we are left with a single unknown ($\veps$-dependent) constant $c_{\phi^2}$ which is essentially the one-point function of $\phi^2$. Evaluating this correlator at $\veps=1$ and assuming that $c_{\phi^2}\big|_{\veps=1}\neq 0$, we obtain a correlator that cannot satisfy the defect bootstrap equations. The inevitable conclusion is that either $c_{\phi^2}\big|_{\veps=1}= 0$ yielding a trivial defect, or that there is no fixed point defect CFT.

In the interacting case, instead, we can compute the correlator up to order $\veps^2$ finding
\begin{align} \label{resultcorrelator}
 F(r,w) 
 = \xi^{-\Delta_\phi}
 +c_{\phi^2} \left(1+ 
   \frac{\veps}{2} \log \frac{4 r}{(1 + r)^2}
   + \frac{5 \veps}{11} \big(1+ \log 2+ H(r, w) \big)
 \right)
 + O(\veps^3)\,,
\end{align}
where $r$ and $w$ are two conformal cross-ratios \eqref{coordinates} and $\xi$ a combination thereof \eqref{xi}.  The function $F(r,w)$ is the correlator up to a kinematical prefactor (see \eqref{correlatorswithprefactor}), whereas $H(r,w)$ is the most complicated part of the result and it is the same function that already appeared in the case of the localized magnetic field in \cite{Gimenez-Grau:2022ebb, Bianchi:2022sbz}. In other words, at this order in perturbation theory, the results for the two line defects (localized magnetic field and magnetic impurity) have the same structural form.
The only major difference is the presence of an extra contribution for the magnetic impurity, namely the logarithm in \eqref{resultcorrelator}, which is necessary to account for the $\hat S^a$ operator in the defect OPE.
Because this logarithm is independent of $w$, it only contributes to spin $s=0$ operators in the defect OPE, and therefore it corresponds to a low-spin ambiguity  that cannot be reconstructed from the dispersion relation.
Expanding \eqref{resultcorrelator} in conformal blocks, it is possible to extract an infinite set of defect CFT data, up to approximate degeneracies. Our results for the defect scaling dimensions and for the bulk to defect couplings $b_{\phi \hat{\Om}}$ are presented in \eqref{dimdefspin}--\eqref{bphiOs}. These results are confirmed by diagrammatic computations in Section \ref{sec:diagrams}, where we also compute the correlator $\braket{\phi^2\phi^2}$ in the interacting theory, which is hard to extract from a bootstrap analysis.

Our results for the two-point function of $\hat{S}^a$ and its anomalous dimension had already been computed in $\veps$-expansion in \cite{sachdev_quantum_1999, Vojta_2000} and can be compared to the Monte Carlo analysis of \cite{H_glund_2007}. The agreement between the two methods is not very good. This is expected because in order to obtain sensible results for $\veps=1$ one has to first resum the perturbative expansion, see e.g. \cite{LeGuillou:1979ixc}. All our other results are new and, to our knowledge, have never been computed neither using $\veps$-expansion nor Monte Carlo.

\section{Magnetic impurities}
Our starting point is the $O(3)$ critical model at the Wilson-Fisher fixed point, which is characterized by the quartic action
\begin{equation}
S=\int d^d x \left[ \frac{1}{2}\left( \partial_\mu \phi_a \right)^2+\frac{\lambda_0}{4!}\left(\phi^a \phi_a\right)^2 \right],
\end{equation}
where $\lambda_0$ is the bare coupling and the fundamental fields $\phi_a$ transform in the adjoint representation of $\mathfrak{su}(2)$.
For $d=4-\veps<4$, the coupling $\lambda$ is relevant and it triggers a renormalization group (RG) flow, which can be studied perturbatively in $\veps$. The $\beta$-function for the quartic coupling reads
\begin{equation}
\beta_{\lambda}=\frac{\partial \lambda}
{\partial \log \mu}=- \veps \lambda +\frac{11 }{48 \pi ^2} \lambda ^2 -\frac{23 }{768 \pi ^4} \lambda ^3 + O(\lambda^4)\,,
\end{equation}
and a non-trivial fixed point is reached for 
\begin{equation}
\lambda_* =\frac{48 \pi^2}{11}\veps+\frac{3312 \pi ^2}{1331} \veps^2 +\mathcal{O}(\veps^3)\,.
\end{equation}
This solution is perturbative in $\veps$ and this is the regime we are going to be interested in for this work.

Following \cite{Cuomo:2022xgw}, the line defect is represented by an extended operator given by the trace of the following path-ordered exponential
\begin{equation}\label{defect}
 \mathcal{D}_j(u,v) = \Pexp\left(\frac{\zeta_0}{\sqrt{\kappa}} \int_{u}^{v} d\tau \, \phi^a(\tau) T_a\right) \,,
\end{equation}
where the factor
\begin{equation}\label{eq:kappa}
 \kappa
 = \frac{\Gamma\!\left(\frac d2\right)}{2\pi^{d/2} (d-2)}\,,
\end{equation}
has been introduced for future convenience.
Throughout the paper, we choose the defect to extend along the imaginary time direction,\footnote{Another interesting observable is the circular loop, which is monotonic under RG flow \cite{Cuomo:2021rkm}. The two configurations are mapped into each other by a conformal transformation and, although at the level of the defect expectation value there might be conformal anomalies \cite{Drukker:2000rr}, all our conclusions about defect correlators can be easily adapted to the circular case. The straight line has a more direct interpretation as an impurity in a condensed-matter system.} and for compactness we write $\phi_a(\tau) \equiv \phi_a(\tau,0,\ldots,0)$.
We will be mostly interested in the infinite defect $\Dm_j \equiv \Dm_j(-\infty,\infty)$, but occasionally it will be useful to work with the finite version \eqref{defect}.

The matrices $T^a$ form a spin-$j$ representation of $\mathfrak{su}(2)$, or equivalently, they are $(2j+1)\times(2j+1)$ matrices.
We normalize them such that the commutation relations and Casimir read
\begin{align}
 [T^a, T^b] = i \epsilon^{abc} T^c \, , \qquad
 T_a T_a = j(j+1) \, .
 \label{eq:comm-rels}
\end{align}
The defect $\Tr \,\Dm_j$ preserves the connected component of the $O(3)$ global symmetry \footnote{Strictly speaking, in the definition of the defect we take the trace over the representations of $SU(2)$ instead of $O(3)$, in order to allow for half-integer values of $j$. At the level of the algebra this makes no difference. We will ignore this subtlety in what follows, since it does not affect our results.}, and as a result, it can be realized in the lattice by inserting a spin-$j$ impurity interacting with other lattice sites through $SU(2)$-preserving interactions.
The coupling $\zeta_0$ is marginally irrelevant in four dimensions, but it is relevant for $d<4$ and the system flows to a non-trivial interacting dCFT in the infrared (IR).
Interestingly, this dCFT is non-trivial even when the bulk is tuned to the free-theory point $\lambda_0=0$.
To see this explicitly, in section \ref{sec:betafun} we summarize the computation of the $\beta$-function, distinguishing the free and the interacting bulk cases. For the free case, we present a new result at order $\zeta^7$ which shows for the first time the dependence of the beta function on the spin $j$.
We also discuss correlation functions in section \ref{corrfuns}, and the discrete symmetries preserved by the defect in section \ref{timereversal}.

\subsection{Defect \texorpdfstring{$\beta$}{beta}-function}
\label{sec:betafun}

The computation of the $\beta$-function for line defects goes back to the work of \cite{Dotsenko:1979wb} for non-abelian gauge theories (see also \cite{Beccaria:2021rmj} for a recent review). For the magnetic impurities of interest here, the $\beta$-function has been computed, up to two loops, in \cite{PhysRevB.61.4041,Sachdev:2001ky}. The idea, as usual, is to select a specific observable and renormalize the coupling $\zeta_0$ such that the result is UV-finite. Since for the purposes of renormalization we are interested in the UV-behaviour of the theory, we can consider a finite line $\tau \in [u,v]$.
Then we impose UV-finiteness of the vertex operator
\begin{align}\label{eq:vertex}
 \Vm(x)
 = \frac{\tr \big\langle \phi_a(x) T^a \, \Dm_j(u,v) \big\rangle}
        {\tr \big\langle \Dm_j(u,v) \big\rangle } \, ,
\end{align}
where $\phi_a(x) T^a$ is inserted in the trace, but it is placed in a point $x$ in the bulk.

\subsubsection{Free bulk}

Let us start from the case of a free bulk, where the bulk operator $\phi_a(x)$ is not renormalized. In this case, all the divergences in \eqref{eq:vertex} can be ascribed to the renormalization of the coupling $\zeta_0$. The details of the computation are spelled out in Appendix \ref{app:detailsbeta}. Here we report only the final result, which we computed for the first time up to order $\zeta^7$ in the minimal subtraction (MS) scheme. The relation between bare and renormalized coupling reads
\begin{align}
 \zeta_0
 = \mu ^{\veps/2} \zeta \left(
   1
 + \frac{\zeta ^2}{\veps}
 - \frac{\zeta ^4}{2 \veps}
 + \frac{3 \zeta ^4}{2 \veps^2}
 + \frac{\zeta^6}{3 \veps}
   \left(2 - \pi ^2 \left(j(j+1)-\frac{1}{3}\right)\right)
 - \frac{11 \zeta ^6}{6 \veps^2}
 + \frac{5 \zeta ^6}{2 \veps^3}
 + O(\zeta^8)
 \right) \, .
 \label{eq:zeta0freeblk}
\end{align}
From the usual condition that the bare coupling does not depend on the renormalization scale $\mu$, namely $d\zeta_0/d\mu = 0$, one extracts the beta function
\begin{align}\label{eq:betafreebulk}
 \beta_{\zeta}
 =
 - \frac{\veps}{2} \, \zeta
 + \zeta^3
 - \zeta^5
 + \left( 2 - \pi^2 \left(j(j+1)-\frac{1}{3}\right)\! \right) \zeta^7
 + \ldots
\end{align}
Observe the interesting fact that, starting at $O(\zeta^7)$, the $\beta$-function depends on the spin $j$. This presents an obstacle towards the resummation of the perturbative series, even for the case of a free bulk. However, when the spin $j$ is large, one can take a double-scaling limit where $\zeta\to0$, $j\to \infty$ and $\zeta^2 j$ is kept fixed. The $\beta$-function in this limit was recently computed in \cite{Cuomo:2022xgw}, and our result is in perfect agreement for large $j$. The fixed point equation $\beta(\zeta)=0$ can be solved perturbatively in $\veps$ leading to a defect fixed point for
\begin{equation}\label{fixedcouplingfree}
 \zeta_*^2=\frac{\veps}{2}+\frac{\veps^2}{4}+\left(j(j+1)-\frac13\right)\frac{\pi^2 \veps^3}{8}+O(\veps^4)\,.
\end{equation}
Notice that the defect coupling is irrelevant in four dimensions, leading to a trivial fixed point at $\veps = 0$. For $\veps>0$ there is a non-trivial fixed point, even though the bulk is free. The existence of this fixed point in three dimensions, \emph{i.e.}\ for $\veps\to1$, was questioned in \cite{Cuomo:2022xgw}, based on a large spin analysis. In Section \ref{sec:bootstrap} we will prove that indeed this fixed point is trivial.

\subsubsection{Interacting bulk}

For an interacting bulk, it is harder to push the calculation at higher orders in perturbation theory due to the presence of diagrams with quartic bulk interactions. In the case of the interacting bulk, we find (see appendix \ref{interactingbetaapp})
\begin{equation}\label{zetaren}
\zeta_0=\mu^{\veps/2}\zeta \left(
   1
 + \frac{\zeta ^2}{\veps}
 - \frac{\zeta ^4}{2 \veps}
 + \frac{3 \zeta ^4}{2 \veps^2}+\frac{5\,  \lambda^2}{72(4\pi)^4 \veps}
 + \frac{\left(j(j+1)-\frac{1}{3}\right) \zeta^2 \lambda}{48 \veps}
 + \ldots
 \right) \, .
\end{equation}
From this we can exrtract the $\beta$-function \cite{Vojta_2000, sachdev_quantum_1999, Sachdev_2001}
\begin{equation}\label{interactingbeta}
 \beta_{\zeta}=- \frac{\veps}{2} \, \zeta
 + \zeta^3
 - \zeta^5
 +\frac{5}{36}\frac{\zeta \lambda^2}{(4\pi)^4}
 +  \left(j(j+1)-\frac{1}{3}\right) \frac{ \zeta^3\lambda}{24}
 + \ldots
\end{equation}
After setting the bulk coupling to the fixed-point value $\lambda_*$, one can solve perturbatively the equation $\beta_{\zeta}(\zeta_*,\lambda_*)=0$, finding
\begin{equation}\label{fixedcouplingint}
 \zeta_*^2
= \frac{\veps}{2}
 + \veps^2 \left( \frac{29}{121} - \frac{\pi^2}{11}\left(j(j+1)-\frac{1}{3}\right) \right)
 + O(\veps^3) \, .
\end{equation}
Notice that in the interacting theory the dependence on $j$ shows up already at order $\veps^2$.

When the bulk and the defect coupling are tuned to their fixed point value, we obtain an interacting defect conformal field theory. While the bulk spectrum is clearly unaffected by the presence of the defect, it is interesting to understand in some depth how to characterize the defect operators. This will be the subject of section \ref{sec:defectop}.

\subsection{Correlation functions}
\label{corrfuns}

In equation \eqref{defect}, we introduced the line defect in terms of the path-ordering operation.
Explicitly, this amounts to the definition
\begin{equation}\label{explicitdefect}
\begin{split}
 \Dm_j(u,v)
 & = \sum_{n=0}^{\infty} \, \frac{\zeta_0^n}{\kappa^{\frac{n}{2}}} \int_{u<\tau_1<\dots<\tau_n<v} \!\!\!\!\!\!\!\!\!\!\!\!\!\!\!\!\!\!\!\!\!\!\!\!\!\!\!\!\! d \tau_1 \dots d \tau_n \
 \phi_{a_1}(\tau_1) \dots \phi_{a_n}(\tau_n) \,
 T^{a_1} \dots T^{a_n} \,.
\end{split}
\end{equation}
In particular, \eqref{explicitdefect} allows to map correlators of the defect theory to those of the homogeneous theory.
The simplest correlation functions in the presence of $\Dm_j$ involve insertions of bulk operators as follows
\begin{equation}\label{defectcorrelators}
\langle \mathcal{O}_1(x_1)\dots \mathcal{O}_n (x_n) \rangle_{\Dm_j} = \frac{\langle \mathcal{O}_1(x_1)\dots \mathcal{O}_n (x_n) \tr \Dm_j  \rangle}{\langle \, \tr \Dm_j \rangle}\,,
\end{equation}
where recall that $\Dm_j = \Dm_j(-\infty,\infty)$ is the infinite length defect.

However, this is not the most general possibility, and in fact one can define operators $\hat\Om(\tau)$ that live on top of the defect.
Throughout this work, we follow the convention of dressing defect operators with a hat in order to distinguish them from bulk operators. Moreover, their argument will be just the one-dimensional coordinate along the defect.
As it will become evident in section \ref{sec:defectop}, defect operators of this model can in principle be matrix-valued (with matrices having the same dimension of the generators $T^a$).
Therefore, for  the most general correlators of defect operators, we have to slice-up the path-ordering as follows
\begin{equation}\label{defectcorrelators2}
\langle \hat{\Om}_1(\tau_1)\dots \hat{\Om}_n (\tau_n) \rangle_{\Dm_j}
= \frac{\langle \tr \big[ \Dm_j(-\infty,\tau_1) \hat\Om(\tau_1) \Dm_j(\tau_1,\tau_2) \hat\Om(\tau_2) \ldots \hat\Om(\tau_n) \Dm_j(\tau_n,\infty) \big] \rangle}{\langle \, \tr \Dm_j \rangle} \, .
\end{equation}
Note that defect operators are not necessarily matrix-valued. 
For example, with a single fundamental field $\phi$ we can build two distinct operators
\begin{align}
 \hat \Om_1^a(\tau) = \phi^a(\tau,0,\ldots) \, , \qquad
 \hat \Om_2(\tau) = \phi^a(\tau,0,\ldots) T_a \, .
\end{align}
When we insert $\hat \Om_1^a$ in correlation functions we can factor it outside the trace, while $\hat \Om_2$ is a matrix and therefore it interacts non-trivially with the trace.
Many examples of defect operators are provided in section \ref{sec:defectop}.

An important point when dealing with defect correlators, is that they depend on the coordinate of a defect operator $\hat{\Om}(\tau)$ also trough the endpoints of the neighbouring defect operators $\Dm_j(\cdot,\tau)$ and $\Dm_j(\tau,\cdot)$, as it is clear by looking at the right hand side of \eqref{defectcorrelators2}.
For this reason, it is convenient to introduce a defect covariant derivative
\begin{equation}\label{Covariantderivativedef}
\Dm_j (u, \tau) \, D_\tau \hat{\Om} (\tau)\, \Dm_j (\tau, v) \equiv \frac{d}{d\tau} \big( \Dm_j (u, \tau) \,\hat{\Om} (\tau)\, \Dm_j (\tau, v)\big),
\end{equation}
From this one readily finds
\begin{equation}\label{Covariantderivative}
D_\tau \hat{\Om}(\tau) = \partial_\tau \hat{\Om}(\tau)+ \frac{\zeta_0}{\sqrt{\kappa}} \phi^a (\tau) \big[ T_a, \hat{\Om}(\tau)\big]\,.
\end{equation}
This covariant derivative is really analogous to the one introduced in the case of Wilson lines (see \emph{e.g.} \cite{defectABJM}).

Finally, it is possible to consider correlators that contain a mix of bulk and defect operators, and their correlators are given by the obvious generalization of \eqref{defectcorrelators} and \eqref{defectcorrelators2}.

\subsection{Discrete symmetries}
\label{timereversal}

It is interesting to look at which discrete symmetries are preserved by the defect, because they imply selection rules in correlation functions, and they also help in the classification of defect operators.
The bulk theory, both in the free and in the interacting case, is invariant under time reversal symmetry\footnote{
In the context of defects, the inversion of the defect coordinate is also known as $\mathcal{S}$-parity \cite{Gaiotto:2013nva, Billo:2013jda}.
} and under a global $\mathbb{Z}_2$ symmetry
\begin{align}
 T_{t}:\phi^a(\tau,x_\perp) \mapsto \phi^a(-\tau,x_\perp) \, , \qquad
 T_{\mathbb{Z}_2}:\phi^a(\tau,x_\perp) \mapsto -\phi^a(\tau,x_\perp) \, .
\end{align}
By a symmetry of the defect theory, one usually means that the correlators in the left hand side of \eqref{defectcorrelators} are invariant under the action of the symmetry generators.
 A sufficient condition for this to happen is that the generator of a symmetry of the homogeneous theory also leaves invariant the trace of the defect operator $\Tr \, \Dm_j$.
 This is exactly what happens to the $SU(2)$ global symmetry.
  On the contrary, the generators $T_{\mathbb{Z}_2}$ and $T_{t}$ clearly modify the defect. 
It is straightforward to see that the net effect of $T_{\mathbb{Z}_2}$ is to change the sign of the defect coupling constant \cite{Liu_2021}
 \begin{equation}
 T_{\mathbb{Z}_2}  \, \Dm_j^{\,\zeta}=   \Dm_j^{-\zeta}\,,
 \end{equation}
 where $ \Dm_j^{\,\zeta}$ is the defect  extended operator with coupling constant $\zeta$.
 On the other hand, $T_{t}$ flips the signs of the arguments of all the fields in \eqref{explicitdefect}.
 However, by a convenient change of integration variables and name redefinitions, this is equivalent to reversing the order of the generators inside the trace. 
 For generators of representations of $\mathfrak{su}(2)$, the following relation holds\footnote{
 This is due to the facts that the generators $T^{a}$ are taken to be Hermitean and that the  $\mathfrak{su}(2)$ representation given by the complex conjugated generators $(T^{a})^*$ is equivalent to the original one, so that $(T^{a})^T=P \,T^{a} \,P^{-1}$ for some matrix $P$.
 }
\begin{equation}
\Tr \left(T^{a_n} \dots   T^{a_1} \right) = (-1)^n\, \Tr \left(T^{a_1} \dots   T^{a_n} \right)\,.
\end{equation}
From this it follows that also $T_{t}$ is tantamount to a change in the sign of the defect coupling constant 
 \begin{equation}
 T_{t} \,  \tr \Dm_j^{\,\zeta} = \tr \Dm_j^{-\zeta}\,.
 \end{equation} 
At this point we can define a modified time reversal symmetry for the defect theory by asking that the fundamental fields are odd under this symmetry
\begin{equation}
\bar{T}_t=T_{\mathbb{Z}_2} \circ T_{t}:\phi^a(\tau,x_\perp) \mapsto -\phi^a(-\tau,x_\perp)\,.
\end{equation}
Now $\bar{T}_t$ is both a symmetry of the homogeneous theory and leaves $\tr \Dm_j^{\,\zeta}$ invariant (it changes the sign of $\zeta$ two times).
Therefore, it is a symmetry of the defect theory as well. 
To get some useful selection rules, we need to understand how this symmetry acts on defect operators. This will be briefly discussed in section  \ref{timerevdef}, after we have acquired a general understanding of the defect operators of this model.

\section{The defect spectrum}\label{sec:defectop}

In this section we study the spectrum of operators that live on top of the magnetic impurity defect.
Our motivation is that, to efficiently apply bootstrap techniques, it is helpful to know what defect operators can contribute to different OPE decompositions.
It is useful to start from the free-bulk theory, because the spectrum is simpler and Ward identities protect several defect operators.
When the bulk interaction is turned on in the $\veps$-expansion, the dimension of these operators will get modified by additional terms proportional to powers of $\lambda_*$, which is perturbatively small.
Our approach allows us to understand the perturbative definition of these operators, which is surprisingly non-trivial in certain cases. This in turn will shed light on how to list all the possible defect operators in perturbation theory.

\subsection{The defect spin operator}\label{defectspinoperator}

As pointed out in \cite{Cuomo:2022xgw}, an interesting Ward identity is obtained by considering the shift of the fields $\phi_a(x) \rightarrow \phi_a(x) + c_a$ for some constants $c_a$. 
This is a symmetry of the free-bulk theory without the defect.
The Noether currents for these symmetries are $J^\mu_a(x)=-\partial^\mu \phi_a(x)$, and their conservation is equivalent to the equations of motion since $0=\partial_\mu J^\mu_a(x) = -\Box \phi_a(x)$.
The defect interaction breaks explicitly the shift symmetry, so the conservation equation is modified by a term localized on the defect
\begin{equation}\label{ward1}
\partial_\mu J^{\mu\,a}(0,x_\perp)=-\frac{\zeta_0}{\sqrt{\kappa}}\,\hat{S}_0^a(0)\, \delta^{d-1}(x_\perp)\,,
\end{equation}
where the minus sign is introduced for future convenience. 
In this section and in the following ones we will often assume that the coordinate parallel to the defect of bulk operators is zero thanks to translational invariance along the defect.
Note that the bulk fundamental fields $\phi_a$ do not renormalize since the bulk is free. 
If we introduce the renormalization factors such that $\hat{S}_0^a= Z_{\hat{S}} \, \hat{S}^a$ and $\zeta_0=\mu^\frac{\veps}{2}Z_{\zeta}\,\zeta$, then it follows that in the MS scheme
\begin{equation}\label{Z}
Z_{\hat{S}}=Z_{\zeta}^{-1}\,,
\end{equation} 
at all orders in perturbation theory, since the right hand side of \eqref{ward1} must be finite. 
In particular, \eqref{ward1} holds also if we substitute renormalized quantities instead of the bare ones.\footnote{More precisely, for renormalized quantities we would have $\partial_\mu J^{\mu\,a}(0,x_\perp)=-\frac{\mu^{\frac{\veps}{2}}\zeta}{\sqrt{\kappa}}\,\hat{S}^a(0)\, \delta^{d-1}(x_\perp)$. We will often forget about the scale factor $\mu$ and set it to one, as is customary in the CFT literature, because we are ultimately interested in correlation functions at the fixed point and they depend on $\mu$ in a trivial way.}
The operator $\hat{S}_a$ responsible for the symmetry breaking is a defect primary operator at the fixed point, and we shall call it the \emph{defect spin} operator.
As argued in \cite{Cuomo:2022xgw}, the above Ward identities protect their dimension to $\hat{\Delta}_{\hat{S}}=\veps / 2$.\footnote{
It is also possible to derive this result using diagrammatic considerations, as is it was originally done in \cite{Vojta_2000}.
}
The explicit form of the defect spin operator $\hat{S}_a$ in the perturbative setup can be derived via the Schwinger-Dyson equations.
To do this, it is convenient to think of the defect as contributing an extra term to the full action $S = S_{\text{bulk}} + S_{\text{defect}}$, where
\begin{equation}\label{defectaction}
S_{\text{defect}}=- \log \Tr \, \Dm_j\,.
\end{equation}
Inside correlation functions it must hold
\begin{equation}\label{defectspinexplicit}
\Box \phi_a(\tau,x_\perp) = \frac{\delta S_{\text{defect}}}{\delta \phi_a(\tau,x_\perp)}=-\frac{\zeta_0}{\sqrt{\kappa}}\,\delta^{d-1}(x_\perp) \frac{\Tr  \big( \Dm_j (-\infty, \tau) \, T_a \, \Dm_j (\tau, \infty)\big)}{\Tr\ \Dm_j }\,.
\end{equation}
Therefore, comparing with \eqref{ward1} one finds that correlators involving a defect spin operator $\hat{S}_0^a(\tau)$ inserted at a point $\tau$ lying on the defect satisfy
\begin{equation}\label{correlatorS}
\langle \mathcal{O}_1(x_1)\dots \hat{S}_0^a(\tau)\dots\mathcal{O}_n(x_n)\rangle_{\Dm_j} =-\langle \mathcal{O}_1(x_1)\dots T^a(\tau)\dots\mathcal{O}_n(x_n)\rangle_{\Dm_j}\,,
\end{equation}
where  the right hand side has to be interpreted in the same sense as \eqref{defectcorrelators2}.
In the following, we will simply write
\begin{equation}\label{STdef}
\hat{S}_a(\tau)=-Z_{\hat{S}}^{-1} \, T_a(\tau)\,.
\end{equation}
In this sense, the $\hat{S}_a$ operators in perturbation theory are just normal matrices which acquire an anomalous dimension once they are inserted into the defect.\footnote{Similar non trivial constant operators have already appeared in the literature, see for example \cite{Gorini:2022jws, Drukker:2022txy}. In our case, this unfamiliar situation could be avoided by considering an equivalent representation of the defect in terms of one dimensional fermions, such as \eqref{fermionicrep}. From that point of view $\hat{S}_a$ can be realized as a regular fermion bilinear operator.}

Another interesting consequence of \eqref{ward1} is that we can rewrite it as
\begin{equation}\label{boxphifree}
\Box \phi_a(0, x_\perp) = \frac{\zeta}{\sqrt{\kappa}} \, \hat{S}_a(0) \, \delta^{d-1}(x_\perp)\,,
\end{equation}
and this equation can be easily inverted 
\begin{equation}\label{fielddec}
\phi_a(0, x_\perp)=  \sqrt{\kappa}\,\zeta \int d\tau \, \frac{\hat{S}_a(\tau)}{\left( |x_\perp|^2+\tau^2 \right)^{1-\frac{\veps}{2}}}  + \phi_a^{\text{\,free}}(0, x_\perp)\,,
\end{equation}
where without loss of generality we assumed the defect coordinate of $\phi_a$ to be zero and $\phi_a^{\text{\,free}}$ is just a free field which does not interact with the defect. 
In particular, correlators involving fundamental fields and its orthogonal derivatives (both in the bulk and on the defect) can be reduced to defect integrals of correlators involving $\hat{S}_a$ (not necessarily at the fixed point), as it will be shown in sections \ref{corrdefectspin} and \ref{corrhatphi}. 

It is important to understand what are the conformal descendants of the operator $\hat{S}_a$ at the fixed point.
Such descendants are obtained by acting with the defect covariant derivative defined in \eqref{Covariantderivative}.
In the case of the $\hat{S}_a$ operator, we get
\begin{equation}\label{Sdescendant}
D_\tau \hat{S}^a (\tau) = -i \frac{\zeta_0}{\sqrt{\kappa}} \epsilon^{abc} \phi_b T_c (\tau)\,,
\end{equation}
where the generator on the right hand side has to be inserted inside the path ordering, similarly to \eqref{correlatorS}.
This example shows that in this setup the question of whether an operator is a primary or not can be hard to address, because even though \eqref{Sdescendant} contains no $\partial_\tau$ derivatives it is in fact a descendant.

Once the bulk quartic interaction is turned on, the shift symmetry is explicitly broken in the bulk, hence the above analysis does not apply. 
Nevertheless, it still makes sense to consider the $\hat{S}_a$ operators defined by \eqref{correlatorS}.
The dimension of these operators is no longer protected, and since it is classically vanishing, we have
\begin{equation}
\hat{\Delta}_{\hat{S}} =\left.\left( \beta_{\zeta}\frac{\partial \log Z_{\hat{S}}}{\partial \zeta}+\beta_\lambda \frac{\partial \log Z_{\hat{S}}}{\partial \lambda} \right)\right|_{\zeta_*,\lambda_*},
\end{equation}
where now $Z_{\hat{S}}$ depends also on the bulk coupling constant $\lambda$.
Interestingly, we observed that up to two loops in perturbation theory $Z_{\hat{S}}$ does not receive any divergent corrections from the bulk interaction.\footnote{
One way to see this is noting that diagrammatically no propagator can be attached to the operator $\hat{S}_a$, since in the definition \eqref{STdef} there are no fundamental fields. This implies that any contribution that involves both the defect and the bulk interaction either comes from a correction to bulk propagators and is at least of order $O(\zeta^2 \lambda^2)$, or has at least four internal legs attached to the defect and is at least of order $O(\zeta^4 \lambda)$.   }
Therefore, in the interacting case we can still write
\begin{equation}
Z_{\hat{S}}=\left(\left.Z_{\zeta}\right|_{\lambda=0}\right)^{-1}+O(\zeta^2 \lambda^2, \zeta^4 \lambda)\,,
\end{equation}
This is sufficient to compute the first correction to the dimension $\Delta_{\hat{S}}$ using only the result for the $\beta$-function in the interacting case, without having to do any further diagrammatic computation \cite{Vojta_2000}
\begin{equation}\label{deltaSint}
\hat{\Delta}_{\hat{S}} =\left.\beta_{\zeta}\frac{\partial \log Z_{\hat{S}}}{\partial \zeta}\right|_{\zeta_*,\lambda_*}\!\!\!\!+O(\veps^3)= \frac{\veps}{2}- \veps^2 \left[\frac{5}{484}+\frac{\pi ^2}{11}  \left(j (j+1)-\frac{1}{3}\right)\right]+ O(\veps^3) \,.
\end{equation}

\subsubsection{Correlators of defect spin operators in perturbation theory}\label{corrdefectspin}
Once the explicit form of the defect spin operator in perturbation theory is known, it is possible to evaluate correlators using standard diagrammatic techniques.
This section is devoted to the computation of the two-point function $\langle \hat{S}_a(\tau_1) \hat{S}_b(\tau_2) \rangle_{\Dm_j}$ at two loops, both in the free and interacting bulk cases.
The overall normalization of the two-point function in free theory has a physical meaning, since the normalization of $\hat S$ is fixed by its definition \eqref{ward1}, and in fact this normalization will be useful later.

Neglecting for the moment the renormalization factors, this two-point function is the expectation value of the defect with generators $T_a$ and $T_b$ inserted at $\tau_1$ and $\tau_2$, respectively. Since in \eqref{defectcorrelators2} one needs to divide by the defect expectation value, we can normalize traces by dividing by $2j+1$, which is the classical expectation value.
Moreover, it is convenient to define the ``connected part'' of a diagram as what remains after one subtracts all contributions that are products of lower order diagrams, or pieces that contain ``defect bubbles''.
Using this terminology, the defect correlator is the sum of all connected diagrams.

The leading order term is given by the following diagram
\begin{equation}
  	\begin{tikzpicture}[scale=0.5]
	 \draw[double,thick,blue] (-3,0)--(3,0);
	 \node[above] at (-1.5,0) {$\hat{S}_a$};
	 \node[below] at (-1.5,0) {$\tau_1$};
	 \fill[blue] (-1.5,0) circle (4pt);
	  \node[above] at (1.5,0) {$\hat{S}_b$};
	  \node[below] at (1.5,0) {$\tau_2$};
	  	 \fill[blue] (1.5,0) circle (4pt);
	\end{tikzpicture}
	\end{equation}
	where the blue line represents the defect and the blue points indicate that a generator has to be inserted into the trace.
	Since there are no lower order diagrams, this diagram is already connected, and it gives
	\begin{equation}
	I_c^{(0)}(\tau_1,\tau_2)= \frac{1}{2j+1}\Tr \left(T_a T_b \right) = \frac{j(j+1)}{3}\delta_{ab}.
	\end{equation}
	At one loop there are only two diagrams contributing to the connected term (all other diagrams exactly factor into an order zero diagram times a piece of a one-loop bubble and they have to be subtracted)
	\begin{equation}\label{oneloopSS1}
  	\begin{tikzpicture}[scale=0.5]
	 \draw[double,thick,blue] (-4,0)--(2.5,0);
	 \node[above] at (-1.5,0) {$\hat{S}_a$};
	 \node[below] at (-1.5,0) {$\tau_1$};
	  \draw[thick, black]    (-3,0) to[out=-90,in=-90] (0,0);
	 \fill[blue] (-1.5,0) circle (4pt);
	  \fill[blue] (0,0) circle (4pt);
	   \fill[blue] (-3,0) circle (4pt);
	  \node[above] at (1.5,0) {$\hat{S}_b$};
	  \node[below] at (1.5,0) {$\tau_2$};
	  	 \fill[blue] (1.5,0) circle (4pt);
	\end{tikzpicture}
	\hspace{1 cm}
	\begin{tikzpicture}[scale=0.5]
	\draw[thick, black]    (0,0) to[out=-90,in=-90] (3,0);
	 \draw[double,thick,blue] (-2.5,0)--(4,0);
	   \fill[blue] (0,0) circle (4pt);
	   \fill[blue] (3,0) circle (4pt);
	 \node[above] at (-1.5,0) {$\hat{S}_a$};
	 \node[below] at (-1.5,0) {$\tau_1$};
	  
	 \fill[blue] (-1.5,0) circle (4pt);
	  \node[above] at (1.5,0) {$\hat{S}_b$};
	  \node[below] at (1.5,0) {$\tau_2$};
	  	 \fill[blue] (1.5,0) circle (4pt);
	\end{tikzpicture}
	\end{equation}
	where the additional blue points indicate interactions with a generator insertion and the black line represents a free propagator.
	In $d=4-\veps$ dimensions the free propagator reads
	\begin{align}
	\left.\langle \phi_a(x_1) \phi_b(x_2) \rangle \right|_{\lambda=0}
	= \frac{\kappa \, \delta_{ab}}{|x_1-x_2|^{2-\veps}} \, ,
	\end{align}
	where $\kappa$ was defined in equation \eqref{eq:kappa}.
	The interactions have to be integrated along the defect, but without crossing any other generator insertions.
	These two diagrams have the same color factor, given by
	\begin{equation}
	I^{(1)} \sim \frac{1}{2j+1}\Tr \left( T_c T_a T_c T_b \right)=\frac{j(j+1)(j(j+1)-1)}{3}\, \delta_{ab}\,.
	\end{equation}
	From these diagrams, we still need to subtract the product of the order zero diagram times pieces of one-loop ``defect bubbles'', which have the same kinematical integral but color factor given by
	\begin{equation}
		I^{(0)}_c \times \text{bubbles}^{(1)} \sim \frac{1}{(2j+1)^2}\Tr \left( T_a T_b  \right) \, \Tr \left( T_c T_c \right)=\frac{j^2(j+1)^2}{3}\, \delta_{ab}\,.
	\end{equation}
	Therefore, we get
	\begin{equation}
	I_c^{(1)}(\tau_1,\tau_2)=-\zeta_0^2\frac{j(j+1)}{3}\delta_{ab}\left( \int\displaylimits_{\,-\infty<\tau<\tau_1<\tau'<\tau_2} \!\!\!\!\frac{d\tau \, d\tau'}{|\tau-\tau'|^{2-\veps}}\ \ +\!\!\!\!\!\!\int\displaylimits_{\tau_1<\tau<\tau_2<\tau'<+\infty} \!\!\!\!\frac{d\tau \, d\tau'}{|\tau-\tau'|^{2-\veps}}\right)\,.
	\end{equation}
	After performing the trivial integrals, one finds
	\begin{equation}\label{SSoneloop}
	I_c^{(1)}(\tau_1,\tau_2)=-\frac{2\,\zeta_0^2\,j(j+1)}{3(1-\veps)\veps }|\tau_1-\tau_2|^\veps \, \delta_{ab}\,.
	\end{equation}
	As expected, this contribution has a pole for $\veps\rightarrow 0$, since we are computing the bare two-point function. 
	At the next order there are many diagrams that contribute to this two-point function, but the computation goes on in a similar way, and it is carried out explicitly in the appendix \ref{twoptspinapp}. 
    It is interesting to note that the same diagrams contribute to the free and interacting bulk cases. The reason is that, at the order we are working, the only new diagram in the interacting case would be a mass correction to the bulk propagator, which is set to zero.
	Once all the diagrams are evaluated, one introduces the wavefunction renormalization coefficient $Z_{\hat{S}}$ and rewrite the bare coupling constant in term of the renormalized one, while keeping in mind that $Z_{\hat{S}}=Z_{\zeta}^{-1}$. Then, imposing finiteness of $Z_{\hat{S}}^{-2} \langle \hat{S}_0^a(\tau_1) \hat{S}_0^b(\tau_2) \rangle_{\Dm_j}$ at this order in the coupling constant yields
	\begin{equation}
	Z_{\hat{S}}=1-\frac{\zeta^2}{ \veps}-\frac{\zeta^4}{2 \veps^2}+\frac{\zeta^4 }{2\veps}+ O(\zeta^6).
	\end{equation}
	Putting everything together, the renormalized two-point function evaluated at the free bulk fixed point \eqref{fixedcouplingfree} is
	\begin{equation}\label{defectspintwopoint}
 \langle \hat{S}_a(\tau_1) \hat{S}_b(\tau_2) \rangle_{\Dm_j}=  \frac{\Nm_{\hat{S}}}{ |\tau_1-\tau_2|^{2\hat{\Delta}_{\hat{S}}}}\cdot \frac{\delta_{ab}}{3}\,,
	\end{equation}
	where $\hat{\Delta}_{\hat{S}}= \veps/2$ and
	\begin{equation}\label{defectspincoefficient}
	\Nm_{\hat{S}}=j (j+1) \left( 1-\veps+ \veps^2 \frac{12+\pi ^2}{24}  \right) +O(\veps^3)\,.
	\end{equation}
Clearly, by conformal symmetry and by the fact that $\hat{S}_a$ is protected, we already knew that \eqref{defectspintwopoint} holds at the non-perturbative level. The above computation is nevertheless necessary to determine the constant $\Nm_{\hat{S}}$.\footnote{
Note that the normalization of $\hat{S}_a$ is already fixed from the bulk through the Ward identity \eqref{boxphifree}.}

We can use this result together with \eqref{fielddec} to compute the bulk-to-defect two-point function between $\phi^a$ and $\hat{S}^b$
\begin{equation}\label{2ptphiS}
    \langle \phi^a (0,x_\perp) \hat{S}^b(0) \rangle_{\Dm_j} = \sqrt{\kappa}\,\zeta\int d\tau  \frac{\braket{\hat{S}^a (\tau) \hat{S}^b (0)}_{\Dm_j}}{(\tau^2+|x_\perp|^2)^{1-\frac{\veps}{2}}}\,,
 \end{equation}
 which is exact in the free-bulk theory.\footnote{Since we are interested in the correlator at the fixed point, it is enough to evaluate it with vanishing parallel distance between the operators. The kinematics is already fixed by conformal symmetry.}
 Using \eqref{defectspintwopoint}, solving the integral and evaluating at the fixed point yields
\begin{equation}
   \langle \phi^a (0,x_\perp) \hat{S}^b(0) \rangle_{\Dm_j} =\frac{\delta^{ab}}{3\, |x_\perp|}\cdot\frac{\sqrt{\kappa }\, \zeta_* \, \Nm_{\hat{S}} \, \sqrt{\pi}\, \Gamma \left( \frac{1-\veps}{2}\right)}{\Gamma \left( 1-\frac{\veps}{2} \right)}  \equiv\frac{\delta^{ab}}{3\, |x_\perp|}\,b_{\phi \hat{S}}\,.
 \end{equation}
 Interestingly, the above correlator contains a factor $\Gamma \left( \frac{1-\veps}{2}\right)$ that diverges in the $\veps \rightarrow 1$ limit. At this stage, it is still unclear whether the divergence could be cured by the $\veps$-dependent term $\zeta_*  \Nm_{\hat{S}}$. Nevertheless, this should be taken as a hint that the theory is sick for $\veps=1$, \emph{i.e.} in three dimensions, as it will be proved in section \ref{Bootstrapofphiphi} . 
 
When the bulk interaction is turned on, using \eqref{fixedcouplingint} and \eqref{deltaSint} we obtain
\begin{equation}\label{defectspininteracting}
\begin{split}
    &\Nm_{\hat{S}}=j (j+1) \left[ 1-\veps+ \veps^2 \left(\frac{1512-55 \pi ^2 }{2904}+ \frac{2 \pi ^2 j (j+1)}{11}\right)\right] +O(\veps^3)\,. \\
\end{split}
\end{equation}

\subsection{The displacement operator and the defect stress-energy tensor} 

In a similar way, one can consider the Ward identity given by the translational invariance of the bulk theory.
The defect explicitly breaks this symmetry and the conservation of the bulk stress-energy tensor is also modified by a term localized on the defect \cite{Cuomo:2021rkm, Billo:2016cpy}
	\begin{equation}\label{displacement}
	\partial_\mu T^{\mu \nu}(0,x_\perp)= - \left( \delta_i^\nu \hat{D}^i (0) + \partial_\tau x^\nu (0) \, \partial_\tau \hat{T}_{\Dm_j}(0) \right) \delta^{d-1}(x_\perp)\,.
	\end{equation}
    where $x^\nu (\tau)$ is the embedding function that describes the defect and $\tau$ is the coordinate that parametrizes the line.
	$\hat{D}^i$ is called the \emph{displacement} operator and it is a primary operator. 
	By the above Ward identity, it has protected dimension $	\hat{\Delta}_D=2$. 
	The explicit expression for the bare displacement operator can be derived by considering the variation of the action with respect to  $x^i (\tau)$
	\begin{equation}
	\hat{D}_0^i(x(\tau))=\frac{1}{|\dot{x}(\tau)|}\frac{\delta S_{\text{defect}}}{\delta x_i(\tau)}.
	\end{equation}
	Computing this functional derivative\footnote{
	Note that one needs to first reintroduce the arc length element $ | \dot{x}(\tau)|$ in the integral of the defect action \eqref{defectaction} since a generic variation of the embedding spoils the unit speed parametrization.
	}  and at the end evaluating at unit speed parametrization of the flat defect one finds
	\begin{equation}\label{displacementexplicit}
	\hat{D}_0^i(\tau)=\frac{\zeta_0}{\sqrt{\kappa}} \partial^i \phi_a(\tau)\,\frac{\Tr  \big( \Dm_j (-\infty, \tau) \, T_a \, \Dm_j (\tau, \infty)\big)}{\Tr\ \Dm_j }\,.
	\end{equation}
In terms of correlators, the bare displacement operator inserted at a point $\tau$ lying on the defect satisfies
\begin{equation}
\langle \mathcal{O}_1(x_1)\dots \hat{D}_0^i(\tau)\dots\mathcal{O}_n(x_n)\rangle_{\Dm_j} =\frac{\zeta_0}{\sqrt{\kappa}}\, \langle \mathcal{O}_1(x_1)\dots \partial^i \phi^a(\tau)T_a\dots\mathcal{O}_n(x_n)\rangle_{\Dm_j}\,.
\end{equation}
We will just rewrite this as
\begin{equation}
\hat{D}_i(\tau) \sim  \partial_i\phi^a T_a(\tau)\,.
\end{equation}
Note that this analysis holds regardless of whether the bulk is interacting or not, since the bulk stress-energy tensor is nevertheless conserved.

The other operator that appears in the Ward identity \eqref{displacement} is the \emph{defect stress-energy tensor} $\hat{T}_{\mathcal{D}_j}$.
By the Ward identity, it has protected dimension $\hat{\Delta}_{\hat{T}_{\mathcal{D}_j}} =1$.
The existence of such operator breaks conformal invariance on the line defect, therefore it must vanish at the fixed point.
In our case the defect stress-energy tensor reads\footnote{For a generic line defect with a Lagrangian of the form $\mathcal{L}_{\text{defect}} = g \hat{\mathcal{O}}$, the defect stress tensor reads $\hat{T} = \beta_g \hat{\mathcal{O}}$. This follows from the more general result $\partial_\nu T_\mu ^\nu x^\mu = \beta_i \frac{\partial \mathcal{L}}{\partial g_i} $, which is a consequence of Noether's theorem applied to the renormalized Lagrangian in the case of scale transformations.}
\begin{equation}\label{TD}
    \hat{T}_{\mathcal{D}_j}  \, (\tau) = \frac{\beta_\zeta}{\sqrt{\kappa}} \hat{\Phi}(\tau)\,,
 \end{equation}
 where for future convenience we defined $\hat{\Phi}(\tau)=\phi_a T^a (\tau)$.\footnote{
 Here and in the rest of this paper we will assume that defect operators with generator insertions have to be interpreted in the sense of \eqref{correlatorS} and \eqref{displacementexplicit}.  }
Using the definition of conformal dimension $\mu \frac{\partial \hat{\mathcal{O}}}{\partial \mu}= -\hat{\Delta}_{\hat{\mathcal{O}}}\hat{\mathcal{O}} $ and the fact that $\hat{T}_{\mathcal{D}_j}$ is protected, we obtain
 \begin{equation}\label{DeltaPhi}
     \hat{\Delta}_{\hat{\Phi}} = 1+\frac{\partial \beta_{\zeta}}{\partial \zeta} + \frac{\beta_\lambda}{\beta_\zeta} \frac{\partial \beta_\zeta}{\partial \lambda}\,.
 \end{equation}
 This formula is exact and holds both for the case of free and interacting bulk theories.  \\
  A consequence of the last equation and the definition of the anomalous dimension of $\hat{\Phi}$ in terms of the wave function normalization of the operator is that in free theory
 \begin{equation}\label{ZPhi}
     Z_{\hat{\Phi}} = -\frac{2 \beta_{\zeta}}{\veps \, \zeta \, Z_\zeta}\,.
 \end{equation}
 Finally, using the expression for the beta function in the free bulk theory \eqref{eq:betafreebulk} and the value of $\zeta$ at the critical point, one obtains the conformal dimension of the defect operator $\hat{\Phi}$
 \begin{equation}\label{dimPhifree}
     \hat{\Delta}_{\hat{\Phi}} = 1+ \veps - \frac{ \veps^2 }{2} + 
\frac{\veps^3}{2} \left[1 - \pi^2 \left(j\left(j+1\right)-\frac{1}{3}\right)\right] + O(\veps^4)\,.
 \end{equation}
 One can do the same in the interacting case, where
 \begin{equation}\label{dimPhiint}
     \hat{\Delta}_{\hat{\Phi}} = 1+ \veps -\veps ^2 \left[\frac{257}{484} -\frac{4 \pi ^2}{11}  \left(j (j+1)-\frac{1}{3}\right)\right]+O(\veps^3)\,.
 \end{equation}

\subsubsection{Correlators of $\hat{\Phi}$}
In this section we compute the one-loop two-point function of $\hat{\Phi}$ both in the free bulk and in the interacting bulk case. This computation, besides providing a sanity check of the arguments of the previous section, it also shows how correlators of operators composed both of generators insertions and of fundamental fields are evaluated in practice.

At tree level there is only one diagram 
\begin{equation}
\begin{tikzpicture}[baseline={([yshift=-.5ex]current bounding box.center)},scale=0.5]
  	\draw[thick, black]   (-1.5,0) to[out=90,in=90] (1.5,0);
	 \draw[double,thick,blue] (-3,0)--(3,0);
	 \draw[blue, fill=white] (1.5,0) circle (4pt);
	 \node[below] at (-1.5,0) {$\hat{\Phi}$};
	  \node[below] at (1.5,0) {$\hat{\Phi}$};
	   \draw[blue, fill=white] (-1.5,0) circle (4pt);
	   \end{tikzpicture} \vspace{0.1cm} = j(j+1) \, \frac{\kappa}{|\tau_1-\tau_2|^{2-\veps}}\,.
	   \end{equation}
	   
	   At one loop one finds two kind of connected diagrams: the two operators can be either connected by a free bulk propagator, or they interact with the defect. Note that even in the interacting bulk case there are no other diagrams, since bulk interactions contribute only at the next order. The first kind of diagrams are
	   \begin{equation}
	      		  \begin{tikzpicture}[scale=0.6]
	 \draw[double,thick,blue] (-3.75,0)--(0.75,0);
	  \draw[thick, black]    (-3,0) to[out=-90,in=180](-2,-1)  to[out=00,in=-90]  (-1,0);
	   \draw[thick, black]    (-2,0) to[out=90,in=180](-1,1)  to[out=00,in=90]  (0,0);
	 \draw[blue, fill=white] (-2,0) circle (4pt);
	  \draw[blue, fill=white] (0,0) circle (4pt);
	  \draw[blue, fill=blue] (-3,0)circle (4pt);
	  \draw[blue, fill=blue] (-1,0) circle (4pt);
	\end{tikzpicture}
	\hspace{1cm}
		  \begin{tikzpicture}[scale=0.6]
	 \draw[double,thick,blue] (-3.75,0)--(0.75,0);
	  \draw[thick, black]    (-3,0) to[out=90,in=180](-2,1)  to[out=00,in=90]  (-1,0);
	   \draw[thick, black]    (-2,0) to[out=-90,in=180](-1,-1)  to[out=00,in=-90]  (0,0);
	  \draw[blue, fill=white] (-1,0) circle (4pt);
	   \draw[blue, fill=white] (-3,0) circle (4pt);
	    \draw[blue, fill=blue]  (-2,0)circle (4pt);
	  \draw[blue, fill=blue] (0,0) circle (4pt);
	\end{tikzpicture}
	   \end{equation}
	   the computation of these integrals is analogous to the one for the operators $\hat{S}_a$ in section \ref{corrdefectspin},  with the only difference that now everything is multiplied by a free propagator. The result is
	   \begin{equation}
	   	I_1^{(1)}(\tau_1,\tau_2)=-\frac{\zeta _0^2\,j(j+1)\Gamma\left(2-\frac{\veps}{2}\right)}{\pi^{2-\frac{\veps}{2}}(2-\veps)(1-\veps)\,\veps \,|\tau_1-\tau_2|^{2-2\veps}} \,.
	   \end{equation}
	  The other diagrams are those where the two operators interact with the defect. There are twelve of them and they come with two different color structures: eight diagrams with $\Tr \left( T_a T_a T_b T_b \right) \sim j^2 (j+1)^{2}$ and the remaining four with $\Tr \left( T_a T_a T_b T_b \right) \sim j(j+1)(j^2 + j -1)$. The sum of all the diagrams contains a piece proportional to $j^2 (j+1)^{2}$ which is a sum of the ordered integral of two propagators over all possible orders, hence giving $\kappa^2\int d \sigma_1 \,|\sigma_1-\tau_1|^{-2+\veps} \int d \sigma_2 \,|\sigma_2-\tau_2|^{-2+\veps}$ which is vanishing in our regularization. Hence we are left with the evaluation of the following four diagrams
	    \begin{equation}
	   \begin{split}
	      		&\begin{tikzpicture}[baseline={([yshift=-.5ex]current bounding box.center)},scale=0.5]
	 \draw[double,thick,blue] (-3.75,0)--(0.75,0);
	  \draw[thick, black]    (-3,0) to[out=-90,in=180](-2,-1)  to[out=00,in=-90]  (-1,0);
	   \draw[thick, black]    (-2,0) to[out=90,in=180](-1,1)  to[out=00,in=90]  (0,0);
	 \draw[blue, fill=white] (-2,0) circle (4pt);
	  \draw[blue, fill=blue] (0,0) circle (4pt);
	  \draw[blue, fill=white] (-3,0)circle (4pt);
	  \draw[blue, fill=blue] (-1,0) circle (4pt);	    
	   \end{tikzpicture}
	\hspace{0.1cm}  =-\frac{\zeta_0^2 \, j(j+1)\,\Gamma\left(2-2\veps\right)\Gamma\left(1-\frac{\veps}{2}\right)\Gamma\left(\veps\right)}{4\pi^{2-\frac{\veps}{2}}(1-\veps)\Gamma\left(2-\veps\right)\,|\tau_1-\tau_2|^{2-2\veps}}\,,	\\
		  &\begin{tikzpicture}[baseline={([yshift=-.5ex]current bounding box.center)},scale=0.5]
	 \draw[double,thick,blue] (-3.75,0)--(0.75,0);
	  \draw[thick, black]    (-3,0) to[out=90,in=180](-2,1)  to[out=00,in=90]  (-1,0);
	   \draw[thick, black]    (-2,0) to[out=-90,in=180](-1,-1)  to[out=00,in=-90]  (0,0);
	  \draw[blue, fill=blue] (-2,0) circle (4pt);
	  \draw[blue, fill=white] (0,0) circle (4pt);
	  \draw[blue, fill=white] (-3,0)circle (4pt);
	  \draw[blue, fill=blue] (-1,0) circle (4pt);	 
	  \end{tikzpicture} \hspace{0.1cm}=-\frac{\zeta_0^2 \, j(j+1)\,\Gamma\left(1-\frac{\veps}{2}\right)\left(\Gamma\left(\veps\right)^2-\Gamma\left(2\veps-1\right)\right)}{8\pi^{2-\frac{\veps}{2}}(1-\veps)^3\Gamma\left(-2+2\veps\right)\,|\tau_1-\tau_2|^{2-2\veps}}\,, \\
		&\begin{tikzpicture}[baseline={([yshift=-.5ex]current bounding box.center)},scale=0.5]
	 \draw[double,thick,blue] (-3.75,0)--(0.75,0);
	  \draw[thick, black]    (-3,0) to[out=90,in=180](-2,1)  to[out=00,in=90]  (-1,0);
	   \draw[thick, black]    (-2,0) to[out=-90,in=180](-1,-1)  to[out=00,in=-90]  (0,0);
	\draw[blue, fill=white] (-2,0) circle (4pt);
	  \draw[blue, fill=blue] (0,0) circle (4pt);
	  \draw[blue, fill=blue] (-3,0)circle (4pt);
	  \draw[blue, fill=white] (-1,0) circle (4pt);	 
	  \end{tikzpicture}
 \hspace{0.1cm} =-\frac{\zeta_0^2 \,j(j+1)\,\Gamma\left(1-\frac{\veps}{2}\right)}{4\pi^{2-\frac{\veps}{2}}(1-\veps)^2\,|\tau_1-\tau_2|^{2-2\veps}}\,,\\
		  &\begin{tikzpicture}[baseline={([yshift=-.5ex]current bounding box.center)},scale=0.5]
	 \draw[double,thick,blue] (-3.75,0)--(0.75,0);
	  \draw[thick, black]    (-3,0) to[out=90,in=180](-2,1)  to[out=00,in=90]  (-1,0);
	   \draw[thick, black]    (-2,0) to[out=-90,in=180](-1,-1)  to[out=00,in=-90]  (0,0);
	  \draw[blue, fill=blue] (-2,0) circle (4pt);
	  \draw[blue, fill=white] (0,0) circle (4pt);
	  \draw[blue, fill=blue] (-3,0)circle (4pt);
	  \draw[blue, fill=white] (-1,0) circle (4pt);	 
	  	\end{tikzpicture} \hspace{0.1cm }=-\frac{\zeta_0^2\,j(j+1)\,\Gamma\left(2-2\veps\right)\Gamma\left(1-\frac{\veps}{2}\right)\Gamma\left(\veps\right)}{4\pi^{2-\frac{\veps}{2}}(1-\veps)\Gamma\left(2-\veps\right)\,|\tau_1-\tau_2|^{2-2\veps}}\,.
	\end{split}
	   \end{equation}
	Summing all the contributions, introducing the wavefunction renormalization coefficient $Z_{\hat{\Phi}}$ and imposing finiteness of $   Z_{\hat{\Phi}}^{-2}\langle \hat{\Phi}(\tau_1) \hat{\Phi}(\tau_2)\rangle_{\Dm_j}$ at one loop, we get
	   \begin{equation}
	   \begin{split}
	  & Z_{\hat{\Phi}}=1-\frac{3 \,\zeta^2}{\veps}+O\left(\zeta^4,\zeta^2 \lambda, \lambda^2\right)\,, \\
	    & \left.\gamma_{\hat{\Phi}}\right|_{{\zeta_*,\lambda_*}}=\left.\beta_{\zeta}\frac{\partial \log Z_{\hat{\Phi}}}{\partial \zeta}\right|_{\zeta_*,\lambda_*}\!\!\!\!+O(\veps^2)=\frac{3}{2}\,\veps+O(\veps^2)\,.
	   \end{split}
	   \end{equation}
	 The renormalized two-point function evaluated at the fixed point is
	\begin{equation}\label{tilttwopoint}
\langle \hat{\Phi}(\tau_1) \hat{\Phi}(\tau_2)\rangle_{\Dm_j}=  \frac{\Nm_{\hat{\Phi}}}{ |\tau_1-\tau_2|^{\hat{\Delta}_{\hat{\Phi}}}}\,,
	\end{equation}
	where both in the free bulk and in the interacting bulk case
	\begin{equation}
	\begin{split}
	&\Nm_{\hat{\Phi}}=\frac{j(j+1)}{4\pi^2}\left( 1+\veps\left(-2+\frac{\gamma_E}{2}+\frac{\log \pi}{2} \right)\right)+O(\veps^2)\,, \\
	&\hat{\Delta}_{\hat{\Phi}}=1+\veps+O(\veps^2)\,.
	\end{split}
	\end{equation}

\subsection{General defect operators}\label{Other defect operators}

The defect spin and the displacement operators arose as defect corrections to Ward identities. 
It is natural to ask whether there are other defect operators with protected dimensions that can be obtained in this way. 
In particular, in the bulk-free theory there exists an infinite tower of conserved higher spin currents of the schematic form \cite{Belitsky:2007jp , Giombi:2016hkj}
\begin{equation}\label{highercurrent}
   \mathcal{J}_{\mu_1 \dots \mu_{s+1}}^{ab}(x) \sim \sum_{k=0}^s c_{s,k} \,  \partial_{\{ \mu_1} \dots \partial_{\mu_{k}} \phi^a\, \partial_{\mu_{k+1}} \dots \partial_{\mu_{s+1}\}} \phi^b (x)\,,
\end{equation}
where brackets denote traceless symmetrization, and $s \geq 1$.\footnote{
For $s=0$, up to an antisymmetric tensor one just get the Noether current associated to the $SU(2)$ global symmetry: $\Jm^a_\mu \sim \epsilon^{abc} \phi_b \partial_\mu \phi_c$, which is conserved also in the defect theory.}
Their dimension is $\Delta_{J_{s+1}}=s+1-\veps$. 
From the modified Ward identity
\begin{equation}\label{spinsdefect}
\partial^{\nu} \mathcal{J}_{\nu \mu_1 \dots \mu_s}^{ab}(0,x_\perp) =\frac{\zeta_0}{\sqrt{\kappa}} \,\hat{\mathcal{J}}^{ab}_{\mu_1\dots\mu_s}(0) \, \delta^{d-1}(x_\perp)\,,
\end{equation}
one immediately finds a tower of defect operators that at the fixed point have protected dimension $\hat{\Delta}_{\hat{\mathcal{J}}_s}=s+1 \in \mathbb{N}$. 
Their explicit form can be determined by a computation analogous to the one of the defect spin operator.
In equation \eqref{spinsdefect}, one gets defect primary operators only when all the free spatial indices are taken to be orthogonal to the defect, since derivatives parallel to the defect give rise to descendants.
Therefore we can just consider $\hat{\mathcal{J}}^{ab}_{i_1\dots i_s}$, which clearly has orthogonal spin $s$. 
As for color indices, it is convenient to think in terms of $\mathfrak{so(3)}$ rather than $\mathfrak{su(2)}$.
For even $s$, the two color indices must be in the antisymmetric representations; this is equivalent to the vector representation $\hat{\mathcal{J}}^{a}_{i_1\dots i_{s}}$.
For odd $s$, the representations can be the traceless symmetric $\hat{\mathcal{J}}^{\{ab\}}_{i_1\dots i_{s}}$ and the singlet $\hat{\mathcal{J}}_{i_1\dots i_{s}}$.
Like in the case of the defect spin operators, when the bulk interaction is turned on, these higher spin currents are weakly broken and their dimensions will get corrections starting at second order in $\veps$.

It is possible to obtain more information on the defect spectrum by looking at Ward identities for particular correlators. Following \cite{Lauria:2020emq}, consider the bulk-to-defect two-point function of $\phi$ and $\hat{\phi}$, which by conformal symmetry must take the form
\begin{equation}
\langle \phi_a(0,x_\perp) \hat{\phi}_b(0) \rangle_{\Dm_j} = \frac{b_{\phi \hat{\phi}}}{|x_\perp|^{\Delta_\phi-\hat{\Delta}_{\hat{\phi}}}|x_\perp|^{2\hat{\Delta}_{\hat{\phi}}}}\delta_{ab}\,.
\end{equation}
Specializing to the free-bulk case and acting with the Laplacian $\Box_x$ at a point away from the defect $x$, we find
\begin{equation}
0=\langle \Box \phi_a(0, x) \hat{\phi}_b(0) \rangle_{\Dm_j}=(\hat{\Delta}_{\hat{\phi}}+\Delta_\phi- 1)(\hat{\Delta}_{\hat{
\phi}}-\Delta_\phi)\frac{b_{\phi \hat{\phi}}}{|x_\perp|^{\Delta_\phi-\hat{\Delta}_{\hat{\phi}}+2}|x_\perp|^{2\hat{\Delta}_{\hat{\phi}}}}\delta_{ab}\,,
\end{equation}
and since we know from an immediate tree-level computation diagram that $\hat{\Delta}_{\hat{\phi}}=1+O(\veps)$ and that $b_{\phi \hat{\phi}}\neq 0$, it must be that $\hat{\Delta}_{\hat{\phi}}=\Delta_\phi$ holds at the non-perturbative level. 
The same argument can be applied to the transverse spin-$s$ operators $\hat{\mathcal{O}}_{i_1 \dots i_s}^{\,a} \sim \partial_{i_1} \dots \partial_{i_s} \hat{\phi}^a$. 
One readily finds that their exact dimension is $\hat{\Delta}_s=\Delta_\phi + s$.
Again, those dimensions will receive correction in the interacting bulk case, starting always at second order in $\veps$.

In the previous sections, we have seen that in this theory there are some defect operators such as the defect spin and the displacement operator that include in their definition the insertion of a generator $T_a$, and thus are matrix-valued.
As anticipated in section \ref{corrfuns}, this suggests that a generic local defect operator is a $2j+1 \times 2j+1$ Hermitian matrix, with entries that are composite operators made of fundamental fields and their derivatives. Their correlators are given by \eqref{defectcorrelators2}. Clearly, when the matrix is proportional to the identity, one recovers the case of operators that can be factored outside the trace of the path-ordering, such as the fundamental fields $\hat{\phi}_a$. In order to be able to construct and identify all the possible defect operators, it is useful to choose a convenient basis for the space of these matrices.

In the simplest situation where $j=\frac{1}{2}$, \emph{i.e.}\ in the fundamental representation of $\mathfrak{su}(2)$, the three generators together with the identity span the whole real vector space of $2 \times 2$ Hermitian matrices.
In particular, a defect operator with an arbitrary Hermitian matrix insertion in the defect can be decomposed into operators with insertions that are at most linear in the generators $T^a$. For the more general case of spin $j>\frac{1}{2}$ representations, the space of possible Hermitian matrix insertions has real dimension $(2j+1)^2$.
We can span this space by taking Hermitian combinations of products of the generators $T^a$.
A natural choice of basis is given by the totally symmetrized traceless product of generators $T^{ \{ a_1 }\dots T^{ a_k \} }$, with $k=0,\dots, 2j$.\footnote{
To show that these symmetrized traceless products constitute a basis, note that they are $\sum_{k=0}^{2j} (2k+1) = (2j+1)^2$ and that they are orthogonal with respect to the trace inner product.
}
In particular, there are $4j(j+1) $ primary defect operators defined by the basis elements $\hat{S}^{\{a_1\dots a_k\}}(x) \equiv T^{ \{ a_1 }\dots T^{ a_k \} }(x)$ for $k \geq 1$ inserted in the path-ordered exponential, without any fundamental field. 
These operators are expected to be among the lightest operators of the theory, since their classical dimension is zero. Moreover, there cannot be any mixing between them for representation theory reasons. For operators composed also of powers of the fundamental fields and their derivatives, it still makes sense to organize operators according to their color index structure. 
However, in general there will be several operators in the same representation and with the same classical dimension, so that one should worry about their mixing.

Finally, it is important to pay attention to the fact that defect descendants are given by the defect covariant derivative \eqref{Covariantderivative} and not by the ordinary one. As an example, as we found out in \eqref{Sdescendant}, the defect operator defined by $\epsilon^{abc}\phi_b T_c (\tau)$ is not a new primary, but it is a descendant.

\subsubsection{Correlators of $\hat{\phi}_a$ and $\hat{\mathcal{O}}_{i_1 \dots i_s}^{\,a} $ }\label{corrhatphi}

Among all the defect operators we have just discussed, there are some interesting exact relations that holds between correlators when the bulk is free.
As an example, consider the defect operator $\hat{\phi}_a$, which is just the fundamental field placed on the defect.
Using the analogous of \eqref{fielddec} for $\hat{\phi}_a$ (\emph{i.e.}\ when $x_\perp=0$), we can compute its two-point function in terms of the one of the defect spin operator
\begin{equation}
\langle \hat{\phi}_a(\tau_1)\hat{\phi}_b(\tau_2)\rangle_{\Dm_j}=\frac{\kappa \,j(j+1) \, \delta_{ab}}{3 \, |\tau_1-\tau_2|^{2-\veps}}+\kappa \, \zeta^2 \int d\sigma_1 \int d\sigma_2 \,  \frac{\langle \hat{S}(\sigma_1) \hat{S}(\sigma_2)\rangle_{\Dm_j} }{\left(|\tau_1-\sigma_1|\,|\tau_2-\sigma_2|\right)^{2-\veps}}\,,
\end{equation}
which holds at all orders in perturbation theory. In particular, evaluating at the fixed point we get
\begin{equation}\label{twoptphihat}
\langle \hat{\phi}_a(\tau_1)\hat{\phi}_b(\tau_2)\rangle_{\Dm_j}=\frac{ \, \delta_{ab}}{3 \, |\tau_1-\tau_2|^{2-\veps}}\left(\kappa \, j(j+1)-\frac{\zeta^2_*\,\Nm_{\hat{S}}\,\Gamma(1-\veps)\,\Gamma\left(\frac{\veps-1}{2}\right) \sin \left( \frac{\pi \veps}{2}\right)}{2^{2-\veps}\,\pi^{\frac{3-\veps}{2}}}\right)\,.
\end{equation}
From this we can check that $\hat{\phi}_a$ has a vanishing anomalous dimension, as we saw through Ward identities in section \ref{Other defect operators}.
Similarly, we can do the same for the two-point function of one operator in the bulk and one placed on the defect.
At the fixed point we find
\begin{equation}\label{twoptphiphihat}
\langle \phi_a(0,x_\perp)\hat{\phi}_b(\tau)\rangle_{\Dm_j}=\frac{ \delta_{ab}}{3\, \left(|x_\perp|^2+\tau^2 \right)^{1-\frac{\veps}{2}}}\left(\kappa \,j(j+1)-\frac{\zeta^2_*\,\Nm_{\hat{S}}\,\Gamma(1-\frac{\veps}{2})\tan \left( \frac{\pi \veps}{2}\right)}{\pi^{1-\frac{\veps}{2}}(\veps-1)}\right)\,.
\end{equation}
Note that the above two-point function depends only on the four-dimensional distance between the bulk field and the defect field because they have exactly the same conformal dimension.
The same argument applies to correlators involving $\hat{\mathcal{O}}_{i_1 \dots i_s}^{\,a} \sim \partial_{i_1} \dots \partial_{i_s} \hat{\phi}^a$, in which case one only need to take orthogonal derivatives in \eqref{fielddec} before setting $x_\perp=0$.
Finally, one could do the same to get the two-point function of two bulk fields, as we shall see in Section \ref{sec:diagrams}.

Obviously, in the interacting bulk case there will be corrections to the correlators \eqref{twoptphihat} and \eqref{twoptphiphihat} starting at order $\veps^2$.

\subsubsection{Time reversal symmetry for defect operators}\label{timerevdef}

We are finally in the position to extend the discussion of time reversal symmetry to generic defect operators.
Their parity under this symmetry will be an useful tool to classify such operators.
It is clear that defect operators without any insertions, \emph{i.e.}\ those composed only of fundamental fields and their derivatives, behave just like bulk operators under this symmetries.
On the other hand, the very same argument used in section \ref{timereversal}  can be repeated in the presence of operators with insertions into the defect.
 A careful analysis shows that the effect of $T_{t}$ on generator insertions is $T_{t}: T^a(\tau) \mapsto - T^a (-\tau)$.
  Since $T_{\mathbb{Z}_2}$ does not act on generators, it follows that also $T^a$ is odd under $\bar{T}_t$. 
  This can also be seen, for example, by the Ward identity
\begin{equation}
\Box \phi_a(0,x_\perp) = \frac{\zeta_0}{\sqrt{\kappa}} \, \hat{S}_a(0) \, \delta^{d-1}(x_\perp)\,.
\end{equation}
When there are more than just one generator inserted at the same point (only for $j\geq 1$), an analogue analysis shows that the effect of time reversal is not only a factor $(-1)$ for each generator, but also an inversion in the order of the insertions. For this reason, it is convenient to express insertions in the basis given at the end of section \ref{Other defect operators}. Indeed one can clearly see that this basis is diagonal under time reversal, and that $\bar{T}_t: T_{\{a_1}\dots T_{a_k\}}(\tau)\mapsto (-1)^k \, T_{\{a_1}\dots T_{a_k\}}(-\tau)$.

This symmetry imposes useful constraints on correlators. For example, sometimes it can be used to lift some degeneracies, since two defect operators with different parities must have vanishing two-point function at the non-perturbative level. The same holds for the two-point function of a bulk operator and a defect operator, giving useful selection rules for the coefficients of the defect block expansion \eqref{defectexpansion}. Finally, one must be careful that this conclusion does not generalize to correlators with more defect operators. In fact, in the case of one-dimensional defects, the three-point function of three defect operators can be antisymmetric \cite{Lauria:2020emq}.\footnote{
Indeed given any two points on an ordered straight line it is possible to invert their order through a special conformal transformation that preserves the line. But the same cannot be done for three points.} For example, one can check that $\langle \hat{S}_a(\tau_1)\hat{S}_b(\tau_2)\hat{S}_c(\tau_3) \rangle \propto i \epsilon_{abc}$.

\subsubsection{Classification of low-lying defect operators}\label{defectopclass}

In this section we conveniently collect together all the information about the low-lying spectrum of the defect that we gained so far through various tools. Defect operators are classified according to their transverse spin $s$, their $\mathfrak{su}(2)$ representation (which we characterize through their dimensions), their parity under the time reversal symmetry $\bar{T}_t$ and their classical dimension. Note that some of these operators only exist for sufficiently high values of $j$, where $j$ specifies the $\mathfrak{su}(2)$-representation of the generators $T_a$ in the definition of the defect \eqref{defect}. 

Obtaining a complete list of defect operators at twist zero is straightforward.\footnote{
Recall that the defect twist $\hat{\tau}$
of a defect operator with dimension $\hat{\Delta}$ and orthogonal spin $s$ is defined as $\hat{\tau} \equiv \hat{\Delta}-s$.} At twist one, it is sufficient to construct all the possible composite operators using only one fundamental field $\phi_a$, an arbitrary number of generators $T_a$ and orthogonal derivatives $\partial_i$. Then one need to decompose them into irreducible representations of $\mathfrak{su}(2)$. Finally, since the defect covariant derivative increases the twist by one, one need to exclude all the descendants of the twist-zero primaries. Clearly, it is in principle possible to continue the classification by considering higher twist operators, which can be constructed by using an arbitrary number of fundamental fields and also orthogonal Laplacians $\Box_\perp$. Again, one need to exclude all the descendants of lower-twist primaries.
The number of primary operators grows combinatorially with the defect twist.

In table \ref{table:1}, we list all the defect twist-zero and defect twist-one primary operators, together with their quantum numbers and their scaling dimensions at the fixed point (both for the free bulk and for the interacting bulk cases).
In table \ref{table:2}, we show the explicit definition of these operators in perturbation theory.
{\renewcommand*{\arraystretch}{1.3}
\begin{table}[H]
\begin{center}
\begin{tabular}{||c| c |c |c  |c|c||} 
 \hline
 $\hat{\mathcal{O}}$ & $\;\; s \;\;$ & $\dim R_{\mathfrak{su(2)}}$ & $\;\;\;\bar{T}_t\;\;\;$ & $\quad\quad\hat{\Delta}_{\hat{\mathcal{O}}}\big|_{\lambda=0}\quad\quad$ & $\hat{\Delta}_{\hat{\mathcal{O}}}\big|_{\lambda_*}$ \\ [0.5ex]
 \hline\hline
  $\hat{S}^a$ & $0$ & $3$ & $-$ &$\frac{\veps}{2}$ & \eqref{deltaSint} \\
 \hline
  $\hat{S}^{\{a_1 \dots a_k \}}$ & $0$ & $2k+1$ & $(-)^k$  &$O(\veps)$ & $O(\veps)$\\
  \hline
  $\hat{\phi}^a$ & $0$ & $3$ & $-$ &  $1-\frac{\veps}{2}$ & $1-\frac{\veps}{2}+O(\veps^2)$  \\
  \hline
  $\hat{\Phi}$ & $0$ & $1$ & $+$ & \eqref{dimPhifree} & \eqref{dimPhiint}  \\
  \hline
  $\hat{D}_i$ & $1$ & $1$ & $+$ &  $2$ & $2$ \\
  \hline
   $\hat{\mathcal{J}}^{a}_{i_1\dots i_s}$ & $s$ & $3$ & $+$ &$s+1$ & $s+1+O(\veps^2)$\\
   \hline
    $\hat{\mathcal{J}}^{\{ab\}}_{i_1\dots i_s}$ & $s$ & $5$ & $+$ &$s+1$ & $s+1+O(\veps^2)$ \\
   \hline
    $\hat{\mathcal{J}}_{i_1\dots i_s}$ & $s$ & $1$ & $+$ &$s+1$ & $s+1+O(\veps^2)$ \\
   \hline
   $\hat{\Om}^a_{i_1 \dots i_s}$ & $s$ & $3$ & $-$ &  $s+1-\frac{\veps}{2}$ & $s+1-\frac{\veps}{2}+O(\veps^2)$ \\
   \hline
    $\hat{U}^{\{a_1 \dots a_k \}}_{i_1 \dots i_s}$ & $s$ & $2k+1$ & $(-)^k$ &  $s+1+O(\veps)$ & $s+1+O(\veps)$   \\
  \hline
   $\hat{V}^{\{a_1 \dots a_k \}}_{i_1 \dots i_s}$ & $s$ & $2k+1$ & $(-)^k$ &  $s+1+O(\veps)$ & $s+1+O(\veps)$  \\
  \hline
   $\hat{W}^{\{a_1 \dots a_k \}}_{i_1 \dots i_s}$ & $s$ & $2k+1$ & $(-)^{k+1}$ & $s+1+O(\veps)$ & $s+1+O(\veps)$  \\
  \hline
\end{tabular}
\caption{Defect twist-zero and twist-one primary operators with their quantum numbers and scaling dimensions.}
\label{table:1}
\end{center}
\end{table}}
{\renewcommand*{\arraystretch}{1.3}
\begin{table}[H]
\begin{center}
\begin{tabular}{||c| c | c ||} 
\hline
Operator  & Perturbative definition & Existence \\ [0.5ex]
 \hline\hline
 \hline
 $\hat{S}^a$ & $T^a$ &\\
 \hline
  $\hat{S}^{\{a_1 \dots a_k \}}$ & $T^{\{a_1}\dots T^{a_k\}} $& $2 \leq k \leq 2j$\\
  \hline
  $\hat{\phi}^a$ & $\phi^a$ & \\
  \hline
  $\hat{\Phi}$ & $\phi^a T_a$  & \\
  \hline
  $\hat{D}_i$ & $\partial_i \phi^a T_a$ & \\
  \hline
   $\hat{\mathcal{J}}^{a}_{i_1\dots i_s}$ & $\epsilon^{abc} \partial_{i_1}\dots \partial_{i_s} \phi_b T_c$ & even $s$, $s \geq 2$ \\
   \hline
    $\hat{\mathcal{J}}^{\{ab\}}_{i_1\dots i_s}$ & $\partial_{i_1}\dots \partial_{i_s} \phi^{\{a} T^{b\}}$  & odd $s$, $s \geq 1$\\
   \hline
    $\hat{\mathcal{J}}_{i_1\dots i_s}$ & $\partial_{i_1}\dots \partial_{i_s} \phi^{a} T_a$ & odd $s$, $s \geq 3$ \\
   \hline
   $\hat{\Om}^a_{i_1 \dots i_s}$ &   $\partial_{i_1}\dots \partial_{i_s} \phi^a$ &  $s \geq 1$ \\
   \hline
   $\hat{U}^{\{a_1 \dots a_k \}}_{i_1 \dots i_s}$ & $\partial_{i_1} \dots \partial_{i_s} \phi_{b} T^{\{b} T^{a_1}\dots T^{a_k\}}$ & $1 \leq k\leq 2j-1, s \geq 0$ \\
  \hline
   $\hat{V}^{\{a_1 \dots a_k \}}_{i_1 \dots i_s}$ &  $\partial_{i_1} \dots \partial_{i_s} \phi^{\{a_1} T^{a_2} \dots T^{a_k\}}$& $3-\delta_{0,s} \leq k\leq 2j+1, s \geq 0$ \\
  \hline
   $\hat{W}^{\{a_1 \dots a_k \}}_{i_1 \dots i_s}$ &  $\partial_{i_1} \dots \partial_{i_s} \phi^{b}\epsilon^{ b c \{a_1} T^c T^{a_2} \dots T^{a_k\}}$ & $2 \leq k\leq 2j, s \geq 1$ \\
  \hline
\end{tabular}
\caption{Schematic perturbative definition of defect twist-zero and twist-one operators.}
\label{table:2}
\end{center}
\end{table}}
Note that for some of these operators the form is just schematic. Indeed, beyond tree level mixing among operators sharing the same quantum numbers can occur, and  an orthogonalization with respect to the two-point functions must be performed.
For example, the explicit form of the operator $\hat{U}^a=$``$\phi_b T^{\{b}T^{a\}}$'' is correct only at tree-level, and at higher loops one must make sure that this operator is orthogonalized with respect to $\hat{\phi}_a$.

Further results about defect operators, their dimensions and coefficients of their correlators will be obtained in section \ref{sec:bootstrap} trough analytic bootstrap techniques.

\section{Analytic bootstrap calculation}
\label{sec:bootstrap}
In this section, we study the magnetic impurity with analytic bootstrap techniques.
Although a variety of inversion formulas have been used for defect CFT \cite{Lemos:2017vnx,Liendo:2019jpu,Gimenez-Grau:2021wiv}, here we employ the dispersion relation developed in \cite{Bianchi:2022ppi, Barrat:2022psm}, which significantly streamlines calculations.
To help the reader, in appendix \ref{app:kinematics} we present our conventions for cross-ratios, conformal blocks, etc.

The main result of this section is the computation of the two-point functions of $\phi_a$ and $\phi^2$ at the conformal fixed point
\begin{align}\label{correlatorswithprefactor}
\langle \phi_a (x) \phi_b (y) \rangle _{\mathcal{D}_j} = \frac{\delta_{ab} F_{\phi \phi} (r,w)}{|x_\perp| ^{\Delta_\phi} |y_\perp| ^{\Delta_\phi}}  \,, \qquad
 \langle \phi^2 (x) \phi^2 (y) \rangle _{\mathcal{D}_j} = \frac{F_{\phi^2 \phi^2} (r,w) }{|x_\perp| ^{\Delta_{\phi^2}} |y_\perp| ^{\Delta_{\phi^2}}} \,.
\end{align}
Here $r$ and $w$ are the two conformal cross ratios expressed in radial coordinates \eqref{coordinates}.
Note that throughout this section we unit-normalize all operators, a convention that differs the rest of the paper, but which is standard in the CFT literature.
After computing these correlators, we also extract the bulk and defect CFT data by expanding the correlators using either the bulk\footnote{We write this and subsequent expressions for $F_{\phi \phi} (r,w)$, but analogous results apply to $F_{\phi^2 \phi^2}$.}
\begin{equation}\label{bulkexpansion}
     F_{\phi \phi} (r,w) = \xi^{-\Delta_\phi}\sum_{\mathcal{O}}\lambda_{\phi \phi \mathcal{O}} \,a_{\mathcal{O}} \, f_{\Delta,\ell}(r,w) \,, \quad \xi=\frac{(1-r w)(w-r)}{r w} \,,
\end{equation}
or the defect conformal block expansion
\begin{equation}\label{defectexpansion}
     F_{\phi \phi}(r,w) = \sum_{\hat{\mathcal{O}}} b^2 _{\phi \hat{\mathcal{O}}} \, \hat{f}_{\hat{\Delta}, s}(r,w) \,.
\end{equation}
The former is a consequence of the usual Operator Product Expansion (OPE) valid in CFT, while the latter is obtained by fusing one of the external operators with the defect.
In the expressions above, $f_{\Delta, \ell}(r, w)$ and $\hat{f}_{\hat{\Delta}, s}(r, w)$ are respectively the bulk and defect conformal blocks. 
They are special functions that depend on the dimensions and spins of the bulk or defect operators exchanged in the two OPE channels. Their explicit expressions can be found in \eqref{defectblock} and \eqref{bulkblock}.
The coefficients in the bulk expansion are the product of one-point functions $a_\mathcal{O}$ and three-point functions coefficients $\lambda_{\phi \phi \mathcal{O}}$ , whereas in the defect channel the coefficients of the expansion are the squares of the bulk-defect couplings $b_{\phi \hat{\mathcal{O}}}$.
We refer to appendix \ref{app:kinematics} for their precise definitions.

In order to compute the two-point functions, we are going to use the defect dispersion relation \cite{Bianchi:2022ppi, Barrat:2022psm}
\begin{equation}\label{dispersion}
     F_{\phi \phi}(r,w)=\int_{0}^{r} \frac{d w'}{2\pi i}\left(\frac{1}{w'-w}+\frac{1}{w'-\frac{1}{w}}-\frac{1}{w'}\right) \text{Disc}F_{\phi \phi}(r,w') \,.
\end{equation}
This formula reconstructs the full correlator from its discontinuity  through the cut running from $w=0$ to $w=r$
\begin{equation}\label{discontinuity}
 \text{Disc} F_{\phi \phi}(r,w)=F_{\phi \phi}(r,w+i0)-F_{\phi \phi}(r,w-i0) \,.
\end{equation}
The dispersion relation is powerful because in perturbation theory the discontinuity is a much simpler object than the full correlator.
More precisely, at any given order in perturbation theory, the discontinuity depends only on a subset of known bulk data.
To see this, it is convenient to take the discontinuity of the bulk expansion
\begin{equation}
\label{eq:compute-disc}
     \text{Disc} F_{\phi \phi}(r,w)= \sum_{\mathcal{O}} c_\Om  \, \text{Disc} \left[\xi^{-\Delta_\phi}f_{\Delta,\ell}(r,w)\right] \,.
\end{equation}
Here and below we introduce the shorthand notation $c_\Om = \lambda_{\phi \phi \mathcal{O}} a_{\mathcal{O}}$.
Although the sum in \eqref{eq:compute-disc} runs over all operators in the theory, let us focus for simplicity only on twist-two operators\footnote{Recall that the twist $\tau$ of a bulk operator with dimension $\Delta$ and spin $\ell$ is defined as $\tau \equiv \Delta - \ell$.}.
Later we will argue that these are the only operators needed for the calculations in this work, at least at the order in perturbation theory we consider.
Twist-two operators have the schematic form
\begin{equation}
    \mathcal{J}_{\ell}
    \sim \phi^a \partial_{\mu_1} \ldots \partial_{\mu_\ell} \phi_a \, ,
\end{equation}
and we can expand their CFT data in perturbation theory as
\begin{equation}
   \Delta_{\Jm_{\ell}}=2 \Delta_\phi+ \ell +\varepsilon \gamma_{\ell}^{(1)} +O(\varepsilon^2) \,,\quad
   c_{\ell}
   =  c_{\ell} ^{(0)}
   + \varepsilon c_{\ell} ^{(1)}
   + \ldots \, .
\end{equation}
Here $c_{\ell}^{(0)}$ might itself be of order $O(\veps^n)$, and in fact, in the examples below $c^{(0)} \sim O(\veps)$.
Now we can compute the discontinuity of each individual block.
Using the formulas in appendix \ref{app:kinematics}, one sees that the discontinuity vanishes at tree-level, and the leading term is proportional to the anomalous dimension
\begin{equation}
\label{eq:disc-dt}
    \left.\text{Disc} F_{\phi \phi}(r, w)\right|_{\text{double-twist}}
    = \varepsilon  \sum_{\mathcal{O}} c_{\mathcal{O}}^{(0)} \gamma_{\mathcal{O}}^{(1)} \Disc \left[ \xi^{-\Delta_\phi} \partial_\Delta f_{2 \Delta_\phi+\ell, \ell}(r, w) \right] \, + \ldots \, ,
\end{equation}
where $\ldots$ stand for contributions at higher order in $\veps$.
This depends only on tree-level coefficients and one-loop bulk anomalous dimensions.
The tree-level coefficients can be extracted from knowledge of free theory, while the bulk anomalous dimensions are independent of the defect and are often known from other means.
As a result, the discontinuity can be computed in perturbation theory, which is the main reason we can employ the dispersion relation to do bootstrap.\footnote{In practice, the biggest challenge in applying this reasoning to high order in perturbation theory is the existence of operators with nearly-degenerate scaling dimension. One then needs to consider the individual contribution of each of these operators to the OPE, which is generally non-trivial.}

There is an important subtlety to the dispersion relation \eqref{dispersion},
namely that it works provided
\begin{align}
    F_{\phi \phi}(r,w) \sim w^{s^*}, \quad w \rightarrow \infty\,, \quad s^* < 0 \,.
\end{align}
If instead $s^* \ge 0$, the formula reproduces the two-point function up to the contribution of low spin defect operators
\begin{equation}
\label{eq:lowspin}
    F_{\phi \phi}(r,w) = \text{dispersion relation} +  \sum_{s=0}^{s^*} \sum_{\hat{\Delta}} b^2 _{\phi \hat{\mathcal{O}}_{\hat{\Delta},s}} \hat{f}_{\hat{\Delta}, s}(r, w) \,.
\end{equation}
A deeper discussion on the origin of this subtlety, along with a more complete treatment of the dispersion relation and related analytic bootstrap techniques, can be found in  \cite{Lemos:2017vnx, Bianchi:2022ppi, Barrat:2022psm }.

\subsection{Analytic bootstrap for the free bulk}

Let us start applying this machinery to the magnetic impurity \eqref{defect} in a free bulk theory in $d=4-\veps$ dimensions, a case for which we find certain results exactly in $\veps$.
This is an interesting result because even when the bulk is interacting, a situation to be considered in section \ref{sec:boot-int-blk}, the exact free-theory result captures an infinite subset of diagrams that contribute to the interacting correlator.

\subsubsection{Bootstrap of $\langle \phi \phi \rangle $}\label{Bootstrapofphiphi}

We first study the two-point function of the order parameter $\phi$.
In this case, we know that the OPE of $\phi$ contains only the identity and twist-two operators,\footnote{Intuitively, the reason why only twist-two operators appear in free theory is the following: since there are no interactions, only operators with two $\phi$'s appear in the OPE, namely $\phi \times \phi\sim \phi \partial_{\mu_1}\ldots \partial_{\mu_\ell} \Box^n \phi$. Those are exactly the double-twist operators. However, because of the equation of motion $\Box \phi = 0$, only the $n=0$ family of twist-two operators contributes. More rigorously, one should compute the four-point function of free fields $\phi_a$, and observe that its block expansion contains only twist-two operators.} namely
\begin{equation}
    \phi_a \times \phi_a = \mathds{1}+ \sum_\ell \mathcal{J}_{\ell}\,,
\end{equation}
with
\begin{equation}
    \Delta_{\ell}
    = 2\Delta_\phi +\ell, \quad \Delta_\phi = 1-\frac{\varepsilon}{2}\,.
\end{equation}
Because the exact OPE contains only twist-two operators, the discussion around \eqref{eq:disc-dt} applies here directly.
As we argued there, the discontinuity of twist-two operators is proportional to their anomalous dimension.
Since in free theory twist-two operators do not have anomalous dimension, the discontinuity only receives a contribution from the identity operator
\begin{equation} 
    \text{Disc} F_{\phi\phi} = \text{Disc} \ \xi^{-\Delta_\phi} = 2i\sin(\pi \Delta_{\phi}) (-\xi)^{-\Delta_\phi}\, .
\end{equation}
Notice that this equation is correct to all orders $\veps$, and not only at leading order.
Using the dispersion relation \eqref{dispersion}, and adding a low-spin ambiguity as in \eqref{eq:lowspin}, we obtain
\begin{align}
  F_{\phi\phi}(r,w)
 = \xi^{-\Delta_\phi}
 + \text{low-spin ambiguity} \, .
 \label{eq:Ft-free}
\end{align}
In order to fix the low-spin ambiguity, we can leverage the results of section \ref{sec:defectop} on what operators may appear in the defect channel.
As we saw there, the equation of motion $\Box \phi_a = 0$ constrains the dimensions of the defect operators that couple to it.
In particular, following \cite{Lauria:2020emq}, from
\begin{equation}
     \Box \langle \phi_a(x) \hat{\mathcal{O}_b}(\tau) \rangle = 0 \,,
\end{equation}
we can infer the existence of two families of operators:
\begin{itemize}
 \item Modes $\hat{\mathcal{O}}^a_{0,s} \sim \left(\partial_\perp \right)^s \phi^a$ with $s \ge 0$ and $\hat{\Delta}_{0,s} = \Delta_{\phi}+ s = 1 - \varepsilon/2 + s$.
 \item An operator $\hat{S}^a$, with $s = 0$ and $\hat{\Delta} = \varepsilon/2$.\footnote{More generally, we would find spinning operators with $\hat{\Delta} = \varepsilon/2-s$, but they break unitarity for $s>0$.}
 This is precisely the spin operator of section \ref{sec:defectop}, which is defined by \eqref{correlatorS}.
\end{itemize}

It is a known fact that the defect-channel expansion of the term $\xi^{-\Delta_\phi}$ contains the $\hat{\mathcal{O}}^a_{0,s}$ operators \cite{Lemos:2017vnx}, but it does not include any operator with the quantum numbers of $\hat S^a$.
As a result, we conclude that the dispersion relation fails to reproduce the contribution for spin $s=0$, and therefore we need to correct all possible $s=0$ operators.
The result is that the most general ansatz for the correlator is
\begin{equation}
\label{eq:ansatz-free}
 F_{\phi\phi}(r,w)
 = \xi^{-\Delta_\phi}
 + k_1 \hat{f}_{1-\varepsilon/2, 0}(r,w)
 + k_2 \hat{f}_{\varepsilon/2, 0} (r,w) \, ,
\end{equation}
where the extra terms are the defect blocks associated with the low-spin ambiguities.
The coefficients $k_1$ and $k_2$ are not arbitrary.
The reason is that for arbitrary $k_1$ and $k_2$, it is generically not possible to expand \eqref{eq:ansatz-free} in the bulk channel.
To be more precise, consider changing from radial coordinates $(r,w)$ to lightcone coordinates $(z,\bar{z})$, defined in \eqref{coordinates}.
In this coordinate system, one can see that the expansion of \eqref{eq:ansatz-free} around $|1-z| \ll |1-\zb| \ll 1$ contains spurious powers $(1-z)^n (1-\zb)^{-m}$ for $m \ge 2$, and spurious logarithms $\log(1-\zb)$ which are not accompanied by $\log(1-z)$.
These terms are incompatible with an expansion in terms of bulk-channel conformal blocks, and therefore we must choose the relative size of $k_1$ and $k_2$ to make sure they are absent.
After carrying out this procedure, we find that the free correlator is
\begin{align}
 F_{\phi\phi}(r, w)
 & = \xi^{-\Delta_\phi}
 + c_{\phi^2} J_\varepsilon(r) \, ,
 \label{eq:free-Fpp}
\end{align}
where we introduced\,\footnote{Note that this function does not depend on $w$ because it is a sum of two $s=0$ blocks \eqref{defectblock}.}
\begin{align}\label{Jepsexplicit}
 J_\varepsilon(r)
 = \frac{\Gamma  \left(\frac{1-\varepsilon}{2}\right)}
        {\sqrt{\pi} \Gamma  \left(\frac{2-\varepsilon}{2}\right)}
        \hat{f}_{\varepsilon/2,0}(r,w)
 + \frac{\Gamma  \left(\frac{\varepsilon-1}{2}\right)}
        {\sqrt{\pi} \Gamma  \left(\frac{\varepsilon}{2}\right)}
        \hat{f}_{1-\varepsilon/2,0}(r,w) \, .
\end{align}
Let us stress that this correlation function is exact to all orders in $\veps$.
However, it depends on one parameter $c_{\phi^2}$ that cannot be fixed by the bootstrap. Since \eqref{eq:free-Fpp} is exact in $\veps$, it is possible to investigate the properties of the fixed point in three dimensions by simply setting $\veps=1$. Even though in \eqref{Jepsexplicit} there are some divergent factors, one can check that $J_\veps (r) $ is perfectly finite in the $\veps \rightarrow 1$ limit.
We are left with two possibilities: either $ c_{\phi^2}\big|_{\veps=1}=0$, or $ c_{\phi^2}\big|_{\veps=1}\neq 0$. In the first case, $F_{\phi\phi}$ is just a free correlator. This is sufficient to show that $\phi$ satisfies the free-field equations of motion, and therefore all its correlators are those of the free theory. Instead if $ c_{\phi^2}\big|_{\veps=1}$ is a finite non-zero number, we can try to expand the correlator in the defect channel by taking $r \ll 1$. However, this expansion contains terms with factors of $\log r$ that cannot be reproduced by the defect blocks. Therefore, this correlator does not obey the defect bootstrap equation. The inevitable conclusion is that in three dimensions and for a free bulk there is no non-trivial magnetic impurity.

For $0<\veps < 1$, instead, the function $J_\varepsilon(r)$ is a truncated solution of crossing,\footnote{This is an analog of the solutions of crossing with finite support in spin of \cite{Alday:2016jfr}, which play an important role in the $\varepsilon$--expansion bootstrap for four-point functions \cite{Alday:2017zzv,Henriksson:2018myn,Henriksson:2020fqi,Henriksson:2021lwn,Gimenez-Grau:2021wiv}.} because it has sensible bulk and defect expansions on its own, and it involves finitely many transverse spins (in this case, $s=0$ only).

As an interesting exercise, from the full two-point function we can still extract the CFT data for $\veps < 1$ in both OPE channels as a function of $c_{\phi^2}$. Let us start with the defect expansion, which from the discussion above takes the form
\begin{align}
 F_{\phi\phi}(r, w)
 = b_{\phi \hat{\mathcal{O}}_{0,0}}^2 \hat{f}_{1-\varepsilon/2, 0}(r, w)
 + b_{\phi \hat{S}}^2 \hat{f}_{\varepsilon/2, 0}(r, w)
 + \sum_{s=1}^\infty b_{\phi \hat{\mathcal{O}}_{0,s}}^2 \hat{f}_{\Delta_\phi+s, s}(r, w) \, .
\end{align}
Using the formula for conformal blocks \eqref{defectblock} it is not hard to extract the CFT data
\begin{align}
 b_{\phi \hat{\mathcal{O}}_{0,0}}^2 = 1 +\frac{\Gamma \! \left(\frac{\varepsilon-1}{2}\right)}
        {\sqrt{\pi} \Gamma \! \left(\frac{\varepsilon}{2}\right)} c_{\phi^2}\, , \qquad
b_{\phi \hat{S}}^2 = \frac{\Gamma \! \left(\frac{1-\varepsilon}{2}\right)}
        {\sqrt{\pi} \Gamma \! \left(\frac{2-\varepsilon}{2}\right)} c_{\phi^2} \, , \qquad
b_{\phi \hat{\mathcal{O}}_{0,s}}^2 =  \frac{2^s (\Delta_\phi)_s}{s!} \, .
\end{align}
Similarly, using the formulas for the bulk blocks \eqref{bulkblock} we find the expansion
\begin{align}
  F_{\phi\phi}(r, w)
 = \xi^{-\Delta_\phi}
 + \xi^{-\Delta_\phi}
   \sum_{\ell=0}^\infty \lambda_{\phi\phi\mathcal{J}_{\ell}}
   a_{\mathcal{J}_{\ell}} f_{2\Delta_\phi +\ell, \ell}(r, w) \, .
   \label{eq:freebulkexp}
\end{align}
The bulk expansion contains only twist-two operators, as we expect in the bulk-free theory.
Here $\lambda_{\phi\phi\mathcal{J}_{\ell}}$ are thee-point OPE coefficients of the $O(3)$ model at the free fixed point, which are known exactly
\begin{equation} \label{lambdafree}
   \lambda_{\phi\phi\mathcal{J}_{\ell}}
 =\sqrt{\frac{2}{3}} \frac{2^{\frac{\ell}{2}} (\Delta_\phi)_\ell}{\sqrt{ \ell! (2 \Delta_\phi + \ell - 1)_\ell}}\,.
\end{equation}
As a result, from the block expansion \eqref{eq:freebulkexp}, we obtain a prediction for all the one-point functions of twist-two operators:
\begin{align}
\label{eq:afreeboot}
 a_{\phi^2}
 = \sqrt{\frac{3}{2}} \, c_{\phi^2} \, , \qquad
 a_{\mathcal{J}_\ell}
 = \frac{(1-\varepsilon)_\ell \left(\frac{\ell-\varepsilon+2}{2}\right)_{\frac{\ell}{2}} \sqrt{\ell! (\ell-\varepsilon+1)_\ell}}{2^{5\ell/2} \left(\frac{\ell}{2}!\right)^2 \left(\frac{\ell-\varepsilon+1}{2} \right)_{\frac{\ell}{2}} \left(1-\frac{\varepsilon}{2}\right)_\ell}
 a_{\phi^2} \, .
\end{align}
The particular case $\ell=2$ corresponds to the stress tensor $a_{\mathcal{J}_2} \sim \langle T_{\mu \nu} \rangle_{\mathcal{D}_j}$.\footnote{In the $\mathcal{N} = 4$ SYM literature this observable is usually called the Bremsstrahlung function.}
This has been conjectured to be positive \cite{Lemos:2017vnx}, hence in bulk-free theory we should have $a_{\phi^2} > 0$.

At the end of the day, we see that the two-point function of $\phi$ and the CFT data are entirely fixed by the bootstrap analysis up to an undetermined constant, which corresponds to the value of the one-point function $a_{\phi^2}$ of the unit-normalized $\phi^2$ operator at the critical point.
We shall compute this to order $O(\veps^3)$ in equation \eqref{1ptphi2} below.

\subsubsection{Bootstrap of $\langle \phi^2 \phi^2 \rangle$}

Let us now calculate the two-point function of $\phi^2$. 
Once again the bulk OPE is known exactly
\begin{equation}
    \phi^2 \times \phi^2 = \mathds{1}+ \sum_{\ell} \mathcal{J}_{\ell}+ \sum_{n>1}\sum_{\ell} \mathcal{O}_{n,\ell}\,,
\end{equation}
where
\begin{equation}
    \mathcal{O}_{n, \ell}
    \sim \phi^a \square ^n \partial_{\mu_1} \ldots \partial_{\mu_\ell} \phi_a \, ,
\end{equation}
with
\begin{equation}
    \Delta_{n,\ell} = 2\Delta_{\phi^2}+2n +\ell, \quad \Delta_{\phi^2} = 2\Delta_\phi= 2-\veps\,.
\end{equation}
The bulk expansion reads
\begin{align}
 F_{\phi^2\phi^2}(r, w)
 = \xi^{-2\Delta_\phi}
 + \xi^{-2\Delta_\phi}\sum_{\ell=0}^\infty  \lambda_{\phi^2 \phi^2 \mathcal{J}_{\ell}} a_{\mathcal{J}_{\ell}} f_{2\Delta_\phi+\ell, \ell}
 + \xi^{-2\Delta_\phi}\sum_{n,\ell=0}^\infty c_{n,\ell} f_{4\Delta_\phi+\ell+2n, \ell} \, .
\end{align}
In this case the discontinuity receives contributions from the identity and $\mathcal{J}_{\ell}$ operators, but it kills all of the $\Om_{n,l}$ operators.
To compute the discontinuity of the $\mathcal{J}_{\ell}$ operators, notice that although $\xi^{-2\Delta_{\phi}} f_{2\Delta_\phi +\ell, \ell}(r, w)$ has a branch cut, the combination $\xi^{-\Delta_\phi} f_{2\Delta_\phi +\ell, \ell}(r, w)$ does not have a discontinuity, see \eqref{bulkblock}.
As a result, the discontinuity reads
\begin{align}
 \Disc F_{\phi^2\phi^2}(r, w)
 & = \Disc \Big[
     \xi^{-2\Delta_\phi} \Big] +\sum_{\ell=0}^\infty \lambda_{\phi^2 \phi^2 \mathcal{J}_{\ell}} a_{\mathcal{J}_{\ell}} \xi^{-\Delta_\phi} f_{2\Delta_\phi +\ell, \ell}(r, w) \Disc \Big[\xi^{-\Delta_\phi}
 \Big] \, .
\end{align}
Furthermore, one can check that in free theory $\lambda_{\phi^2 \phi^2 \mathcal{J}_{\ell}} = 2 \lambda_{\phi\phi\mathcal{J}_{\ell}}$.
Therefore, the sum that appears in the discontinuity gives the same result as in the $\langle \phi \phi \rangle_{\mathcal{D}_j} $ correlator, up to a factor of $2$.
All in all:
\begin{align}
 \Disc F_{\phi^2\phi^2}(r, w)
 & = \Disc \Big[
     \xi^{-2\Delta_\phi} \Big] +2 c_{\phi^2}  J_\varepsilon(r) \Disc \Big[\xi^{-\Delta_\phi}
 \Big] \, .
\end{align}
Both terms get mapped to themselves by the dispersion relation \eqref{dispersion} because they are of the form $\xi^{\alpha} f(r)$, and the dispersion relation only involves $w$.
There is an obvious spin $s=0$ contribution from the defect identity $a_{\phi^2}^2$ that is missed by the dispersion relation. In principle there could be other low-spin ambiguities.
Contrary to the previous case, to our knowledge there are no powerful Ward identities that can constrain the form of the low spin ambiguities in $\langle \phi^2 \phi^2 \rangle_{\mathcal{D}_j}$, therefore the conclusion of our bootstrap analysis is
\begin{align}\label{bootsphi2}
 F_{\phi^2\phi^2}(r, w)
 & = \xi^{-2\Delta_\phi}
   + 2 c_{\phi^2} \xi^{-\Delta_\phi} J_\varepsilon(r)
   + a_{\phi^2}^2 \, + \text{low-spin ambiguity}\,.
\end{align}
From a diagrammatic computation we shall see that, at next-to-leading order, there is indeed an extra term which corresponds, in terms of CFT data, to the contribution of the operator $\hat{\Phi}$ and an infinite family of integer twist operators, both with spin $s=0$.
This means that the bootstrap result is not complete, but it can still be used to extract defect CFT data with $s>0$.
We refer to the discussion around \eqref{2ptphi2} for further details on the low spin contribution and move on to the extraction of the CFT data in the defect.
In the defect channel we find two families of operators
\begin{align}
 F_{\phi^2\phi^2}(r, w)
 = \sum_{n,s=0}^\infty \left(
   b_{\phi \hat{\mathcal{O}}_{n,s}}^2\,  \hat{f}_{\hat{\Delta}_{n,s}, s}
 + b_{\phi \hat{\mathcal{J}}_{n,s}}^2 \, \hat{f}_{\hat{\Delta}_{\hat{\mathcal{J}}_{n,s}}, s}
 \right) \, .
\end{align}
The first family of operators has the interpretation $\hat{\mathcal{O}}_{n,s} \sim (\partial_\bot)^s \Box^n \phi^2$, with
\begin{align}
 \hat{\Delta}_{s,n}
&= 2 \Delta_\phi + s + 2n \, , \\
  b_{\phi \hat{\mathcal{O}}_{n,s}}^2 
&= \textstyle c_{\phi^2} \frac{ 2^s \left(\frac{3}{2}\right)_{n-1} \Gamma \left(\frac{\veps -1}{2}\right) \left(1-\frac{\veps }{2}\right)_n ^2 \left(n-\frac{\veps }{2}+1\right)_s (n+s-\veps +2)_n}{\sqrt{\pi } n! \Gamma \left(\frac{\veps }{2}\right) (n+s)! \left(\frac{3}{2}-\frac{\veps }{2}\right)_n \left(n+s-\frac{\veps }{2}+1\right)_n}
+\frac{2^s \left(1-\frac{\veps }{2}\right)_n (2-\veps )_{2 n+s}}{n! (n+s)! \left(n+s-\frac{\veps }{2}+1\right)_n} \, ,
\end{align}
whereas the second family is $\hat{\mathcal{J}}_{n,s} \sim (\partial_\perp)^s \Box^n 
\phi^a T_a$, with
\begin{align}
  \hat{\Delta}_{\hat{\mathcal{J}}_{n,s}}
& = 1 + s + 2n  \, , \\
  b_{\phi \hat{\mathcal{J}}_{n,s}}^2
& = c_{\phi^2} \frac{ 2^s (-4)^n \left(\frac{3}{2}\right)_{n-1} \Gamma \left(\frac{1-\veps }{2}\right) (2 n+s)! \left(\frac{\veps }{2}\right)_n \left(-n-\frac{\veps }{2}+1\right)_{2 n+s}}{\sqrt{\pi } n! \Gamma \left(1-\frac{\veps }{2}\right) ((n+s)!)^2 (\veps )_{2 n} \left(n+s+\frac{\veps }{2}\right)_n}\, .
\end{align}
Let us stress again that these results are only valid for $s>0$ due to low-spin ambiguities.
Notice that the operators $\hat \Jm_{n,s}$ have integer scaling dimension. In particular, the operators $\hat \Jm_{0,s}$ are related to the higher spin symmetries in the free bulk theory which are broken by the defect, as we discussed in Section \ref{sec:defectop}. Finally, since low $s$ ambiguities potentially contribute to all data in the bulk OPE, we cannot extract any meaningful observable in that channel.

\subsection{Analytic bootstrap for the interacting bulk}\label{sec:boot-int-blk}
\subsubsection{Bootstrap of $\langle \phi \phi \rangle$}
Let us now assume that the bulk theory is the $O(3)$ model at the Wilson-Fisher fixed point in $d=4-\veps$ dimensions, restricting to the first non-trivial order in the perturbative expansion at small $\veps$.
We want to study the two-point function of the order parameter $\phi_a$, namely
\begin{equation}
    \langle \phi_a (x) \phi_a (y) \rangle _{\mathcal{D}_j} = \frac{ F_{\phi \phi} (r,w)}{|x_\perp |^{\Delta_\phi} |y_\perp |^{\Delta_\phi}}\,.
\end{equation}
The same correlator was computed in presence of a different line defect using the analytic bootstrap in a series of recent papers \cite{Gimenez-Grau:2022ebb, Bianchi:2022sbz}.
It turns out that the computation in the present case is very similar.
The reason is very simple: we compute the discontinuity by expanding the two-point function in bulk blocks and evaluating the discontinuity of each block. At first order in perturbation theory, the discontinuity of a block is proportional to the anomalous dimension of the corresponding bulk operator, which is independent of the defect. Finally, it turns out that in the $O(3)$ model all the operators that appear in the $\phi \times \phi$ OPE at order $\veps$ have vanishing anomalous dimensions, except for one operator. 
More precisely, we have the bulk OPE
\begin{equation}
    \phi_a \times \phi_a = \mathds{1}+ \sum_\ell \mathcal{J}_{\ell}+ \sum_{n>1} \sum_{\ell} \mathcal{O}_{n,\ell}\,,
\end{equation}
where, for the leading twist family \cite{Henriksson:2022rnm, Kehrein:1992fn}
\begin{equation}
    \Delta_{\ell} = 2\Delta_\phi +\ell +\veps \frac{5 }{11} \delta_{\ell,0} +O(\veps^2)\,, \quad \Delta_\phi = 1-\frac{\varepsilon}{2}+O(\veps^2)\,,
\end{equation}
and
\begin{equation}
     \lambda_{\phi \phi \mathcal{J}_{\ell}} =  \lambda_{\phi \phi \mathcal{J}_{\ell}}^{(0)} + \veps \lambda_{\phi \phi \mathcal{J}_{\ell}}^{(1)}+ \veps^2 \lambda_{\phi \phi \mathcal{J}_{\ell}}^{(2)} +O(\veps^3)\,,
      \quad a_{_{\mathcal{J}_{\ell}}} =  a_{\mathcal{J}_{\ell}}^{(0)}+  \veps a_{\mathcal{J}_{\ell}}^{(1)}+  \veps^2 a_{\mathcal{J}_{\ell}}^{(2)}  +O\left(\veps^3\right)\,.
\end{equation}
For higher-twist operators, we have 
\begin{equation}
      \Delta_{n,\ell} = 2\Delta_\phi+2n +\ell +\veps \gamma_{n, \ell} ^{(1)} +O(\veps^2)\,,
      \quad \lambda_{\phi \phi \mathcal{O}_{n,\ell}} =  \veps \lambda_{\phi \phi \mathcal{O}_{n,\ell}}^{(1)}+ \veps^2 \lambda_{\phi \phi \mathcal{O}_{n,\ell}}^{(2)} +O(\veps^3) \, .
\end{equation}
Therefore only the bulk identity and $\phi^2$ operators contribute to the discontinuity. All the other operators do not contribute at the order we are working because their anomalous dimension or OPE coefficients are higher order.
At the end of the day, the discontinuity reads
\begin{equation}\label{discint}
   \Disc F_{\phi \phi}(r, w) = \Disc \xi^{-\Delta_\phi}+\veps^2 \frac{5 \pi i}{11} \lambda_{\phi \phi \phi^2}^{(0)} a_{\phi^2}^{(1)}  \xi^{-1} f_{2,0}(r,w) +O(\veps^3)\,.
\end{equation}
A comment is in order: the one-point coefficient $a_{\phi^2}$ which appears in the discontinuity cannot be determined by the bootstrap but, at leading order,  coincides with the tree-level free theory result \cite{Cuomo:2022xgw}, and in particular it turns out that $a_{\phi^2} \sim \veps$.
Therefore the non-trivial correction to the discontinuity starts at order $\veps^2$.
The discontinuity \eqref{discint} is the same that was found in \cite{Gimenez-Grau:2022ebb, Bianchi:2022sbz}, up to a different factor in front of the non-trivial term, which depends on the specific defect through the one-point function coefficient $a_{\phi^2}$.
The other coefficient $\lambda_{\phi \phi \phi^2}$ does not depend on the defect, just like the anomalous dimensions of bulk operators, and has the value \cite{Dey:2016mcs,Alday:2017zzv, Henriksson:2018myn}
\begin{equation}\label{lambdaint}
    \lambda_{\phi \phi \phi^2} = \sqrt{\frac{2}{3}}\left(1-\veps \frac{5}{22}\right) + O(\veps^2)\,.
\end{equation}
In particular, at leading order $\lambda_{\phi \phi \phi^2}^{(0)} = \sqrt{\frac{2}{3}}$.
The discontinuity and the result of the dispersion relation can both be evaluated explicitly in terms of special functions. 
It turns out that the result of the dispersion relation is
\begin{equation}
    F_{\phi \phi}(r,w) = \xi^{-\Delta_\phi}
 + \veps^2 \sqrt{\frac{2}{3}} \frac{5 a_{\phi^2}^{(1)}}{11}  H(r,w)+ \text{low spin}+ O(\veps^3)\,,
\end{equation}
where $H(r,w)$ can be conveniently expressed as
\begin{equation}\label{Hfunction}
     H(r, w)= \xi^{-1} \left(\partial_{\Delta}-1-\log 2\right) f_{2,0}(r, w)\,,
\end{equation}
where as always $ f_{2,0}(r, w)$ is a bulk block.
This function can also be represented in a variety of different ways which are better suited for explicit evaluation or the extraction of the CFT data, for example as a multivariable hypergeometric function.
We refer to \cite{Gimenez-Grau:2022ebb, Bianchi:2022sbz} for more details on the computation of $H(r,w)$.
As in the free bulk case, the result of the dispersion relation may miss low spin contributions.
For example, in the free theory discussed in the previous section, we had to add by hand a truncated solution to crossing $J_\veps(r)$ to the correlator.
Since now we work perturbatively in $\veps$, it is convenient to expand this function
\begin{align}
\begin{split}
 J_\veps(r)
 = 1
 + \frac{\veps}{2} \log \frac{4 r}{(1 + r)^2}
 + O(\veps^2) \, .
 \label{eq:J-epsexp}
\end{split}
\end{align}
We expect a similar correction in the present case. Our goal here is to see if we can find more general truncated solutions to be added to the final interacting correlator.
To be able to make progress we make the simplifying assumption that only the defect twist-one family contributes.
This assumption is motivated by the fact that this is what happens in the free bulk case (and in a variety of other perturbative setups) and we expect the defect spectrums in the free and interacting case to be perturbatively close to each other.
Provided the assumption is true, the most general ansatz for the ambiguity is
\begin{align}
\label{eq:amb-ansatz}
 F_{\text{amb}}(r, w)
 = \left( q_0 + r_0 \partial_{\hat{\Delta}} \right) \hat{f}_{0,0} 
 + \left( q_1 + r_1 \partial_{\hat{\Delta}} \right) \hat{f}_{1,0} 
 + \sum_{s=1}^{s_*} \big( q_{s+1} + r_{s+1} \partial_{\hat{\Delta}} \big) \hat{f}_{s+1,s}  \,,
\end{align}
where $q$ and $r$ are arbitrary constants.
For any finite $s_* \in \mathbb{N}$, the conformal blocks in this sum can be written as polylogarithms using \texttt{HypExp} \cite{Huber:2005yg,Huber:2007dx}.
For example, the lowest lying ones take a very simple form:
\begin{align}
 \hat{f}_{0,0}(r,w) = 1 \, , \quad
 \partial_{\hat{\Delta}} \hat{f}_{0,0}(r,w)
 = \log \frac{r}{1-r^2} \, , \quad
 \hat{f}_{1,0}(r,w)
 = \tanh^{-1} \! r \, , \quad \ldots
\end{align}
Given the ansatz \eqref{eq:amb-ansatz}, one can attempt to expand it in the bulk channel, but generically this is not possible.
The reason can be seen more easily by changing from radial coordinates $(r,w)$ to lightcone coordinates $(z,\bar{z})$, defined in \eqref{coordinates}.
In this coordinate system, one can see that the expansion of \eqref{eq:amb-ansatz} around $|1-z| \ll |1-\zb| \ll 1$ contains spurious powers $(1-z)^n (1-\zb)^{-m}$ for $m \ge 2$, and spurious logarithms $\log(1-\zb)$ which are not accompanied by $\log(1-z)$.
These terms are incompatible with an expansion in terms of bulk-channel conformal blocks \eqref{bulkblock}, since the latter are symmetric in $z$ and $\Bar{z}$.
Therefore, demanding they vanish puts non-trivial constraints in the ansatz.
Although we do not have a general proof, by experimenting with low values $s_* \le 5$, we conclude that the most general truncated solution is
\begin{align}
\label{eq:ambiguity-final}
 F_{\text{amb}}(r, w)
 = q_0 \hat{f}_{0,0}(r,w)
 + r_0 \left( \partial_{\hat{\Delta}} \hat{f}_{0,0}(r,w)
   - 2 \hat{f}_{1,0}(r,w) \right)
 = q_0
 + r_0 \log \frac{r}{(1 + r)^2} \, .
\end{align}
Both equation \eqref{eq:J-epsexp} and \eqref{eq:ambiguity-final} suggest the ambiguities of interest are a constant and a logarithm.
At the end of the day the ansatz for the correlator can be expressed as \\
\begin{equation}
\label{correlatoransatz}
 F_{\phi \phi}(r,w) 
 = \xi^{-\Delta_\phi}
 + \veps^2 \sqrt{\frac{2}{3}}  \frac{5  a_{\phi^2}^{(1)}}{11} H(r, w)
 + q_0
 + r_0 \log \frac{r}{(1 + r)^2}
 + O(\veps^3) \, .
\end{equation}
The constants $ a_{\phi^2}^{(1)}, q_0, r_0$ cannot be fixed from the bootstrap alone. However, it can be shown that they are not independent. One can fix $r_0$ in terms of $ a_{\phi^2}^{(1)}$ by exploiting the analysis on the defect spectrum in Section \ref{sec:defectop}.
Indeed, the defect expansion has the same form as in the case of the free bulk
\begin{align}
 F_{\phi\phi}(r, w)
 = b_{\phi \hat{\mathcal{O}}_{0,0}}^2 \hat{f}_{\hat{\Delta}_{0,0}, 0}(r, w)
 + b_{\phi \hat{S}}^2 \hat{f}_{\hat{\Delta}_{\hat{S}}, 0}(r, w)
 + \sum_{s=1}^\infty b_{\phi \hat{\mathcal{O}}_{0,s}}^2 \hat{f}_{\Delta_\phi+s, s}(r, w) \, .
\end{align}
and in particular it contains the spin operator $\hat{S}^a$. 
As discussed in Section \ref{sec:defectop}, while this operator has no longer protected dimension if the bulk is not free, one can see from the Ward identity that the correction to the anomalous dimension starts at order $\veps^2$.
Therefore, the leading dimension must coincide with the one in the free bulk case. This fixes $r_0 = \veps \sqrt{\frac{2}{3}} \frac{a_{\phi^2}^{(1)}}{2}$.
Finally, one can fix $q_0$ in terms of $ \lambda_{\phi \phi \phi^2} a_{\phi^2}$ just by expanding the correlator in the bulk channel, namely
\begin{align}
\begin{split}
  F(r, w)
 = \xi^{-\Delta_\phi}
 + c_{\phi^2} \xi^{-\Delta_\phi} f_{\Delta_{\phi^2},0}
 + \xi^{-\Delta_\phi} \sum_{\ell=2}^\infty c_{\ell} f_{2\Delta_\phi+\ell,\ell}
 \, .
\end{split}
\end{align}
By comparing with \eqref{correlatoransatz} we fix $q_0 = \lambda_{\phi \phi \phi^2} a_{\phi^2} +\veps^2 \sqrt{\frac{2}{3}} a_{\phi^2}^{(1)}\left(\frac{5}{11}+\frac{16}{11} \log2\right)$.
Therefore, at the end of the day the correlator and all the CFT data are fixed in terms of a single unknown one-point function coefficient $a_{\phi^2}$, more precisely we obtain
\begin{align}
\label{phiintboot}
 F(r,w) 
 = \xi^{-\Delta_\phi}
 +c_{\phi^2} \left(1+ 
   \frac{\veps}{2} \log \frac{4 r}{(1 + r)^2} + \veps \, \frac{5}{11} \left(1+ \log 2+ H(r, w) \right)
 \right)
 + O(\veps^3)\,,
\end{align}
where as always $c_{\phi^2} = \lambda_{\phi \phi \phi^2} a_{\phi^2} =  \sqrt{\frac{2}{3}}(1-\veps \frac{5}{22}) \left( \veps a_{\phi^2}^{(1)}+\veps^2 a_{\phi^2}^{(2)}\right)+O(\veps^3) $.
The CFT data for the defect spin operator reads
\begin{align}
 & \hat{\Delta}_{\hat{S}} =\frac{\veps}{2} + O(\veps^2) \, , \label{dimdefspin} \\
 &  b_{\phi \hat{S}}^2 = c_{\phi^2}+\veps^2 \sqrt{\frac{2}{3}} a_{\phi^2}^{(1)} \left(\frac{5}{11}+\frac{16}{11}\log2 \right)+ O(\veps^3) \,.
\end{align}
Notice that from \eqref{phiintboot} we can extract the defect spin dimension up to $\mathcal{O}(\veps)$ because $b_{\phi \hat{S}}$ is also $\mathcal{O}(\veps)$. However let us stress that the $\mathcal{O}(\veps^2)$ for $\hat{\Delta}_{\hat{S}}$ is known and we reported it in \eqref{deltaSint}.
Moving on to the other operators in the defect channel, we see that for the operator $\hat{\mathcal{O}}_{0,0}$ the CFT data is
\begin{align}
 & \hat{\Delta}_{0,0} = \Delta_\phi + \veps^2 \sqrt{\frac{2}{3}} \frac{10 a_{\phi^2}^{(1)}}{11}  + O(\veps^3) \, , \\
&  b_{\phi \hat{\mathcal{O}}_{0,0}}^2 = 1
  -   \veps^2 \sqrt{\frac{2}{3}} a_{\phi^2}^{(1)} \left( \frac{31}{11}-\frac{20}{11}\log 2 \right)
   + O(\veps^3) \, .
\end{align}
Finally, we find a single infinite family of defect operators  $\hat{\mathcal{O}}_{0,s}$  with
\begin{align}
 \hat{\Delta}_{0,s}
 & = \Delta_\phi
   + s
   + \veps^2 \sqrt{\frac{2}{3}} \frac{5 a_{\phi^2}^{(1)}}{11} \frac{1}{s + 1/2}
   + O(\veps^3) \, , \\
  b_{\phi \hat{\mathcal{O}}_{0,s}}^2 
 & =  2^s \left(\frac{(\Delta_\phi)_s}{s!}
   + \veps^2 \sqrt{\frac{2}{3}} \frac{5 a_{\phi^2}^{(1)}}{11} \left(
   \frac{ H_s-H_{s-1/2} }{s+1/2}
   - \frac{ 1 }{(s+1/2)^2}
 \right)
 + O(\veps^3) \right) \, .\label{bphiOs}
\end{align}
In the bulk channel we have the twist-two operators $\mathcal{J}_{\ell}$ with 
\begin{align}
\begin{split}
 c_{\ell} 
 & = \frac{\Gamma \! \left(\frac{\ell+1}{2}\right) \Gamma (\ell+1)^2} 
        {8^{\ell} \left(\frac{\ell}{2}!\right)^2 
         \Gamma \! \left(\frac{\ell+2}{2}\right) 
         \Gamma \! \left(\frac{2 \ell+1}{2}\right)} \times \\
 & \qquad \quad \times \left[ c_{\phi^2}+ \veps^2
   \sqrt{\frac{2}{3}} a_{\phi^2}^{(1)} \left(\frac{5}{11}\left(1+\log2\right)+\frac{1}{2}\left( H_{\frac{\ell}{2}}-H_{\frac{\ell-1}{2}}+H_{\ell-\frac{1}{2}}-3  H_\ell\right)
    \right)
    + O(\veps^3)
\right] \, .
\end{split}
\end{align}
Notice the absence of double-twist operators with twist higher than $2$. This is consistent with the fact that \cite{Alday:2017zzv, Kehrein:1992fn} $\lambda_{\phi \phi \mathcal{O}_{n,\ell}} \sim \veps$ and $a_{\mathcal{O}_{n,\ell}} \sim \veps^2$.
The latter fact follows immediately by considering tree-level Feynman diagrams.

\section{Diagrammatic computation}
\label{sec:diagrams}
In this section we will outline the diagrammatic computation for the correlators of the bulk fields $\phi_a$ and $\phi^2$ and compare it with the bootstrap results of Section \ref{sec:bootstrap}.
\subsection{One-point function \texorpdfstring{$\langle \phi^2 \rangle$}{< phi2 >}}
\subsubsection{Free bulk}
We start from the computation of the one-point function of $\phi^2$ in the free theory,\footnote{The one-point function of $\phi^a$ is zero because of symmetry.} which was already computed at next-to-leading order in \cite{Cuomo:2022xgw}.
This observable is not accessible by our bootstrap analysis and indeed it is the only information needed to completely fix the two point function of $\phi$.  
Since the bulk is free, we have two ways of doing the computation: we can exploit the Ward identity and write the bulk correlator in terms of an integrated defect correlator using \eqref{fielddec}, or perform a direct computation of the bulk correlator in terms of Feynman diagrams.
In terms of the defect correlator, we have
 \begin{equation}
 \begin{split}
     \langle \phi^2 (0,x_\perp) \rangle_{\mathcal{D}_j} &= \kappa \, \zeta^2 \int d\tau \int d\tau' \frac{\braket{\hat{S}^a (\tau) \hat{S}^a (\tau')}_{\mathcal{D}_j}}{(\tau^2+|x_\perp |^2)^{1-\frac{\veps}{2}} ({\tau'}^2+|x_\perp |^2)^{1-\frac{\veps}{2}}}\,.
 \end{split}
 \end{equation}
At the fixed point, $\langle \hat{S}^a (\tau) \hat{S}^a (\tau') \rangle_{\mathcal{D}_j} $ is given by \eqref{defectspintwopoint}.
By computing the integrals, we find\footnote{Notice that in the bootstrap computation we have taken the operators to be unit-normalized, as is customary in the CFT literature. In the diagrammatic calculation it is convenient to use a different normalization, this is why the one point function here has an extra factor compared to \eqref{1ptdef}. } 
\begin{equation}\label{1ptphi2general}
\begin{split}
    \langle \phi^2 (0,x_\perp) \rangle _{\mathcal{D}_j} &=  \frac{ \kappa \,  \zeta_* ^2 \mathcal{N}_{\hat{S}}}{|x_\perp |^{2-\veps}}  \frac{\pi ^{3/2} \Gamma \left(\frac{1}{2}-\frac{\veps }{2}\right)}{\Gamma \left(1-\frac{\veps }{2}\right)} =\\
     &\equiv \frac{\Nm_{\phi^2} a_{\phi^2}}{|x_\perp| ^{2-\veps}}\,.
\end{split}
\end{equation}
Here $\Nm_{\phi^2}$ is the normalization of the two-point function, which according to our conventions is
\begin{align}
 \langle \phi^2(x) \phi^2(0) \rangle
 = \frac{\Nm_{\phi^2}^2}{|x|^{2\Delta_{\phi^2}}} \, , \qquad
 \Nm_{\phi^2}^2
 = 6 \kappa^2 \, .
\end{align}
If we plug in the value of the coupling at the fixed point \eqref{fixedcouplingfree} and the normalization constant $\mathcal{N}_{\hat{S}}$ \eqref{defectspincoefficient}, we obtain\footnote{Our result for the one-point function at the fixed point differs from the one in equation $2.14$ of \cite{Cuomo:2022xgw} at order $\veps^2$. We believe this is due to the fact that the authors of \cite{Cuomo:2022xgw} used the critical coupling at leading order instead of next-to-leading order, thus missing a contribution of order $\veps^2$ in the one-point coefficient. }
\begin{equation}\label{1ptphi2}
         a_{\phi^2} = \frac{\pi ^2 j (j+1) \veps }{2 \sqrt{6}} \left( 1+\veps \frac{\log4-1}{2}+\veps^2 \frac{2 \pi ^2 j (j+1)+(\log4-2) \log4}{8}\right) + O(\veps^4) \, .
\end{equation}
We checked that this result can be reproduced from Feynman diagrams.

\subsubsection{Interacting bulk}
When we add interactions, it is no longer particularly convenient to express the bulk correlator in terms of a defect one, since the Ward identity \eqref{ward1} gets corrected and the relation between the two correlators becomes more involved.
Therefore we will perform the computation using Feynman diagrams.
The diagrams that contribute to the one-point function $\langle \phi^2 \rangle_{\mathcal{D}_j}$ up to order $\veps^2$ are
\begin{align}
 \langle \phi^2 (0,x_\perp) \rangle_{\mathcal{D}_j}
 \; = \;
 \begin{tikzpicture}[valign]
    \coordinate   (x1) at (-0.0, 0.7);
    \coordinate   (y1) at (-0.5,   0);
	\coordinate   (y2) at (+0.5, 0.0);
    \draw[double,thick,blue] (y1) -- (y2);
    \node         (y3) at (-0.2,   0) [vtx] {};
	\node         (y4) at (+0.2, 0.0) [vtx] {};
	\draw [thick,black]   (x1) to[out=-120,in=90] (y3);
	\draw [thick,black]    (x1) to[out=-60,  in=90] (y4);
	 \draw[blue, fill=white] (0.0,0.7) circle (2pt);
	 \draw[blue, fill=blue] (y3) circle (2pt);
	 \draw[blue, fill=blue] (y4) circle (2pt);
 \end{tikzpicture}
 \; + \;
 \begin{tikzpicture}[valign]
    \coordinate   (x1) at (-0.0, 0.7);
    \coordinate   (y1) at (-0.7,   0);
	\coordinate   (y2) at (+0.7, 0.0);
	\draw [double,thick,blue] (y1) -- (y2);
    \node         (y3) at (-0.5,   0) [vtx] {};
	\node         (y4) at (+0.5, 0.0) [vtx] {};
	\node         (y5) at (-0.2,  0) [vtx] {};
	\node         (y6) at (+0.2,  0) [vtx] {};
	\draw[thick,black]  (x1) to[out=-150,in=90] (y3);
	\draw [thick,black]  (x1) to[out=-30,  in=90] (y4);
	\draw [thick,black] (-0.2,0) to[out=90,in=180] (0, 0.3)to[out=0,in=-90] (0.2,0);
	 \draw[blue, fill=white] (0.0,0.7) circle (2pt);
	 \draw[blue, fill=blue] (y3) circle (2pt);
	 \draw[blue, fill=blue] (y4) circle (2pt);
	 	 \draw[blue, fill=blue] (y5) circle (2pt);
	 \draw[blue, fill=blue] (y6) circle (2pt);
 \end{tikzpicture}
 \; + \;
 \begin{tikzpicture}[valign]
    \coordinate   (x1) at (+0.5, 0.7);
    \coordinate   (y1) at (-0.6,   0);
	\coordinate   (y2) at (+1.0, 0.0);
	\draw [double,thick,blue] (y1) -- (y2);
    \node         (y3) at (+0.3,   0) [vtx] {};
	\node         (y4) at (+0.7, 0.0) [vtx] {};
	\node         (y5) at (-0.4,   0) [vtx] {};
	\node         (y6) at (+0.0, 0.0) [vtx] {};
	\draw [thick,black]   (x1) to[out=-120,in=90] (y3);
	\draw [thick,black]   (x1) to[out=-60,  in=90] (y4);
	\draw [thick,black] (-0.4,0) to[out=90,in=180] (-0.2, 0.3)to[out=0,in=-90] (0,0);
	 \draw[blue, fill=white] (x1) circle (2pt);
	 \draw[blue, fill=blue] (y3) circle (2pt);
	 \draw[blue, fill=blue] (y4) circle (2pt);
	 	 \draw[blue, fill=blue] (y5) circle (2pt);
	 \draw[blue, fill=blue] (y6) circle (2pt);
 \end{tikzpicture}
  \; + \;
  \begin{tikzpicture}[valign]
    \coordinate   (x1) at (-0.0, 0.7);
    \coordinate   (y1) at (-0.7,   0);
	\coordinate   (y2) at (+0.7, 0.0);
	\draw[double,thick,blue] (y1) -- (y2);
    \node         (y3) at (-0.2,   0) [vtx] {};
	\node         (y4) at (+0.2, 0.0) [vtx] {};
	\node         (y5) at (-0.5,   0) [vtx] {};
	\node         (y6) at (+0.5, 0.0) [vtx] {};
	\draw [thick,black]    (x1) to[out=-120,in=90] (y3);
	\draw [thick,black]   (x1) to[out=-60,  in=90] (y4);
	\draw [thick,black]   (-0.5,0) to[out=-90,in=-90] (0.5, 0);
		 \draw[blue, fill=white] (x1) circle (2pt);
	 \draw[blue, fill=blue] (y3) circle (2pt);
	 \draw[blue, fill=blue] (y4) circle (2pt);
	 	 \draw[blue, fill=blue] (y5) circle (2pt);
	 \draw[blue, fill=blue] (y6) circle (2pt);
 \end{tikzpicture}
 \; + \;
 \begin{tikzpicture}[valign]
    \coordinate   (x1) at (-0.0, 0.7);
    \coordinate   (y1) at (-0.5,   0);
	\coordinate   (y2) at (+0.8, 0.0);
	\draw[double,thick,blue] (y1) -- (y2);
    \node         (y3) at (-0.3,   0) [vtx] {};
	\node         (y4) at (+0.3, 0.0) [vtx] {};
	\node         (y5) at (-0.0,   0) [vtx] {};
	\node         (y6) at (+0.6, 0.0) [vtx] {};
	\draw [thick,black]  (x1) to[out=-130,in=90] (y3);
	\draw [thick,black] (x1) to[out=-50,  in=90] (y4);
	\draw [thick,black]  (y5) to[out=-90, in=-90] (y6);
		 \draw[blue, fill=white] (x1) circle (2pt);
	 \draw[blue, fill=blue] (y3) circle (2pt);
	 \draw[blue, fill=blue] (y4) circle (2pt);
	 	 \draw[blue, fill=blue] (y5) circle (2pt);
	 \draw[blue, fill=blue] (y6) circle (2pt);
 \end{tikzpicture}
 \; + \;\;
\begin{tikzpicture}[baseline,valign]
  \draw[double,thick,blue]   ( 0,  0) -- ( 1.2, 0);
  \draw[thick,black](.6, .9) to[out=-120,in=120] (.6, 0.3);
  \draw[thick,black] (.6, .9) to[out=- 60,in= 60] (.6, 0.3);
  \draw[thick,black] (.3, 0) -- (.6, .3) -- (.9, 0);
  \node at (0.3, 0.0) [dcirc] {};
  \node at (0.9, 0.0) [dcirc] {};
  \node at (0.6, 0.3) [bcirc] {};
  		 \draw[blue, fill=white] (.6,.9) circle (2pt);
	 \draw[blue, fill=blue] (.9,0) circle (2pt);
	 \draw[blue, fill=blue] (.3,0) circle (2pt);
	  \draw[black, fill=black] (.6,.3) circle (2pt);
\end{tikzpicture}
 \,,
\end{align}
where in the last equation it is intended that one should consider also the specular version of diagrams such as the third and the fifth, and the black point represents the bulk interaction.
The diagrams without bulk interactions were already computed in the free bulk case in \cite{Cuomo:2022xgw}, but because of the shift in the critical coupling \eqref{fixedcouplingint}, they will give slightly different results in the interacting case. The only diagram with bulk interactions was computed in \cite{Cuomo:2021kfm}. We refer to these papers for the details of the computation.
All in all we find
\begin{align}
   a_{\phi^2}
  = \frac{\pi ^2 j (j+1) \veps}{2 \sqrt{6}} \left(
      1
    - \frac{2\pi^2}{11} \left(j (j+1)-\frac{1}{3} \right) \veps
    - \frac{181 \veps}{242}
    + \frac{6}{11} \veps \log 2
 \right) + O(\veps^3) \, .
 \label{eq:one-pt-phi2-int}
\end{align}

\subsection{Two-point function \texorpdfstring{$\langle \phi \phi \rangle$}{< phi phi >}}
\subsubsection{Free bulk}
Moving on to the two point function of the order parameter $\phi$ in a free bulk, we can play the same trick as before and compute it in terms of an integrated defect two-point function.
In particular, using \eqref{fielddec}, we find
 \begin{equation}
 \begin{split}
     \langle \phi_a (0, x_\perp) \phi_b (0, y_\perp) \rangle_{\mathcal{D}_j} = \textstyle \kappa \, \zeta^2 \int d\tau \, d\tau' \frac{\braket{\hat{S}_a (\tau) \hat{S}_b (\tau')}_{\mathcal{D}_j}}{(\tau^2+|x_\perp| ^2)^{1-\frac{\veps}{2}} ({\tau'}^2+|y_\perp| ^2)^{1-\frac{\veps}{2}}} +\langle \phi_a ^{\text{free}} (0, x_\perp) \phi_b ^{\text{free}} (0, y_\perp) \rangle_{\mathcal{D}_j}.
 \end{split}
 \end{equation}
 At the fixed point, using \eqref{fixedcouplingfree} and \eqref{defectspintwopoint}, we obtain
 \begin{equation} \label{2ptphi}
 \begin{split}
     \langle \phi_a (0, x_\perp) \phi_b (0, y_\perp) \rangle _{\mathcal{D}_j} &= \textstyle \frac{\kappa \zeta_* ^2 \mathcal{N}_{\hat{S}}}{3} \int d\tau \int d\tau' \frac{\delta_{a b}}{|\tau-\tau'|^{2\hat{\Delta}_{\hat{S}}}(\tau^2+r^2)^{1-\frac{\veps}{2}} ({\tau'}^2+1)^{1-\frac{\veps}{2}}} +\frac{\mathcal{N}_{\phi}^2 \delta_{a b} \xi^{-\Delta_\phi}}{|x_\perp| ^{\Delta_\phi}| y_\perp |^{\Delta_\phi} }=\\
      &\equiv \frac{\mathcal{N}_{\phi} ^2 \delta_{a b} F_{\phi \phi}(r,w)}{|x_\perp| ^{\Delta_\phi} |y_\perp| ^{\Delta_\phi} }\,,
 \end{split}
 \end{equation}
 where we exploited symmetry to set the first operator at $x=(0,z,\bar{z},0,0,...)$ and the other one at $(0,1,0,0,...)$ and then expressed everything in radial coordinates \eqref{coordinates} in order to simplify the computation.
 The factor $\xi$ was defined in \eqref{bulkexpansion} and $\mathcal{N}_\phi = \sqrt{\kappa}$.
 The integral can be solved in terms of hypergeometric functions and we obtain
 \begin{equation}
 \begin{split}
      F_{\phi \phi}(r,w) &= \textstyle \xi^{-\Delta_\phi}+\frac{\zeta_* ^2 \mathcal{N}_{\hat{S}}}{3}  \left(\frac{2 \pi  \tan \left(\frac{\pi  \veps }{2}\right) r^{1-\frac{\veps}{2}} \, _2F_1\left(\frac{1}{2},1-\frac{\veps }{2};\frac{3}{2}-\frac{\veps }{2};r^2\right)}{\veps -1}+\frac{  \pi r^{\frac{\veps}{2}} \Gamma \left(\frac{1}{2}-\frac{\veps }{2}\right)^2 \, _2F_1\left(\frac{1}{2},\frac{\veps }{2};\frac{\veps +1}{2};r^2\right)}{\Gamma \left(1-\frac{\veps }{2}\right)^2}\right) =\\
      &= \xi^{-\Delta_\phi} + \lambda_{\phi \phi \phi^2} \, a_{\phi^2} J_{\veps}(r) \, ,
 \end{split}
 \end{equation}
 where the second line was obtained using the expression of the one-point function \eqref{1ptphi2general}, the three-point function coefficient \eqref{lambdafree} and well known identities for the hypergeometric function.
 This result holds for all $\veps$ and perfectly matches the bootstrap prediction \eqref{eq:free-Fpp}, in particular we notice that the non-trivial integral corresponds to  $J_\veps (r)$, the contribution of spin $s=0$ defect operators defined in \eqref{eq:J-epsexp}.

\subsubsection{Interacting bulk}
When we add interactions, it is convenient to use Feynman diagrams, since most of the diagrams have already been evaluated in \cite{Bianchi:2022sbz, Gimenez-Grau:2022ebb}. At the order we are working,  we have
\begin{align}
 \langle \phi_a \phi_a \rangle_{\mathcal{D}_j}
 \;\; = \;\;
 \begin{tikzpicture}[valign]
    \coordinate   (y1) at (-0.5,   0);
	\coordinate   (y2) at (+0.5, 0.0);
    \draw [thick,black]  (-0.4, 0.5) -- (0.4, 0.5);
     \draw[blue, fill=white] (-0.4, 0.5)  circle (2pt);
	 \draw[blue, fill=white] (0.4, 0.5) circle (2pt);
	 \draw[double,thick,blue] (y1) -- (y2);
 \end{tikzpicture}
 \;\; + \;\;
 \begin{tikzpicture}[valign]
    \coordinate   (x1) at (-0.3, 0.7);
    \coordinate   (x2) at (+0.3, 0.7);
    \coordinate   (y1) at (-0.6,   0);
	\coordinate   (y2) at (+0.6, 0.0);
	\draw[double,thick,blue] (y1) -- (y2);
		\draw [thick,black]    (x1) -- (y3);
	\draw [thick,black]   (x2) -- (y4);
	 \draw[blue, fill=white] (x1)  circle (2pt);
	 \draw[blue, fill=white] (x2) circle (2pt);
	 \draw[blue, fill=blue] (y3) circle (2pt);
	 \draw[blue, fill=blue] (y4) circle (2pt);
 \end{tikzpicture}
 \;\; + \;\;
 \begin{tikzpicture}[valign]
    \coordinate   (x1) at (-0.3, 0.7);
    \coordinate   (x2) at (+0.3, 0.7);
    \coordinate   (y1) at (-0.5,   0);
	\coordinate   (y2) at (+0.8, 0.0);
	\draw [double,thick,blue] (y1) -- (y2);
    \node         (y3) at (-0.3,   0) [vtx] {};
	\node         (y4) at (+0.3, 0.0) [vtx] {};
	\node         (y5) at (-0.0,   0) [vtx] {};
	\node         (y6) at (+0.6, 0.0) [vtx] {};
	\draw [thick,black]   (x1) -- (y3);
	\draw [thick,black]   (x2) -- (y4);
	\draw [thick,black]   (y5) to[out=-90, in=-90] (y6);	 
	\draw[blue, fill=white] (x1)  circle (2pt);
	 \draw[blue, fill=white] (x2) circle (2pt);
	 \draw[blue, fill=blue] (y3) circle (2pt);
	 \draw[blue, fill=blue] (y4) circle (2pt);
	  \draw[blue, fill=blue] (y5) circle (2pt);
	 \draw[blue, fill=blue] (y6) circle (2pt);
 \end{tikzpicture}
 \;\;  + \;\;
 \begin{tikzpicture}[valign]
  \coordinate   (y1) at (-0.7,   0);
	\coordinate   (y2) at (+0.7, 0.0);
    \draw [thick,black]  (-0.7, 0.5) -- (0.7, 0.5);
     \draw[blue, fill=white] (-0.7, 0.5)  circle (2pt);
	 \draw[blue, fill=white] (0.7, 0.5) circle (2pt);
	 \draw[double,thick,blue] (y1) -- (y2);
    \draw [thick,black]    (-0.4, 0.5) to[out= 40,in= 140] (0.4, 0.5);
    \draw [thick,black]    (-0.4, 0.5) to[out=-40,in=-140] (0.4, 0.5);
     \draw[black, fill=black] (-0.4,0.5) circle (2pt);
      \draw[black, fill=black] (0.4,0.5) circle (2pt);
 \end{tikzpicture}
 \;\; + \;\;
 \begin{tikzpicture}[valign]
    \coordinate   (x1) at (-0.3, 0.7);
    \coordinate   (x2) at (+0.3, 0.7);
    \coordinate   (y1) at (-0.6,   0);
	\coordinate   (y2) at (+0.6, 0.0);
	\draw [double,thick,blue] (y1) -- (y2);
	\draw [thick,black]   (x1) -- (y4);
	\draw [thick,black]   (x2) -- (y3);
	\draw[blue, fill=white] (x1)  circle (2pt);
	 \draw[blue, fill=white] (x2) circle (2pt);
	 \draw[blue, fill=blue] (y3) circle (2pt);
	 \draw[blue, fill=blue] (y4) circle (2pt);
	  \draw[black, fill=black] (0,0.35) circle (2pt);
 \end{tikzpicture}    
 \,,
\end{align}
where again also the contribution from the specular version of the third diagram is implied.
The first two diagrams are the just the free propagator and the square of the one-point function of $\phi^a$, which is zero. The only non-trivial diagram is the last one, and it was already computed in \cite{Bianchi:2022sbz, Gimenez-Grau:2022ebb} in terms of the function $H(r,w)$ that we introduced in \eqref{Hfunction}.
All in all, we obtain
\begin{align}
 F_{\phi \phi}(r,w)
 = \xi^{-\Delta_\phi}
 + c_{\phi^2} \left(
    1
  + \frac{\veps}{2} \, \log \frac{4r}{(1+r)^2}
  + \frac{5\veps}{11} \big( 1+\log 2 + H(r,w) \big)
 \right) \, ,
\end{align}
where as always $c_{\phi^2} = \lambda_{\phi\phi\phi^2} a_{\phi^2}$, with $\lambda_{\phi \phi \phi^2}$ given by \eqref{lambdaint} and $a_{\phi^2}$ by \eqref{eq:one-pt-phi2-int}.
This result perfectly matches the bootstrap prediction \eqref{phiintboot}.

\subsection{Two-point function \texorpdfstring{$\langle \phi^2 \phi^2 \rangle$}{< phi2 phi2 >}}
\subsubsection{Free bulk}
The two-point function of $\phi^2$ in free theory can once again be computed in terms of defect correlators, namely
 \begin{equation}\label{2ptphi2defect}
      \begin{split}
          \langle \phi^2 \phi^2 \rangle_{\mathcal{D}_j} &= \textstyle \kappa^2 \, \zeta_* ^4 \int d\tau_1 \int d\tau_2  \int d\tau_3  \int d\tau_4 \frac{\langle \hat{S}_a (\tau)\hat{S}^a  (\tau_1)\hat{S}_b (\tau_2) \hat{S}^b (\tau_3)\rangle_{\mathcal{D}_j}}{({\tau_1}^2+r^2)^{1-\frac{\veps}{2}} ({\tau_2}^2+r^2)^{1-\frac{\veps}{2}} ({\tau_3}^2+1)^{1-\frac{\veps}{2}} ({\tau_4}^2+1)^{1-\frac{\veps}{2}}}+ \\
          &+ 2\kappa \, \zeta_* ^2 \int d\tau_1 \int d\tau_2 \frac{\langle \hat{S}^a (\tau_1) \hat{S}^b (\tau_2)\rangle_{\mathcal{D}_j} \langle \phi_a ^{\text{free}} \phi_b ^{\text{free}} \rangle_{\mathcal{D}_j} }{({\tau_1}^2+r^2)^{1-\frac{\veps}{2}} ({\tau_2}^2+1)^{1-\frac{\veps}{2}}} +\langle \phi^2 _{\text{free}} \phi^2 _{\text{free}} \rangle_{\mathcal{D}_j} =\\
          &= \textstyle \kappa^2 \, \zeta_* ^4 \int d^4 \tau \frac{\langle \hat{S}_a (\tau_1)\hat{S}^a  (\tau_2)\hat{S}_b (\tau_3) \hat{S}^b (\tau_4)\rangle_{\mathcal{D}_j}}{({\tau_1}^2+r^2)^{1-\frac{\veps}{2}} ({\tau_2}^2+r^2)^{1-\frac{\veps}{2}} ({\tau_3}^2+1)^{1-\frac{\veps}{2}} ({\tau_4}^2+1)^{1-\frac{\veps}{2}}}+ \\
          &+\frac{1}{|x_\perp|^{\Delta_{\phi^2}} |y_\perp| ^{\Delta_{\phi^2}}} \left( \xi^{-\Delta_{\phi^2}}
 + 2 c_{\phi^2} \xi^{\Delta_\phi} J_\veps (r) \right)\,,
      \end{split}
 \end{equation}
 where we suppressed the explicit dependence on the external coordinates of the free fields $\phi^{\text{free}}$ and used the results of the previous section in order to simplify the expression.
 Contrary to the previous cases, we can't simplify the result further without expanding in $\veps$. The reason is that the four-point function of defect operator is not completely fixed by conformal invariance\footnote{The four-point function at the conformal fixed point reads $\langle \hat{S}_a (\tau_1)\hat{S}_a  (\tau_2)\hat{S}_a (\tau_3) \hat{S}_a (\tau_4)\rangle_{\mathcal{D}_j} = \frac{\mathcal{N}_{\hat{S}} ^2}{\tau_{12}^{2 \hat{\Delta}_{\hat{S}}} \tau_{34}^{2 \hat{\Delta}_{\hat{S}}}} f(z) $ where $f(z)$ is an arbitrary function of the conformal cross-ratio $z= \frac{\tau_{12} \tau_{34}}{\tau_{13} \tau_{24}}$.} and one cannot do the first integral without knowing its explicit expression.
 The four-point function of $\hat{S}^a$ at tree-level is just given by traces of the generators $T^a$, just like the two point function. However, one should be careful of the order of the positions where the generators are inserted, which corresponds to step functions.
 This happens because, as we said in Section \ref{sec:defectop}, the defect spin operator has to be interpreted as a generator inserted at a specific position in the path ordering.
 All in all we get 
 \begin{equation}
    \begin{split}
          \textstyle \langle \hat{S}_a (\tau_1)\hat{S}^a  (\tau_2)\hat{S}_b & (\tau_3) \hat{S}^b (\tau_4)\rangle_{\mathcal{D}_j} = \frac{1}{2j+1}\big[\Tr \left(T_a T_a T_b T_b \right) \left(\theta_{1>2>3>4}+ \text{cyclic perm.} \right)+\\
         &+\Tr \left(T_a T_b T_a T_b \right) \left(\theta_{1>3>2>4} + \theta_{1>4>2>3}+ \text{cyclic perm.} \right)\big]   +O(\veps)= \\
         &=j^2 (j+1)^2 \left(\theta_{1>2>3>4}+\theta_{1>3>2>4}+\theta_{1>4>2>3} +\text{cyclic perm.} \right)+\\
         &- j(j+1) \left(\theta_{1>3>2>4}+\theta_{1>4>2>3}+\text{cyclic perm.}\right)+O(\veps)\,,
    \end{split}
 \end{equation}
 where we indicated the order of the points using theta functions.
 When we plug this result into \eqref{2ptphi2defect} and use the symmetry of the integrand, the term that goes like $\sim j^2 (j+1)^2$ reproduces the square of the one-point function. The other term reduces to
 \begin{align}\label{Wintegral}
  -\int_{\tau_1 > \tau_3 > \tau_2 > \tau_4} \!\!\!\!\!\!\!\!\!\!\!\!\!\!\!\!\!\!\!\!\!\!\! d\tau^4
 \ ({\tau_1}^2+r^2)^{-1} ({\tau_2}^2+r^2)^{-1} ({\tau_3}^2+1)^{-1} ({\tau_4}^2+1)^{-1}
 \equiv \frac{\pi^2}{2 r^2} \, W(r) \, .
\end{align}
where $W(r)$ is
\begin{align}
 W(r)
 = 2 \text{Li}_2\left(\frac{1-r}{2}\right)-\text{Li}_2(1-r)-\text{Li}_2(-r)+\log (r+1) \log \left(\frac{r+1}{4 r}\right) +\log ^2 2 \, .
\end{align}
At the end of the day, the two-point function reads
\begin{equation}
   \langle \phi^2(0,x_\perp) \phi^2(0,y_\perp) \rangle_{\mathcal{D}_j} = \frac{\Nm_{\phi^2}^2 \, F_{\phi^2 \phi^2}(r,w)}{|x_\perp|^{\Delta_{\phi^2}} |y_\perp|^{\Delta_{\phi^2}} } \, .
\end{equation}
where
\begin{align}
\label{2ptphi2}
 F_{\phi^2 \phi^2}(r,w)
&= \xi^{-\Delta_{\phi^2}}
 + a_{\phi^2}^2
 + 2 c_{\phi^2} \xi^{\Delta_\phi} \left(
  1 + \frac{\veps}{2} \, \log \frac{4r}{(1+r)^2}
 \right)
 - \frac{\pi^2 j (j+1)}{6} \veps^2 W(r)
 + O(\veps^3) \, .
\end{align}
Comparing with the bootstrap result \eqref{bootsphi2}, we see that the two results coincide up to the term proportional to $W(r)$.
Indeed the function $W(r)$ corresponds to a  spin $s=0$ ambiguity, in the language of the bootstrap.
It has a simple block expansion in the defect channel
\begin{align}
  W(r)
  = (2 - 4 \log 2) \hat{f}_{1,0}(r)
  - 2 \partial_{\hat{\Delta}} \hat{f}_{1,0}(r)
  + 4 \sum _{n=2}^{\infty}
    \frac{(-1)^{n} (n-2)!}{n \left(\frac{3}{2}\right)_{n-2}}
    \, \hat{f}_{n,0}(r) \, .
\end{align}
Expanding \eqref{2ptphi2} in the defect channel we find that the lowest singlet operator, which we could identify with $\hat{\Phi}$, has 
\begin{equation}
    \begin{split}
         &\hat{\Delta}_{\hat{\Phi}} = 1+\veps + O(\veps^2) \, .
    \end{split}
\end{equation}
This result matches with the expression computed from the beta function \eqref{dimPhifree}. One can also extract the bulk CFT data from \eqref{2ptphi2} but the results are not particularly illuminating. Therefore we do not present them here.

\subsubsection{Interacting bulk}
In the interacting case, the two-point function of $\phi^2$ can only be computed using Feynman diagrams. At the order we are working, the relevant diagrams are
\begin{align}
 \langle \phi^2(0, x_\perp) \phi^2(0,y_\perp) \rangle
 \;\; &= \;\;
 \begin{tikzpicture}[valign]
    \coordinate   (y1) at (-0.5,   0);
	\coordinate   (y2) at (+0.5, 0.0);
	\draw[double,thick,blue] (y1) -- (y2);
    \draw [double, thick,black]    (-0.4, 0.5) -- (0.4, 0.5);
    \draw[blue, fill=white] (-.4,.5) circle (2pt);
    \draw[blue, fill=white] (.4,.5) circle (2pt);
 \end{tikzpicture}
 \;\; + \;\;
 \begin{tikzpicture}[valign]
    \coordinate   (x1) at (-0.3, 0.7);
    \coordinate   (x2) at (+0.3, 0.7);
    \coordinate   (y1) at (-0.6,   0);
	\coordinate   (y2) at (+0.6, 0.0);
	\coordinate   (y3) at (-.3, 0.0);
	\coordinate   (y4) at (+0.3, 0.0);
	\draw [double,thick,blue](y1) -- (y2);
	\draw [thick,black]   (y3) -- (x1) -- (x2) -- (y4);
	   \draw[blue, fill=white] (x1) circle (2pt);
    \draw[blue, fill=white] (x2) circle (2pt);
    	   \draw[blue, fill=blue] (y3) circle (2pt);
    \draw[blue, fill=blue] (y4) circle (2pt);
 \end{tikzpicture}
\;\; + \;\;
 \begin{tikzpicture}[valign]
    \coordinate   (x1) at (-0.3, 0.7);
    \coordinate   (x2) at (+0.3, 0.7);
    \coordinate   (y1) at (-0.5,   0);
	\coordinate   (y2) at (+0.8, 0.0);
	\draw [double,thick,blue] (y1) -- (y2);
    \coordinate         (y3) at (-0.3,   0);
	\coordinate        (y4) at (+0.3, 0.0);
	\coordinate       (y5) at (-0.0,   0) ;
	\coordinate          (y6) at (+0.6, 0.0);
	\draw[thick,black]  (y3) -- (x1) -- (x2) -- (y4);
	\draw [thick,black] (y5) to[out=-90, in=180] (.3,-.2) to[out=0, in=-90] (y6);
		   \draw[blue, fill=white] (x1) circle (2pt);
    \draw[blue, fill=white] (x2) circle (2pt);
    	   \draw[blue, fill=blue] (y3) circle (2pt);
    \draw[blue, fill=blue] (y4) circle (2pt);
       \draw[blue, fill=blue] (y5) circle (2pt);
    \draw[blue, fill=blue] (y6) circle (2pt);
 \end{tikzpicture}
 \;\; + \;\;
 \begin{tikzpicture}[valign]
    \coordinate   (x1) at (-0.35, 0.7);
    \coordinate   (x2) at (+0.35, 0.7);
    \coordinate   (y1) at (-0.7,   0);
	\coordinate   (y2) at (+0.7, 0.0);
	\draw[double,thick,blue](y1) -- (y2);
    \coordinate         (y3) at (-0.5,   0);
	\coordinate        (y4) at (-0.2, 0.0) ;
	\coordinate        (y5) at (+0.2,   0);
	\coordinate          (y6) at (+0.5, 0.0);
	\draw [thick,black]  (y3) -- (x1) -- (y4);
	\draw [thick,black]   (y5) -- (x2) -- (y6);
	\draw[blue, fill=white] (x1) circle (2pt);
    \draw[blue, fill=white] (x2) circle (2pt);
    	   \draw[blue, fill=blue] (y3) circle (2pt);
    \draw[blue, fill=blue] (y4) circle (2pt);
       \draw[blue, fill=blue] (y5) circle (2pt);
    \draw[blue, fill=blue] (y6) circle (2pt);
 \end{tikzpicture}
 \;\; + \;\;
 \begin{tikzpicture}[valign]
    \coordinate   (x1) at (-0.2, 0.7);
    \coordinate   (x2) at (+0.2, 0.7);
    \coordinate   (y1) at (-0.7,   0);
	\coordinate   (y2) at (+0.7, 0.0);
	\draw[double,thick,blue] (y1) -- (y2);
    \coordinate        (y3) at (-0.5,   0) ;
	\coordinate         (y4) at (-0.2, 0.0) ;
	\coordinate         (y5) at (+0.2,   0);
	\coordinate       (y6) at (+0.5, 0.0) ;
	\draw  [thick,black]   (y3) -- (x1) -- (y5);
	\draw  [thick,black]   (y4) -- (x2) -- (y6);
	\draw[blue, fill=white] (x1) circle (2pt);
    \draw[blue, fill=white] (x2) circle (2pt);
    	   \draw[blue, fill=blue] (y3) circle (2pt);
    \draw[blue, fill=blue] (y4) circle (2pt);
       \draw[blue, fill=blue] (y5) circle (2pt);
    \draw[blue, fill=blue] (y6) circle (2pt);
 \end{tikzpicture}
 \;\; + \nonumber \\ &
\;\;  + \;\;
 \begin{tikzpicture}[baseline,valign]
  \draw[double,thick,blue] (-.1,  0) -- ( 1.2, 0);
  \draw[thick,black] (.2, 0) -- (.2, .9) -- (.6, .45);
  \draw[thick,black](1., .9) to[out=-160,in=60] (.6, .45);
  \draw[thick,black] (1., .9) to[out=-110,in=30] (.6, .45);
  \draw[thick,black] (.6, .45) -- (.6, 0);
    \draw[blue, fill=white] (.2,.9) circle (2pt);
    \draw[blue, fill=white] (1,.9) circle (2pt);
    	   \draw[black, fill=black] (.6,.45) circle (2pt);    
    	   \draw[blue, fill=blue] (.6,0) circle (2pt);
    	     \draw[blue, fill=blue] (.2,0) circle (2pt);
\end{tikzpicture}
\;\; + \;\;
 \begin{tikzpicture}[baseline,valign]
  \draw[double,thick,blue]  ( 0,  0) -- ( 1.2, 0);
  \draw[thick,black] (.3, .9) -- (.9, .9);
  \draw[thick,black] (.3, .9) -- (.9, 0);
  \draw[thick,black] (.9, .9) -- (.3, 0);
   \draw[blue, fill=white] (.3,.9) circle (2pt);
    \draw[blue, fill=white] (.9,.9) circle (2pt);
    	   \draw[black, fill=black] (.6,.45) circle (2pt);    
    	   \draw[blue, fill=blue] (.3,0) circle (2pt);
    	     \draw[blue, fill=blue] (.9,0) circle (2pt);     
\end{tikzpicture}
\;\;
\,,
\end{align}
where as before contributions from specular diagrams are implied. The  first diagram stands for corrections to the propagator that come purely from bulk interactions. These have been computed long ago in the theory without the defect and generate corrections to the dimension of $\Delta_{\phi^2}$ in the bulk identity term $\xi^{-\Delta_{\phi^2}}$. We don't draw them explicitly to avoid cluttering.
The only non-trivial diagram is the fifth, which can be computed in terms of $W(r)$ \eqref{Wintegral}. All the other diagrams were already computed in \cite{Gimenez-Grau:2022ebb}, therefore we only write the final result, namely
\begin{align}\label{2ptphi2int}
 F_{\phi^2 \phi^2} (r,w)
&= \xi^{-\Delta_{\phi^2}}
 + a_{\phi^2}^2
 - \frac{\veps^2 \pi^2 j(j+1)}{6} W(r)
 + \frac{\pi^2 j (j+1) \veps}{3 \xi^{1-\frac{\veps }{22}}}
   \Bigg[ 1 +\veps  \log 2 \notag \\
& \quad
 + \frac{\veps}{11} \left(
    - \frac{118}{11}
    + 5 H(r,w)
    - 2 \pi ^2 \left(j (j+1)-\tfrac{1}{3} \right)
    + \frac{1}{2} \log \frac{4r}{(r+1)^2} \right) \Bigg]
 + O(\veps^3) \, .
\end{align}
It would be difficult to compute this result using bootstrap methods, since the discontinuity would receive contributions from all the double-twist operators.

\section*{Acknowledgments} 
We are grateful to M. Billò, M. Meineri, M. Lemos for useful discussions. The research of LB is funded through the MIUR program for young researchers “Rita Levi Montalcini”. The research of DB received partial support through the STFC grant ST/S005803/1 and through the Grant for Internationalization of the University of Torino. DB would also like to thank IHES for hospitality during  completion of this work.
AGG is supported by the Simons Foundation by grants 915279 (IHES) and 733758 (Bootstrap Collaboration).

\appendix
\section{Maps among different realizations of the defect}
\label{app:representations}
Many different representations of the defect \eqref{defect} have appeared in the condensed matter and high energy literature.  
In this appendix we review a few of them, both because we want to help the interested reader navigate the existing literature and because the different representations allow to generalize our setup in different directions. 

\subsection{Coherent state representation}
The first representation we review is a representation in terms of a path integral over constrained bosonic fields, which is commonly known as a coherent state representation in the condensed matter literature \footnote{See \emph{e.g.} \cite{sachdev_2011} or \cite{altland_simons_2010} for a discussion of coherent path integrals in a condensed matter setting.}. 
Following \cite{Clark_1997}, one starts by introducing two copies of bosonic creation and annihilation operators $\hat{a} = (\hat{a}_1, \hat{a}_2)$\footnote{In this section we will not distinguish bare and renormalized operators to avoid cluttering.}
\begin{equation}
    [\hat{a}^i, \hat{a}^\dagger _j]=\delta^i _j \, ,
\end{equation}
that satisfy a constraint on the number operator
\begin{equation}\label{constraint}
     \hat{a}^\dagger \hat{a} = 2j \, ,
\end{equation}
and a set of so called coherent states
\begin{equation}
    \ket{z} = e^{ \hat{a}^\dagger z -\Bar{z} \hat{a}} \ket{0} \, .
\end{equation}
If we define the composite operator
\begin{equation}
    \hat{S}^a  = \hat{a}^\dagger T^a \hat{a} \, ,
\end{equation}
where $T^a$ are the generators of $\mathfrak{so}(3) \cong \mathfrak{su}(2)$\footnote{One can generalize this construction to $\mathfrak{su}(N)$ by changing the generators and considering many copies of creation and annhilation operators.}, namely $T^a = \frac{\sigma^a}{2}$, we see that it satisfies 
\begin{equation}\label{spincommutator}
    [\hat{S}^a, \hat{S}^b] = i \epsilon^{abc} \hat{S}_c \, ,
\end{equation}
and is normalized as
\begin{equation}\label{spincasimir}
    \hat{S}^a \hat{S}^a= j(j+1) \, ,
\end{equation}
It is just the defect spin operator defined in Section 3. 
We then define the Hamiltonian
\begin{equation}
    \hat{H}(\tau) = \frac{\zeta_0}{\sqrt{\kappa}} \hat{S}^a \phi_a (\tau) \, .
\end{equation}
Using the definitions above, we can rewrite the defect \eqref{defect} as an integral over the overcomplete basis $\ket{z}$ as
\begin{equation}
\begin{split}
\Tr \,  \mathcal{D}_j  = \Tr \, \mathcal{P} e^{\int d\tau \ \hat{H}(\tau)}  &= \int \mathcal{D}z \mathcal{D}\Bar{z} \mathcal{D}\mu \bra{z} e^{ \int d\tau \ \hat{H}(\tau)+ i \mu(\Bar{z} z -2j)} \ket{z} =\\ 
    &= \int \mathcal{D}z \mathcal{D}\Bar{z} \mathcal{D}\mu \ e^{2 i j \mu }\bra{z} e^{ \int d\tau \ \hat{H}(\tau)} \ket{e^{-i \mu} z} \, ,
   \end{split} 
\end{equation}
where we introduced the auxiliary field $\mu$ to enforce the constraint \eqref{constraint} and we used that for coherent states\footnote{We refer to Appendix A of \cite{Clark_1997} for a summary of the properties of coherent states.}
\begin{equation}
     e^{-i \mu \hat{a}^\dagger z} \ket{z} = \ket{e^{-i \mu} z} \, .
\end{equation}
In order to simplify the result we introduce a small parameter $\delta$ and discretize 
\begin{equation}
   \mathcal{P} e^{ \int d\tau \, \hat{H}(\tau)} = \prod_{k=1}^{N= 2\pi/\delta} e^{ \delta \, \hat{H}(\tau_k)} \, ,
\end{equation}
then by using
\begin{equation}
        \left\langle z^{\prime} \mid z\right\rangle=e^{\bar{z}^{\prime} z-\frac{1}{2}\left(\bar{z} z+\bar{z}^{\prime} z^{\prime}\right)} \, ,
\end{equation}
and rescaling $\mu \rightarrow \delta \mu$, we obtain
\begin{equation}\label{defectz}
    \begin{split}
  \Tr \,  \mathcal{D}_j   &=\int\left(\prod_{m=1}^N d \mu_m d^2 z\right)\left(\prod_{n=1}^N e^{2 i j \mu_n}\left\langle z_{n+1}\left|e^{ \delta \hat{H}\left(\tau_n\right)}\right| e^{-i \mu_n} z_n\right\rangle\right)=\\
   &=\int\left(\prod_{m=1}^N d \mu_m d^2 z_m\right) e^{\left(\sum_{n=1}^N \delta \left\{2 i j  \mu_n-\bar{z}_n \dot{z}_n-i \mu_n \bar{z}_n z_n+ \hat{H}\left(z_n, \tau_n\right)\right\}\right)} =\\
    &= \int \mathcal{D}z \mathcal{D}\Bar{z} \mathcal{D}\mu  e^{\int d \tau \left(2 i j \mu-\bar{z} \dot{z}-i \mu \bar{z} z+ \hat{H}(z, \tau)\right)}= \\
    &= \int \mathcal{D}z \mathcal{D}\Bar{z} \ \delta(\bar{z} z-2j) e^{\int d \tau \left(-\bar{z} \dot{z}+ \hat{H}(z, \tau)\right)} \, ,
    \end{split}
\end{equation}
where we require that the $z(\tau)$ and $\Bar{z}(\tau)$ obey appropriate boundary condition, \emph{e.g.} they vanish at infinity or obey periodic boundary conditions $z(2\pi) = z(0)$ in the circular case. At the end of the day we see that our defect can be represented by a theory of one dimensional complex bosons with action
\begin{equation}
    S=\int d \tau \left(\bar{z} \dot{z}- \frac{\zeta_0}{\sqrt{\kappa}} \Bar{z} T^a z \phi^a \right) \, ,
\end{equation}
obeying the constraint $\Bar{z} z = 2j$.

\subsection{Dirac monopole representation}
Another interesting representation of the defect can be obtained by integrating over a rescaled spin field in presence of a Dirac monopole  \cite{Vojta_2000}. 
Starting from the previous path integral representation \eqref{defectz}, we rescale 
\begin{equation}
    z \rightarrow j z \, ,
\end{equation}
and define 
\begin{equation}
    n^a =  \bar{z} T^a z \, ,
\end{equation}
Following \cite{Clark_1997}, we introduce spherical coordinates such that
\begin{equation}
    \begin{split}
        n^a & =(\sin \theta \cos \phi, \sin \theta \sin \phi, \cos \theta) \\
z & =e^{i \chi}\left(\cos \left(\frac{\theta}{2}\right) e^{-i \phi}, \sin \left(\frac{\theta}{2}\right)\right) \\
\bar{z} d z & =i d \chi-i \frac{1}{2}(1+\cos \theta) d \phi \\
\epsilon^{a b c} n^a d n^b d n^c & =2 \sin \theta d \theta d \phi
    \end{split}
\end{equation}
where $\epsilon^{abc}$ is the Levi-Civita symbol. Notice that $n^2 =1$. Using the explicit expressions and Stokes' theorem we see that
\begin{equation}
    \frac{1}{2} \int \epsilon^{a b c} n^a d n^b d n^c =- \oint_{\delta D}(1+\cos \theta) d \phi
\end{equation}
and therefore
\begin{equation}
    \begin{split}
        \int d \tau \bar{z} \dot{z} & = \oint_{\delta D} \bar{z} d z \\
& =\oint_{\delta D}\left(i d \chi-\frac{i}{2}(1+\cos \theta) d \phi\right) \\
& =\frac{i}{4} \int_D \epsilon^{a b c} n^a d n^b d n^c +2 \pi i k \\
& = \frac{i}{2} \int d\tau \ (\delta_b ^ \beta \delta_c ^\gamma -\delta_b ^ \gamma \delta_c ^\beta)\frac{\partial A_\beta}{\partial n_\gamma} d n^b \partial_\tau n^c +2 \pi i k \\
& = \int d\tau \ i A_c \partial_\tau n^c +2 \pi i k \, ,
    \end{split}
\end{equation}
where we introduced a Dirac monopole field $A_\beta$ which satisfies
\begin{equation}
    \frac{1}{2} \epsilon^{\alpha \beta \gamma} \frac{\partial A_\beta}{\partial n_\gamma} = n^\alpha \, .
\end{equation}
The equations above imply that one can rewrite the defect as \cite{Vojta_2000}
\begin{equation}\label{defectn}
  \Tr \,  \mathcal{D}_j = \int \mathcal{D} n^a \delta\left(n^2-1\right) e^{-\int d\tau \ i A_c \partial_\tau n^c -\frac{\zeta_0}{\sqrt{\kappa}} n^a \phi^a} \, .
\end{equation}

\subsection{Fermionic representations}
Both \eqref{defectz} and \eqref{defectn} represent the defect in terms of a path integral over bosonic fields, however it is also possible to give a fermionic representation \cite{Gomis:2006im, Gomis:2008jt, Beccaria:2022bcr}. 
Following \cite{Beccaria:2022bcr}, we can write
\begin{equation}\label{spinorrep}
    \textstyle \left[\mathcal{P} \exp \int_{\tau_1}^{\tau_2}\phi^a T^a \right]_{\mathrm{ab}} =e^{\Omega}\left[\frac{\delta^2}{\delta \bar{u}_{\mathrm{a}}\left(\tau_2\right) \delta u_{\mathrm{b}}\left(\tau_1\right)} \exp \left(\int_{\tau_1}^{\tau_2} d \tau \frac{\delta}{\delta u_{\mathrm{c}}(\tau)} \left(T^d \phi^d \right)_{\mathrm{cc}^{\prime}}(\tau) \frac{\delta}{\delta \bar{u}_{\mathrm{c}^{\prime}}(\tau)}\right)\right]_{u=\bar{u}=0} e^{-\Omega} \, ,
\end{equation}
where $u^a$ are complex Grassmann odd fields and
\begin{equation}
   \Omega=\int_{\tau_1}^{\tau_2} d \tau d \tau^{\prime} \bar{u}_{\mathrm{c}}(\tau) u_{\mathrm{c}}\left(\tau^{\prime}\right) \theta\left(\tau-\tau^{\prime}\right) \, ,
\end{equation}
where $\theta(\tau) = (\partial_\tau)^{-1}$ is the step function.
By completing the square we have
\begin{equation}
  \int D \psi D \bar{\psi} \exp \int d \tau\left[\bar{\psi} \partial_\tau \psi+i(\bar{u} \psi+\bar{\psi} u)\right] \sim e^{ \int \bar{u}_{\mathrm{c}} (\partial)^{-1} u_{\mathrm{c}} } = e^{-\Omega} \, ,
\end{equation}
where we dropped an overall constant that depends on the determinant of $\partial_\tau$. At the end of the day we find
\begin{equation}\label{fermionicrep}
  \textstyle Tr \mathcal{P} e^{\frac{\zeta_0}{\sqrt{\kappa}} \int_{\tau_1}^{\tau_2} \phi^a T^a} =\frac{\delta^2}{\delta \bar{u}_a(\tau_2) \delta u_a(\tau_1)}\left[\log \int D \psi D \bar{\psi} \exp \int d \tau\left(\bar{\psi} \partial_\tau \psi-\frac{\zeta_0}{\sqrt{\kappa}} \bar{\psi} T^a \phi_a \psi+i(\bar{u} \psi+\bar{\psi} u)\right)\right]_{u=\bar{u}=0} \, .
\end{equation}
Therefore our defect is created by the insertion of fermionic operators at infinity,
\begin{equation}
    \langle \Tr \, \mathcal{D}_j \rangle = \langle \bar{\psi}(-\infty) \psi(\infty) \rangle \, ,
\end{equation}
where the fermions interact according to the action
\begin{equation}
    S =  \int d \tau \left(\bar{\psi} \partial_\tau \psi-\frac{\zeta_0}{\sqrt{\kappa}} \bar{\psi} T^a \phi_a \psi\right) \, .
\end{equation}
In this approach, the defect spin operator is the fermion bilinear operator
\begin{equation}
    \hat{S}^a = \bar{\psi} T^a \psi \, .
\end{equation}
Another alternative approach to describe the spin impurity, which generalizes more easily to $\mathfrak{so}(N)$ \cite{liu2021magnetic}, for odd $N$, is to introduce $N+1$ one-dimensional Majorana spinors $\sigma_i$.
At the level of operators they satisfy
\begin{equation}\label{majoranaspinor}
    \{ \sigma^i, \sigma_j \} = 2 \delta^i _j \, ,
\end{equation}
They must also satisfy a projection
\begin{equation}\label{projection}
    \prod_{i=0}^N \sigma_i = -1 \, .
\end{equation}
Using the two equations above, if we define
\begin{equation}
    S_{a b} = -\frac{1}{2} i \sigma_a \sigma_b \, ,
\end{equation}
with $a,b = 1,...,N$ and
\begin{equation}
    S^a = \frac{1}{2} \epsilon^{a b c} S_{b c} = -\frac{1}{4} i\sigma_0 \sigma_a \, ,
\end{equation}
 it is a trivial exercise to show that in the case of $\mathfrak{so}(3)$ they satisfy  \eqref{spincommutator} and \eqref{spincasimir} with $j=\frac{1}{2}$. Therefore, $S^a$ represents a spin-$\frac{1}{2}$ operator. The spin impurity defect \eqref{defect} is then equivalent to the path integral with action 
\begin{equation}\label{majoranaaction}
    S =  \int d\tau \ \frac{1}{4} \sigma^i \partial_\tau \sigma^i -\frac{\zeta_0}{\sqrt{\kappa}} S^a \phi^a = \frac{1}{4} \int d\tau \ \sigma^i \partial_\tau \sigma^i - \frac{\zeta_0}{\sqrt{\kappa}}  \ i  \epsilon^{a b c} \sigma_b \sigma_c \phi_a = \frac{1}{4} \int d\tau \ \sigma^i \partial_\tau \sigma^i +\frac{\zeta_0}{\sqrt{\kappa}}  \ i \sigma_0 \sigma_a \phi^a \, .
\end{equation}
Explicitly
\begin{equation}
  \Tr \,  \mathcal{D}_{\frac{1}{2}} = \int \mathcal{D}\sigma^i \ e^{i \frac{1}{4} \int d\tau \  i\sigma^i \partial_\tau \sigma^i +\frac{\zeta_0}{\sqrt{\kappa}}  \ i  \epsilon^{a b c} \sigma_b \sigma_c \phi_a} \, .
\end{equation}
We can integrate out $\sigma_0$, rescale the fields and use that $\epsilon^{abc} = -i \left( T^a \right)^{bc}$, since $\sigma_a$ are in the vector of $\mathfrak{so}(3) \cong \mathfrak{su}(2) $
\begin{equation}\label{Majoranadefect}
    \begin{split}
       \Tr \,  \mathcal{D}_{\frac{1}{2}} &\sim \int \mathcal{D}\sigma^a \   e^{i \int d\tau \ i\sigma^a \partial_\tau \sigma^a + \frac{\zeta_0}{\sqrt{\kappa}}  \ \epsilon^{a b c} \sigma_b \sigma_c \phi_a} \\
        &=\int \mathcal{D}\sigma^a \  e^{i \int d\tau \ i\sigma^a \partial_\tau \sigma^a -i\frac{\zeta_0}{\sqrt{\kappa}}   (T^a)^{b c} \sigma_b \sigma_c \phi_a} \\
        &=\int \mathcal{D}\sigma^a \  e^{i \int d\tau \ \sigma^a \left( i\partial_\tau \delta^{a b} -i \frac{\zeta_0}{\sqrt{\kappa}}   (T^c)^{a b} \phi_c \right) \sigma_b} \, .
    \end{split}
\end{equation}
One can prove that the path integral above indeed reduces to the Wilson loop representation \eqref{defect} of the defect, specialized to $j=\frac{1}{2}$, following the same strategy as in the case of complex spinors. 
In particular, one can introduce real Grassmann odd fields $u^a$ and write
\begin{equation}
    \Omega=\int_{\tau_1}^{\tau_2} d \tau d \tau^{\prime} u_{\mathrm{c}}(\tau) u_{\mathrm{c}}\left(\tau^{\prime}\right) \theta\left(\tau-\tau^{\prime}\right) \, ,
\end{equation}
and 
\begin{equation}
   \int D \sigma^a e^{ \int d \tau\left[ \sigma^a \partial_\tau \sigma^a+i \sigma^a u^a \right] } \sim e^{\int u^a (\partial_\tau)^{-1} u^a} = e^{-\Omega} \, .
\end{equation}
By plugging the above representation in \eqref{spinorrep} we find
\begin{equation}
   \textstyle \Tr \, \mathcal{P}e^{\frac{\zeta_0}{\sqrt{\kappa}} \int_{\tau_1}^{\tau_2} \phi^a T^a} =\frac{\delta^2}{\delta u _a(\tau_2) \delta u_a (\tau_1)}\left[\log \int \mathcal{D}\sigma^a \  \exp \left(i \int_D d\tau \ \sigma^a \left( i\partial_\tau \delta^{a b} -i \frac{\zeta_0}{\sqrt{\kappa}}  (T^c)^{a b} \phi_c \right) \sigma_b+ u^a \sigma^a\right) \right] \, .
\end{equation}

\section{Details on the computation of the \texorpdfstring{$\beta$}{beta}-function}\label{app:detailsbeta}

In this appendix we calculate the three-loop beta function for the interacting Wilson line in the cases of a free and interacting bulk.
The method we use was introduced in \cite{Dotsenko:1979wb}, but see also \cite{Beccaria:2021rmj} for a recent review.

\subsection{Summary of the strategy}

Our goal ultimate goal is to express the bare coupling $\zeta_0$ in terms of the renormalized coupling $\zeta$, given in the MS scheme by
\begin{align}
 \zeta_0 = \mu^{\veps/2} \zeta \left(
  1 + \frac{a_{11} \zeta^2 + a_{12} \zeta^4 + a_{13} \zeta^6}{\veps}
  + \frac{a_{22} \zeta^4 + a_{23} \zeta^6}{\veps^2}
  + \frac{a_{33} \zeta^6}{\veps^3}
  + O(\zeta^8)
 \right) \, .
 \label{eq:renzeta}
\end{align}
Provided this holds, then we can extract the beta function
from the condition that $d\zeta_0/d\mu = 0$, namely
\begin{align}
 \beta(\zeta)
 = \mu \frac{d\zeta}{d\mu}
 =
 - \frac{\veps}{2} \, \zeta
 +   a_{11} \zeta^3
 + 2 a_{12} \zeta^5
 + 3 a_{13} \zeta^7
 + O(\zeta^9) \, .
\end{align}
As usual, one also finds the relation between lower and higher order poles
\begin{align}
 a_{22} = \frac{3}{2} \, a_{11}^2 \, , \qquad
 a_{33} = \frac{5}{2} \, a_{11}^3 \, , \qquad
 a_{23} = \frac{11}{3} \, a_{11} a_{12} \, .
 \label{eq:cons-cond}
\end{align}
Our goal in the rest of this appendix is to compute the vertex $\Vm(x)$ at three loops.

In order to determine the coefficients $a_{ij}$ in equation \eqref{eq:renzeta}, we have to demand that physical observables are finite when expressed in terms of $\zeta$.
As argued in the main text, a good such observable is the vertex function
\begin{align}
 \Vm(x)
 = \frac{\tr \big\langle \phi(x) \Dm_j(0, \tau) \big\rangle}
        {\tr \big\langle \Dm_j(0, \tau) \big\rangle } \, ,
 \label{eq:vertexV}
\end{align}
Note that the field $\phi$ is inserted away from the defect, which extends over the finite interval $[0,\tau]$.
Since $\phi$ itself does not get renormalized because it is free, then $\Vm(x)$ is a finite observable, and requiring this condition will allow us to determine the counterterms $a_{ij}$.

\subsection{Expectation value of defect at three loops}

We start with the expectation value of the defect.
Since $\langle \tr \Dm(\tau) \rangle$ exponentiates, it is convenient to look at its logarithm.
Let us first present the final result, and we shall derive it below:
\begin{align}
 \frac{\log \big\langle \tr \Dm(\tau) \big\rangle}{j(j+1)(2j+1)}
 = \zeta_0^2 \, \textbf{(a1)}
 - \zeta_0^4 \, \textbf{(a2)}
 + \zeta_0^6 \, \big(
      2 \textbf{(a3)}
    + 2 \textbf{(a4)}
    +   \textbf{(a5)}
 \big)
 + \ldots \, .
 \label{eq:prop-ren}
\end{align}
\begin{figure}
 \centering
 \begin{tabular}{c c c c c}
  \;\begin{tikzpicture}
    \draw [defect] (0, 0) -- (1.5, 0);
    \draw [prop]   (0.3, 0) to[out=90,in=90] (1.2, 0);
 \end{tikzpicture}\; &
  \;\begin{tikzpicture}[valign]
    \draw [defect] (0, 0) -- (2.5, 0);
    \draw [prop]   (0.5, 0) to[out=+90,in=+90] (1.5, 0);
    \draw [prop]   (1.0, 0) to[out=-90,in=-90] (2.0, 0);
 \end{tikzpicture}\; \\[0.8em]
 \textbf{(a1)} & \textbf{(a2)}
 \end{tabular}
 \begin{tabular}{c c c}
  \;\begin{tikzpicture}[valign]
    \draw [defect] (0, 0) -- (3.5, 0);
    \draw [prop]   (1.0, 0) to[out=+90,in=+90] (3.0, 0);
    \draw [prop]   (1.5, 0) to[out=+90,in=+90] (2.5, 0);
    \draw [prop]   (0.5, 0) to[out=-90,in=-90] (2.0, 0);
 \end{tikzpicture}\; &
  \;\begin{tikzpicture}[valign]
    \draw [defect] (0, 0) -- (3.5, 0);
    \draw [prop]   (0.5, 0) to[out=-90,in=-90] (1.5, 0);
    \draw [prop]   (1.0, 0) to[out=+90,in=+90] (2.5, 0);
    \draw [prop]   (2.0, 0) to[out=-90,in=-90] (3.0, 0);
 \end{tikzpicture}\; &
  \;\begin{tikzpicture}[valign]
    \draw [defect] (0, 0) -- (3.5, 0);
    \draw [prop]   (0.5, 0) to[out=+90,in=+90] (2.0, 0);
    \draw [prop]   (1.0, 0) to[out=-90,in=-90] (2.5, 0);
    \draw [prop]   (1.5, 0) to[out=+90,in=+90] (3.0, 0);
 \end{tikzpicture}\; \\
  \textbf{(a3)} & \textbf{(a4)} & \textbf{(a5)}
 \end{tabular}
 \caption{One-, two-, and three-loop propagator diagrams.}
 \label{fig:prop-renorm}
\end{figure}
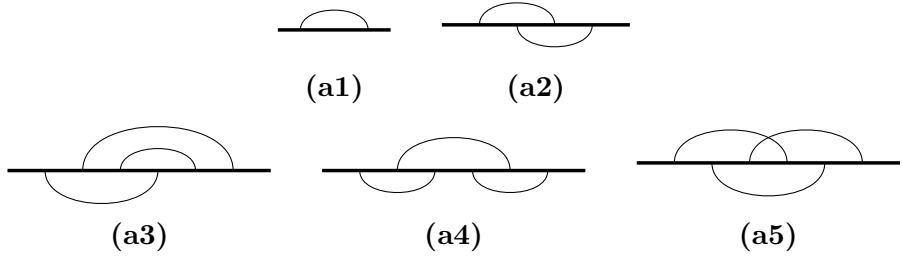%
In this expression, \textbf{(a1)}--\textbf{(a5)} are integrals represented diagrammatically in figure \ref{fig:prop-renorm}.
The thick horizontal line represents the defect, and the thin black lines are $\phi$ propagators.
The integrals are computed in the range $[0,\tau]$, time-ordering is implicit, and propagators are unit normalized.
For example
\begin{align}
  \textbf{(a1)}
&\; = \;
 \begin{tikzpicture}[valign]
    \draw [defect] (-0.7, 0) -- (+1.1, 0);
    \draw [prop]   (-0.3, 0) to[out=90,in=90] (0.7, 0);
 \end{tikzpicture}
 \; = \;
 \int\displaylimits_{0<\tau_1<\tau_2<\tau}
 \frac{d\tau_1 \, d\tau_2}
      {\tau_{21}^{2-\veps}} \, , \\
 \textbf{(a2)}
&\; = \;
 \begin{tikzpicture}[valign]
    \draw [defect] (-0.7, 0) -- (+1.1, 0);
    \draw [prop]   (-0.4, 0) to[out=+90,in=+90] (0.4, 0);
    \draw [prop]   ( 0.0, 0) to[out=-90,in=-90] (0.8, 0);
 \end{tikzpicture}
 \; = \;
 \int\displaylimits_{0<\tau_1<\tau_2<\tau_3<\tau_4<\tau}
 \frac{d\tau_1 \, d\tau_2 \, d\tau_3 \, d\tau_4}
      {(\tau_{31} \tau_{42})^{2-\veps}} \, .
\end{align}
Although these integrals are not hard to compute, we do not need them because their contribution will cancel between numerator and denominator in \eqref{eq:vertexV}.

Now let us explain how we derived \eqref{eq:prop-ren}.
The procedure is tedious, but fortunately it can be automated with the help of \texttt{mathematica}.
The first step is to compute $\langle \Dm(\tau) \rangle$ up to order $\zeta_0^6$ by doing all possible Wick contractions.
Each diagram consists of a product of an integral and a trace.
The traces are of the form $\tr T_{a_1} T_{a_2} \ldots T_{a_l}$, with all indices $a_1$, \ldots, $a_l$ being contracted.
These traces can be computed by applying commutation relations \eqref{eq:comm-rels} repeatedly, a process that we automated with the computer.

Regarding the integrals, we find a large amount of them, many more than the ones shown in figure \ref{fig:prop-renorm}.
However, many of these integrals factorize into products of lower-point integrals.
Let us show an example for the sake of clarity.
Consider the following sum of diagrams
\begin{align*}
 \begin{tikzpicture}[valign]
    \draw [defect] (-0.7, 0) -- (+1.1, 0);
    \draw [prop]   (-0.5, 0) to[out=+90,in=+90] (0.1, 0);
    \draw [prop]   ( 0.3, 0) to[out=-90,in=-90] (0.9, 0);
 \end{tikzpicture}
 \;\;+\;\;
  \begin{tikzpicture}[valign]
    \draw [defect] (-0.7, 0) -- (+1.1, 0);
    \draw [prop]   (-0.4, 0) to[out=+90,in=+90] (0.4, 0);
    \draw [prop]   ( 0.0, 0) to[out=-90,in=-90] (0.8, 0);
 \end{tikzpicture}
 \;\;+\;\;
 \begin{tikzpicture}[valign]
    \draw [defect] (-0.7, 0) -- (+1.1, 0);
    \draw [prop]   (-0.4, 0) to[out=-90,in=-90] (0.8, 0);
    \draw [prop]   (-0.1, 0) to[out=+90,in=+90] (0.5, 0);
 \end{tikzpicture}
 \;\;+\;\;
 \begin{tikzpicture}[valign]
    \draw [defect] (-0.7, 0) -- (+1.1, 0);
    \draw [prop]   (-0.4, 0) to[out=+90,in=+90] (0.8, 0);
    \draw [prop]   (-0.1, 0) to[out=-90,in=-90] (0.5, 0);
 \end{tikzpicture}
 \;\;+\;\;
 \begin{tikzpicture}[valign]
    \draw [defect] (-0.7, 0) -- (+1.1, 0);
    \draw [prop]   (-0.4, 0) to[out=-90,in=-90] (0.4, 0);
    \draw [prop]   ( 0.0, 0) to[out=+90,in=+90] (0.8, 0);
 \end{tikzpicture}
 \;\;+\;\;
 \begin{tikzpicture}[valign]
    \draw [defect] (-0.7, 0) -- (+1.1, 0);
    \draw [prop]   (-0.5, 0) to[out=-90,in=-90] (0.1, 0);
    \draw [prop]   ( 0.3, 0) to[out=+90,in=+90] (0.9, 0);
 \end{tikzpicture}
\end{align*}
We can think of it as the lower diagram ``passing through'' the upper diagram. Since all possible orderings of points are accounted for, this sum is equal to the diagram \textbf{(a1)} squared.
Each diagram appears twice, so we obtain
\begin{align}
 \Big[ \;
 \begin{tikzpicture}
    \draw [defect] (-0.7, 0) -- (0.3, 0);
    \draw [prop]   (-0.5, 0) to[out=+90,in=+90] (0.1, 0);
 \end{tikzpicture} \; \Big]^{\,2}
 \;\; = \;\;2\;\;
  \begin{tikzpicture}[valign]
    \draw [defect] (-0.7, 0) -- (+1.1, 0);
    \draw [prop]   (-0.4, 0) to[out=+90,in=+90] (0.4, 0);
    \draw [prop]   ( 0.0, 0) to[out=-90,in=-90] (0.8, 0);
 \end{tikzpicture}
 \;\;+ \;\;2\;\;
 \begin{tikzpicture}
    \draw [defect] (-0.7, 0) -- (+1.1, 0);
    \draw [prop]   (-0.5, 0) to[out=+90,in=+90] (0.1, 0);
    \draw [prop]   ( 0.3, 0) to[out=+90,in=+90] (0.9, 0);
 \end{tikzpicture}
 \;\;+\;\;2\;\;
 \begin{tikzpicture}
    \draw [defect] (-0.7, 0) -- (+1.1, 0);
    \draw [prop]   (-0.4, 0) to[out=+90,in=+90] (0.8, 0);
    \draw [prop]   (-0.1, 0) to[out=+90,in=+90] (0.5, 0);
 \end{tikzpicture}
 \, .
 \label{eq:factor-int}
\end{align}
Similar expression can be found for other products of diagrams, even when the diagrams being multiplied are different, or there are several copies of each.
The central idea is that a product of diagrams is equal to the sum of all time-ordered diagrams such that the relative order of the legs in each subdiagram is preserved.
Formulated in this way, this is a combinatorial problem that can be automated.
By factorizing sums of diagrams as in this example, we observe that the result exponentiates, and we obtain equation \eqref{eq:prop-ren}.

\subsection{Expectation value of vertex at three loops}

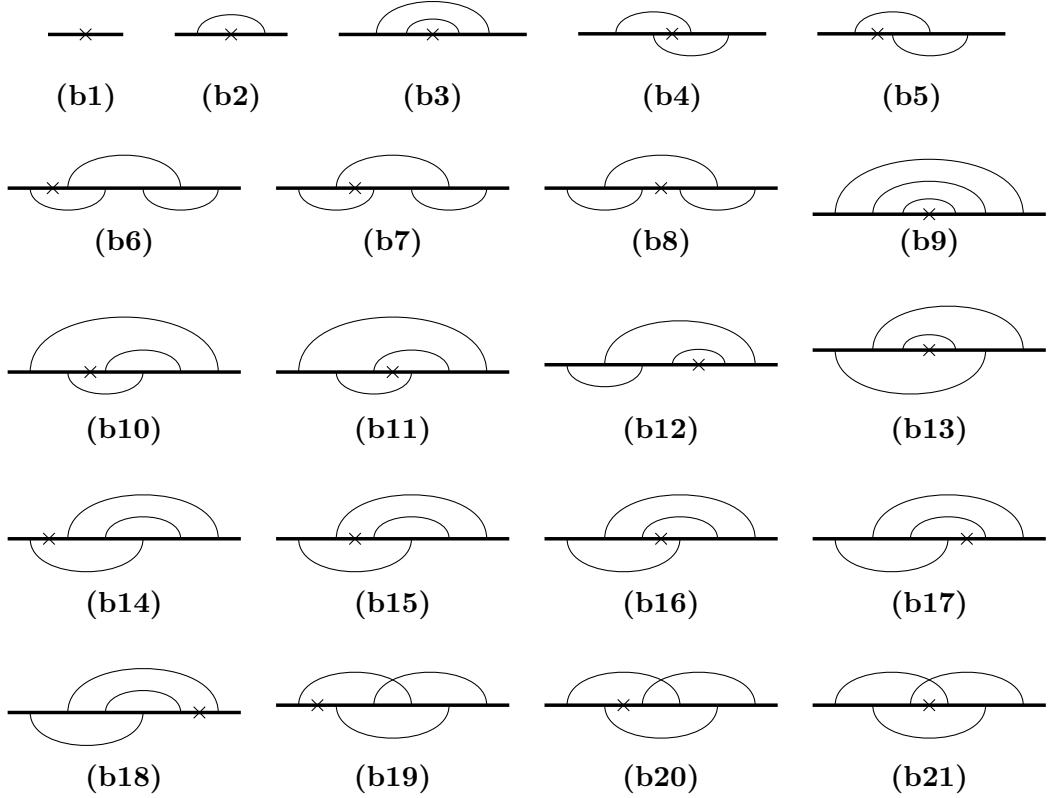
\begin{figure}
 \centering
 \begin{tabular}{c c c c c}
  \;\begin{tikzpicture}
    \draw [defect] (0, 0) -- (1.0, 0);
    \node [cross] at (0.5, 0) {};
 \end{tikzpicture}\; &
  \;\begin{tikzpicture}
    \draw [defect] (0, 0) -- (1.5, 0);
    \draw [prop]   (0.3, 0) to[out=90,in=90] (1.2, 0);
    \node [cross] at (0.75, 0) {};
 \end{tikzpicture}\; &
  \;\begin{tikzpicture}
    \draw [defect] (0, 0) -- (2.5, 0);
    \draw [prop]   (0.5, 0) to[out=+90,in=+90] (2.0, 0);
    \draw [prop]   (0.9, 0) to[out=+90,in=+90] (1.6, 0);
    \node [cross] at (1.25, 0) {};
 \end{tikzpicture}\; &
  \;\begin{tikzpicture}[valign]
    \draw [defect] (0, 0) -- (2.5, 0);
    \draw [prop]   (0.5, 0) to[out=+90,in=+90] (1.5, 0);
    \draw [prop]   (1.0, 0) to[out=-90,in=-90] (2.0, 0);
    \node [cross] at (1.25, 0) {};
 \end{tikzpicture}\; &
  \;\begin{tikzpicture}[valign]
    \draw [defect] (0, 0) -- (2.5, 0);
    \draw [prop]   (0.5, 0) to[out=+90,in=+90] (1.5, 0);
    \draw [prop]   (1.0, 0) to[out=-90,in=-90] (2.0, 0);
    \node [cross] at (0.8, 0) {};
 \end{tikzpicture}\; \\[0.8em]
 \textbf{(b1)} & \textbf{(b2)} & \textbf{(b3)} & \textbf{(b4)} &
 \textbf{(b5)}
 \end{tabular} \\[.7em]
 \begin{tabular}{c c c c}
  \begin{tikzpicture}[valign]
    \draw [defect] (0.2, 0) -- (3.3, 0);
    \draw [prop]   (0.5, 0) to[out=-90,in=-90] (1.5, 0);
    \draw [prop]   (1.0, 0) to[out=+90,in=+90] (2.5, 0);
    \draw [prop]   (2.0, 0) to[out=-90,in=-90] (3.0, 0);
    \node [cross] at (0.8, 0) {};
 \end{tikzpicture} &
  \begin{tikzpicture}[valign]
    \draw [defect] (0.2, 0) -- (3.3, 0);
    \draw [prop]   (0.5, 0) to[out=-90,in=-90] (1.5, 0);
    \draw [prop]   (1.0, 0) to[out=+90,in=+90] (2.5, 0);
    \draw [prop]   (2.0, 0) to[out=-90,in=-90] (3.0, 0);
    \node [cross] at (1.25, 0) {};
 \end{tikzpicture} &
  \begin{tikzpicture}[valign]
    \draw [defect] (0.2, 0) -- (3.3, 0);
    \draw [prop]   (0.5, 0) to[out=-90,in=-90] (1.5, 0);
    \draw [prop]   (1.0, 0) to[out=+90,in=+90] (2.5, 0);
    \draw [prop]   (2.0, 0) to[out=-90,in=-90] (3.0, 0);
    \node [cross] at (1.75, 0) {};
 \end{tikzpicture} &
 \begin{tikzpicture}[valign]
    \draw [defect] (0.2, 0) -- (3.3, 0);
    \draw [prop]   (0.5, 0) to[out=+90,in=+90] (3.0, 0);
    \draw [prop]   (1.0, 0) to[out=+90,in=+90] (2.5, 0);
    \draw [prop]   (1.4, 0) to[out=+90,in=+90] (2.1, 0);
    \node [cross] at (1.75, 0) {};
 \end{tikzpicture} \\
 \textbf{(b6)}  & \textbf{(b7)}  & \textbf{(b8)}  & \textbf{(b9)} \\[1em]
 \begin{tikzpicture}[valign]
    \draw [defect] (0.2, 0) -- (3.3, 0);
    \draw [prop]   (0.5, 0) to[out=+90,in=+90] (3.0, 0);
    \draw [prop]   (1.0, 0) to[out=-90,in=-90] (2.0, 0);
    \draw [prop]   (1.5, 0) to[out=+90,in=+90] (2.5, 0);
    \node [cross] at (1.3, 0) {};
 \end{tikzpicture} &
  \begin{tikzpicture}[valign]
    \draw [defect] (0.2, 0) -- (3.3, 0);
    \draw [prop]   (0.5, 0) to[out=+90,in=+90] (3.0, 0);
    \draw [prop]   (1.0, 0) to[out=-90,in=-90] (2.0, 0);
    \draw [prop]   (1.5, 0) to[out=+90,in=+90] (2.5, 0);
    \node [cross] at (1.75, 0) {};
 \end{tikzpicture} &
  \begin{tikzpicture}[valign]
    \draw [defect] (0.2, 0) -- (3.3, 0);
    \draw [prop]   (0.5, 0) to[out=-90,in=-90] (1.5, 0);
    \draw [prop]   (1.0, 0) to[out=+90,in=+90] (3.0, 0);
    \draw [prop]   (1.9, 0) to[out=+90,in=+90] (2.6, 0);
    \node [cross] at (2.25, 0) {};
 \end{tikzpicture} &
 \begin{tikzpicture}[valign]
    \draw [defect] (0.2, 0) -- (3.3, 0);
    \draw [prop]   (0.5, 0) to[out=-90,in=-90] (2.5, 0);
    \draw [prop]   (1.0, 0) to[out=+90,in=+90] (3.0, 0);
    \draw [prop]   (1.4, 0) to[out=+90,in=+90] (2.1, 0);
    \node [cross] at (1.75, 0) {};
 \end{tikzpicture} \\
 \textbf{(b10)}  & \textbf{(b11)}  & \textbf{(b12)}  & \textbf{(b13)} \\[1em]
 \begin{tikzpicture}[valign]
    \draw [defect] (0.2, 0) -- (3.3, 0);
    \draw [prop]   (0.5, 0) to[out=-90,in=-90] (2.0, 0);
    \draw [prop]   (1.0, 0) to[out=+90,in=+90] (3.0, 0);
    \draw [prop]   (1.5, 0) to[out=+90,in=+90] (2.5, 0);
    \node [cross] at (0.75, 0) {};
 \end{tikzpicture} &
  \begin{tikzpicture}[valign]
    \draw [defect] (0.2, 0) -- (3.3, 0);
    \draw [prop]   (0.5, 0) to[out=-90,in=-90] (2.0, 0);
    \draw [prop]   (1.0, 0) to[out=+90,in=+90] (3.0, 0);
    \draw [prop]   (1.5, 0) to[out=+90,in=+90] (2.5, 0);
    \node [cross] at (1.25, 0) {};
 \end{tikzpicture} &
  \begin{tikzpicture}[valign]
    \draw [defect] (0.2, 0) -- (3.3, 0);
    \draw [prop]   (0.5, 0) to[out=-90,in=-90] (2.0, 0);
    \draw [prop]   (1.0, 0) to[out=+90,in=+90] (3.0, 0);
    \draw [prop]   (1.5, 0) to[out=+90,in=+90] (2.5, 0);
    \node [cross] at (1.75, 0) {};
 \end{tikzpicture} &
 \begin{tikzpicture}[valign]
    \draw [defect] (0.2, 0) -- (3.3, 0);
    \draw [prop]   (0.5, 0) to[out=-90,in=-90] (2.0, 0);
    \draw [prop]   (1.0, 0) to[out=+90,in=+90] (3.0, 0);
    \draw [prop]   (1.5, 0) to[out=+90,in=+90] (2.5, 0);
    \node [cross] at (2.25, 0) {};
 \end{tikzpicture} \\
 \textbf{(b14)}  & \textbf{(b15)} & \textbf{(b16)} & \textbf{(b17)} \\[1em]
 \begin{tikzpicture}[valign]
    \draw [defect] (0.2, 0) -- (3.3, 0);
    \draw [prop]   (0.5, 0) to[out=-90,in=-90] (2.0, 0);
    \draw [prop]   (1.0, 0) to[out=+90,in=+90] (3.0, 0);
    \draw [prop]   (1.5, 0) to[out=+90,in=+90] (2.5, 0);
    \node [cross] at (2.75, 0) {};
 \end{tikzpicture} &
  \begin{tikzpicture}[valign]
    \draw [defect] (0.2, 0) -- (3.3, 0);
    \draw [prop]   (0.5, 0) to[out=+90,in=+90] (2.0, 0);
    \draw [prop]   (1.0, 0) to[out=-90,in=-90] (2.5, 0);
    \draw [prop]   (1.5, 0) to[out=+90,in=+90] (3.0, 0);
    \node [cross] at (0.75, 0) {};
 \end{tikzpicture} &
  \begin{tikzpicture}[valign]
    \draw [defect] (0.2, 0) -- (3.3, 0);
    \draw [prop]   (0.5, 0) to[out=+90,in=+90] (2.0, 0);
    \draw [prop]   (1.0, 0) to[out=-90,in=-90] (2.5, 0);
    \draw [prop]   (1.5, 0) to[out=+90,in=+90] (3.0, 0);
    \node [cross] at (1.25, 0) {};
 \end{tikzpicture} &
 \begin{tikzpicture}[valign]
    \draw [defect] (0.2, 0) -- (3.3, 0);
    \draw [prop]   (0.5, 0) to[out=+90,in=+90] (2.0, 0);
    \draw [prop]   (1.0, 0) to[out=-90,in=-90] (2.5, 0);
    \draw [prop]   (1.5, 0) to[out=+90,in=+90] (3.0, 0);
    \node [cross] at (1.75, 0) {};
 \end{tikzpicture} \\
 \textbf{(b18)} & \textbf{(b19)} & \textbf{(b20)} & \textbf{(b21)}
 \end{tabular}
 \caption{Vertex diagrams up to three loops.}
 \label{fig:vtx-3loop}
\end{figure}

We now consider the full vertex $\Vm(x)$.
Once again, let us first present the final result, and we derive it below:
\begin{align}
&\frac{\Vm(x)}{\zeta_0 \sqrt{\kappa } j(j+1)}
 = \textbf{(b1)}
 - \textbf{(b2)} \zeta_0^2
 + \big( \textbf{(b3)}+2 \textbf{(b4)}+2 \textbf{(b5)} \big) \zeta_0^4
   \notag \\
& \qquad
 + \Big(
 - 2 \textbf{(b6)}
 - 4 \textbf{(b7)}
 -   \textbf{(b8)}
 -   \textbf{(b9)}
 - 2 \textbf{(b10)}
 - 2 \textbf{(b11)}
 - 2 \textbf{(b12)}
 - 2 \textbf{(b13)} \notag \\
& \qquad \qquad
 - 2 \textbf{(b14)}
 - 4 \textbf{(b15)}
 + 2 (2 j(j+1)-5) \textbf{(b16)}
 + 4 (j(j+1)-2) \textbf{(b17)}
 - 2 \textbf{(b18)} \notag \\
& \qquad \qquad
 - 4 \textbf{(b19)}
 + 2 (2 j(j+1)-5) \textbf{(b20)}
 + (2 j(j+1)-7) \textbf{(b21)}
 \Big) \zeta_0^6 + \ldots \, .
 \label{eq:vertex-terms}
\end{align}
The diagrams \textbf{(b1)}--\textbf{(b21)} are presented in figure \ref{fig:vtx-3loop}.
In this figure, the cross $\times$ represents the point where the bulk field $\phi(x)$ is connected to the defect.
For example, if the bulk field sits at $x = (w, x_\perp)$, where $x_\perp$ are directions orthogonal to the defect, then
\begin{align}
 \textbf{(b5)}
 \;\; = \;\;
\begin{tikzpicture}[valign]
    \draw [defect] (0, 0) -- (2.5, 0);
    \draw [prop]   (0.5, 0) to[out=+90,in=+90] (1.5, 0);
    \draw [prop]   (1.0, 0) to[out=-90,in=-90] (2.0, 0);
    \node [cross] at (0.8, 0) {};
 \end{tikzpicture}
 \;\; = \;\;
 \int\displaylimits_{0<\tau_1<\tau_2<\tau_3<\tau_4<\tau_5<\tau}
 \frac{d\tau_1 \, d\tau_2 \, d\tau_3 \, d\tau_4 \, d\tau_5}
      {\tau_{41}^{2-\veps} \tau_{53}^{2-\veps}
      (|x_\perp|^{\,2} + (w-\tau_2)^2)^{1-\frac\veps2}} \, .
 \label{eq:diagb5}
\end{align}
The value of these integrals is important in obtaining the beta function, and we explain in section \ref{sec:app-integrals} how to calculate them.

Regarding the derivation of equation \eqref{eq:vertex-terms}, it proceeds as before.
First we generate all Wick contractions that contribute to $\tr \langle \phi(x) \Dm(\tau) \rangle$.
For each term, we compute the traces using the commutation relations \eqref{eq:comm-rels}, and then factorize the integrals using relations analogous to \eqref{eq:factor-int}.
One observes that the result is proportional to the right-hand side of \eqref{eq:vertex-terms} multiplied by $\tr \langle \Dm(\tau) \rangle$ in \eqref{eq:prop-ren}.
These steps are tedious, but we have automated them with a computer.

\subsection{Integrals}
\label{sec:app-integrals}

To obtain the beta function, all is left is to extract the divergent part of the integrals in equation \eqref{eq:vertex-terms}.
For illustration, we compute in detail diagram \textbf{(b5)} in \eqref{eq:diagb5}.
We then explain how the same idea can be generalized and automated to all other diagrams.

The first observation is that since we are only interested in the divergent part of $\Vm(x)$, it is convenient to take the $\phi(x)$ insertion to be far away from the defect.
More precisely, if $x = (w,x_\perp)$, then we take $|x_\perp| \gg \tau$, where $\tau$ is the length of the defect operator $\Dm(\tau)$.
In this limit, the $\vec x$, $w$ and $\tau_2$ dependence drops out:
\begin{align}
 \overline{\textbf{(b5)}}
 \; \equiv \;
 \lim_{|x_\perp| \to \infty} \, |x_\perp|^{2-\veps} \, \textbf{(b5)}
 =
 \int\displaylimits_{0<\tau_1<\tau_2<\tau_3<\tau_4<\tau_5<\tau}
 \frac{d\tau_1 \, d\tau_2 \, d\tau_3 \, d\tau_4 \, d\tau_5}
      {\tau_{41}^{2-\veps} \tau_{53}^{2-\veps}} \, .
\end{align}
An important observation is that we can choose to integrate the variables in whichever order we like.
Since $\tau_2$ does not appear in the integrand, it is convenient to perform its integral first
\begin{align}
 \overline{\textbf{(b5)}}
 =
 \int\displaylimits_{0<\tau_1<\tau_3<\tau_4<\tau_5<\tau}
 d\tau_1 \, d\tau_3 \, d\tau_4 \, d\tau_5
 \int_{\tau_1}^{\tau_3}
 \frac{d\tau_2}{\tau_{41}^{2-\veps} \tau_{53}^{2-\veps}}
 =
 \int\displaylimits_{1345}
 \frac{\tau_{31}}{\tau_{41}^{2-\veps} \tau_{53}^{2-\veps}}
 \, .
\end{align}
In the second equality we introduced shorthand notation $\int\displaylimits_{ij\ldots k} = \int\displaylimits_{0<\tau_i<\tau_j<\ldots<\tau_k<\tau}$, and we omit $d\tau_i$ for conciseness.
The strategy is to continue choosing the simplest variable to integrate next.
For example, since $\tau_4$ appears only once, it has a simple integral
\begin{align}
 \overline{\textbf{(b5)}}
 =
 \int\displaylimits_{135}
 \int_{\tau_3}^{\tau_5} d\tau_4 \,
 \frac{\tau_{31}}{\tau_{41}^{2-\veps} \tau_{53}^{2-\veps}}
 = \frac{1}{\veps-1} \int\displaylimits_{135} \Big[
   \tau_{51}^{\veps}   \tau_{53}^{\veps-2}
 - \tau_{51}^{\veps-1} \tau_{53}^{\veps-1}
 - \tau_{31}^{\veps}   \tau_{53}^{\veps-2}
 \Big] \, .
 \label{eq:tmp-int}
\end{align}
In the right-hand side we used $\tau_{31} = \tau_{51} - \tau_{53}$ to simplify the result.
Now we observe an important point: as we integrate we generate many terms, and each of them requires a different order of integration to minimize complexity.
For the first two terms in \eqref{eq:tmp-int}, we should integrate first $\tau_1$ and $\tau_3$, and only then $\tau_5$.
In this way, all integrals are elementary
\begin{align}
 \int\displaylimits_{135} \Big[
   \tau_{51}^{\veps}   \tau_{53}^{\veps-2}
 - \tau_{51}^{\veps-1} \tau_{53}^{\veps-1}
 \Big]
&= \int_0^\tau d\tau_5 \int_0^{\tau_5} d\tau_3 \int_0^{\tau_3} d\tau_1
 \Big[
   \tau_{51}^{\veps}   \tau_{53}^{\veps-2}
 - \tau_{51}^{\veps-1} \tau_{53}^{\veps-1}
 \Big] \notag \\
&= \frac{\tau^{2 \veps+1}}{2 (\veps-1) \veps^2 (2 \veps+1)} \, .
\end{align}
Instead, for the last term in \eqref{eq:tmp-int} it is better to integrate first $\tau_1$ and $\tau_5$, and only then $\tau_3$:
\begin{align}
 \int\displaylimits_{135} \tau_{31}^{\veps} \tau_{53}^{\veps-2}
 = \int_0^\tau d\tau_3 \int_{\tau_3}^\tau d\tau_5 \int_0^{\tau_3} d\tau_1
 \,\tau_{31}^{\veps} \tau_{53}^{\veps-2}
&= \int_0^\tau d\tau_3 \, \frac{\tau_3^{\veps+1} (\tau -\tau_3)^{\veps-1}}{(\veps-1) (\veps+1)} \notag \\
&= \frac{2\Gamma (\veps-1) \Gamma (\veps+2)}{\Gamma (2 \veps+3)}
   \, \tau ^{2 \veps+1} \, .
\end{align}
The last $\tau_3$ integral is the so-called Euler integral.
By choosing smartly the order of integrations, we encountered this somewhat harder integral only at the last step.
If instead we had chosen the order of integration poorly, intermediate results would contain hypergeometric functions, and only at the end we would see the results simplify.

For completeness, the value of the diagram of interest is
\begin{align}
 \overline{\textbf{(b5)}}
 \; = \;
 \left(\frac{1}{2\veps^2(2 \veps+1)}-\frac{2 \Gamma (\veps) \Gamma (\veps+2)}{\Gamma (2 \veps+3)}\right)
 \frac{\tau^{2 \veps+1}}{(\veps-1)^2} \, .
\end{align}

We now can apply the lessons we learned computing diagram \textbf{(b5)} to all other integrals.
To summarize, we first take the limit $|x_\perp| \gg \tau$, which simplifies the form of the integral significantly.
Then we start integrating the variables $\tau_i$ that appear at most once in the integrand.
This process generates many terms, and for each of them we might need to pick different orderings to minimize complexity.
Sometimes relations of the form $\tau_{ij} = \tau_{ik} + \tau_{kj}$ are convenient to simplify intermediate expressions.
At the end of the day, we encounter only two integrals that are not elementary:
\begin{align}
 H_1 = \int_0^\tau du \, u^a (\tau-u)^b \, , \qquad
 H_2 = \int_0^\tau du \int_0^u dv \, u^a (u-v)^b (\tau-v)^c \, .
\end{align}
These integrals might appear in the last integration step, or as a subdiagram of a larger diagram.
Fortunately, these integrals can be evaluated straightforwardly
\begin{align}
 H_1
&= \frac{\Gamma (a+1) \Gamma (b+1)}{\Gamma (a+b+2)} \,
   \tau^{a+b+1} \, , \\
 H_2
&= \frac{\Gamma (a+b+2) \Gamma (b+c+2)}{(b+1) \Gamma (a+2 b+c+4)}
   \, _3F_2 \bigg( \begin{array}{*{20}{c}}
    {b+1,a+b+2,b+c+2} \\
    {b+2,a+2 b+c+4}
    \end{array}; 1 \bigg) \,
   \tau^{a+b+c+2} \, .
\end{align}
Implementing this algorithm in \texttt{mathematica} we managed to compute all integrals in figure \ref{fig:vtx-3loop} in closed form.
The expressions are not particularly illuminating, so we present them in an ancillary notebook.

\subsection{Computation of the \texorpdfstring{$\beta$}{beta}-function}

All that is left is to combine all the ingredients.
We compute all integrals in figure \ref{fig:vtx-3loop} using the method of section \ref{sec:app-integrals}.
We insert these values in equation \eqref{eq:vertexV} for the vertex $\Vm(x)$.
Demanding the result to be finite, gives the bare coupling in terms of the renormalized one, the result being presented in \eqref{eq:zeta0freeblk}.
As an important sanity check, the higher-order poles satisfy the consistency conditions \eqref{eq:cons-cond}.
The bare coupling leads to the beta function \eqref{eq:betafreebulk}.
In the limit of large quantum number $j \to \infty$, the beta function agrees with that of \cite{Cuomo:2022xgw}, providing another sanity check.

\subsection{\texorpdfstring{$\beta$}{beta}-function in the interacting bulk case}\label{interactingbetaapp}

In this section, we extend the previous results to calculate the $\beta(\zeta)$ in the case of an interacting bulk.
We perform this calculation to order $\lambda \zeta^3$, where a single Feynman diagram contributes to the vertex renormalization.
The result of this computation was presented without derivation in \cite{Sachdev_2001}, while here we explicitly derive it.

At the order we are interested in, most contributions to the beta function come from diagrams without bulk interaction or from corrections to the bulk propagator.
The only exception is the diagram
\begin{equation}
		  \begin{tikzpicture}[scale=0.4]
	 \draw[double,thick,blue] (0,0)--(6,0);
	  \draw[thick, black]    (1.5,0) to (3,2)  to (4.5,0);
	  \draw[thick, black]    (3,0) to (3,2) ;
	 \draw[blue, fill=blue] (1.5,0) circle (5pt);
	  \draw[blue, fill=blue] (3,0) circle (5pt);
	   \draw[blue, fill=blue] (4.5,0) circle (5pt);
	  \draw[blue, fill=blue] (3,2) circle (5pt);
	    \draw[thick, black]    (3,2) to (3,4)  ;
	    \draw[blue, fill=white] (3,4) circle (5pt);
	    	 \node[above] at (3,4) {$\phi_a T_a$};
	    	 \node[above] at (0,0) {$0$};
	    	 \node[above] at (6,0) {$1$};
	    	  \draw[black, fill=black] (0,0) circle (3pt);
	    	  \draw[black, fill=black] (6,0) circle (3pt);
	\end{tikzpicture}
\end{equation}
where we took the length of the defect operator to be one since the integral is homogeneous.

Let us first compute the symmetry factor of this diagram.
It is important to remember that $T_a$ also participates in the trace, as in the definition of the vertex \eqref{eq:vertex}.
One of the three legs attached to the defect carries a generator $T_a$ with the same index as the external field $\phi_a$, whereas the other two legs carry generators $T_b$ with contracted indices. Of the three possible channels, in two of them the contracted generators $T_b$ are inserted next to each other, and hence simplify to $j(j+1) T_a$. The other channel reads $T_b T_a T_b = (j(j+1)-1)T_a$. Hence we obtain an overall contribution $\left(j(j+1)-\frac{1}{3}\right)T_a$.
Also note that the integral is path-ordered, but by permutation symmetry we can divide it by $3!$ and instead compute the unordered integral. Therefore, the contribution of this diagram to the vertex is
\begin{equation}\label{1loopvertex}
\frac{\Vm_\lambda(x)}{\zeta_0 \sqrt{\kappa} j(j+1)}=-\frac{\lambda_0 \, \zeta_0^2 \, \kappa^2 }{6}\, \left(j(j+1)-\frac{1}{3}\right)\,  I(x)  \,,
\end{equation}
where the integral is
\begin{equation}
I(x)=\int  \frac{d^{4-\veps} y  }{\left((x_\parallel-y_\parallel)^2+|x_\perp-y_\perp|^2)\right)^{\frac{2-\veps}{2}}}\left(\int_0^1\frac{dt}{\left(|y_\perp|^2+(t-y_\parallel)^2\right)^{\frac{2-\veps}{2}}}\right)^3 \,.
\end{equation}
For our purposes, it is sufficient to extract the leading contribution in the $|x_\perp| \rightarrow \infty$ limit that is also divergent in the $\veps \rightarrow 0$ limit. Note that divergencies $\frac{1}{\veps}$ only arise for small $|y_\perp|$ and when $y_\parallel$ lies near the interval $[0,1]$. Indeed, the only other region where the integrand is unbounded is when $y$ approaches the external point $x$, but for any dimension $d$ the integral is finite
\begin{equation}
\int_{|x-y|<l} \frac{d^d y}{|x-y|^{d-2}}
= \frac{\Omega_{d-1} l^2}{2} \,,
\end{equation}
where $\Omega_{d-1}$ is the volume of the $d-1$-dimensional sphere.

Without loss of generality, we can set $x_\perp = \left(\frac{1}{2},L,0,\dots\right)$, and then pass to cylindrical coordinates $\left(y_\parallel, y_\perp\right)\rightarrow \left(y_\parallel, \rho, \theta, \dots\right)$:
\begin{equation}
\begin{split}
I(L)= \int_0^\infty d\rho \int_{-\infty}^{+\infty} dy_\parallel \int_0^\pi d\theta \int d\Omega_{1-\veps} \frac{\rho^{2-\veps}\, \left(\sin \theta\right)^{1-\veps} }{\left(\left(\frac{1}{2}-y_\parallel\right)^2+\rho^2+L^2-2\rho L \cos \theta\right)^{\frac{2-\veps}{2}}}\,\times \\
\times \left(\int_0^1\frac{dt}{\left(\rho^2+(t-y_\parallel)^2\right)^{\frac{2-\veps}{2}}}\right)^3 \,.
\end{split}
\end{equation}
We are interested in the leading term as $L\rightarrow \infty$. As we remarked above, for the purposes of extracting the divergent part we can consider $\rho$ and $y_\parallel$ to be bounded. Therefore, we have
\begin{equation}
\begin{split}
I(L) \sim \frac{1}{L^{2-\veps}}  \int_0^\delta d\rho \int_{-\delta}^{1+\delta} dy_\parallel \int_0^\pi d\theta \int d\Omega_{1-\veps} \ \rho^{2-\veps}\, \left(\sin \theta\right)^{1-\veps} \, \times \\ 
\times \left(\int_0^1\frac{dt}{\left(\rho^2+(t-y_\parallel)^2\right)^{\frac{2-\veps}{2}}}\right)^3 +O(\veps^0)\,,
\end{split}
\end{equation}
where $\delta>0$ is some arbitrarily small parameter. The integrals over $dt$, $d\Omega_{1-\veps}$ and $d\theta$ are easily performed and one finds
\begin{equation}
\begin{split}
I(L) \sim \frac{1}{L^{2-\veps}} \frac{2 \pi^{\frac{3-\veps}{2}}}{\Gamma \left(\frac{3-\veps}{2}\right)} \int_0^\delta d\rho \int_{-\delta}^{1+\delta} dy_\parallel \,\rho^{-4+2\veps} \left( (1-y_\parallel)\,{}_{2}F_1\left(\tfrac{1}{2},1-\tfrac{\veps}{2};\tfrac{3}{2};-\tfrac{(1-y_\parallel)^2}{\rho^2}\right) + \right. \\
+\left.  y_\parallel\,{}_{2}F_1\left(\tfrac{1}{2},1-\tfrac{\veps}{2};\tfrac{3}{2};-\tfrac{y_\parallel^2}{\rho^2}\right) \right)^3 \,.
\end{split}
\end{equation}
Computing this integral in full generality is hard, but since $\rho<\delta$ is small, we can simply expand the integrand.
The key point is that only powers $\rho^{-1+c \, \veps }$ give divergent contributions, since
\begin{equation}\label{rhoexpansion}
\int_0^\delta d \rho \, \rho^{-1+c \, \veps} = \frac{1}{c\, \veps} + O(\veps^0) \, .
\end{equation}
All in all, only one term contributes to the divergence, and the remaining $d\rho$ and $dy_\parallel$ integrations are elementary:
 \begin{align}
 I(L)
& \sim \frac{1}{L^{2-\veps}} \frac{2 \pi^{\frac{3-\veps}{2}}}{\Gamma \left(\frac{3-\veps}{2}\right)}
 \frac{\pi^{\frac{3}{2}}\Gamma \left( \frac{1-\veps}{2}\right)^3}{8\,\Gamma \left(1- \frac{\veps}{2}\right)^3}
 \int_0^\delta d\rho \int_{-\delta}^{1+\delta} dy_\parallel \, \rho^{-1+2\veps} \, \left( \sgn (1-y_\parallel )+\sgn (y_\parallel)\right)^3 \notag \\
& = \frac{1}{L^{2-\veps}} \frac{2 \pi^4}{\veps} + O(\veps^0)\,.
 \end{align}
Inserting this into \eqref{1loopvertex} we finally get
 \begin{equation}
\frac{\Vm_\lambda(L)}{\zeta_0 \sqrt{\kappa}j(j+1)}\sim - \frac{1}{L^2}\frac{\lambda_0\zeta_0^2}{48 \,\veps}  \, \left(j(j+1)-\frac{1}{3}\right) \,,
 \end{equation}
which when combined with the free-theory contributions and with the two-loop correction to the bulk propagator, leads to \eqref{zetaren}.

\section{Kinematics and conformal blocks}
\label{app:kinematics}

In this appendix we review the structure of correlators in defect CFT, and we provide the expressions for conformal blocks needed in this work.
The literature on defect CFT kinematics is extense, for more details see \emph{e.g.} \cite{Billo:2016cpy,Lauria:2017wav,Lauria:2018klo,Herzog:2020bqw}.

The simplest correlators in presence of a defect are the one-point function of bulk operators, and the bulk-defect two-point functions.
For the case when both operators are scalars they read \cite{Billo:2016cpy}
\begin{equation}\label{1ptdef}
    \langle \mathcal{O} (x) \rangle_{\mathcal{D}_j} = \frac{a_{\mathcal{O}}}{|x_\perp|^{\Delta}} \ , \quad \langle \mathcal{O} (x) \hat{ \mathcal{O} }(y) \rangle_{\mathcal{D}_j} = \frac{b_{ \mathcal{O}  \hat{ \mathcal{O} }}}{|x_{ \perp}|^{\Delta-\hat{\Delta}}(|x_{ \perp}|^2+y^2)^{\hat{\Delta}}} \, ,
\end{equation}
where $\Delta$ and $\hat{\Delta}$ are the conformal dimensions of bulk and defect operators respectively, and $a_\mathcal{O}$ and $b_{\mathcal{O}\hat{\mathcal{O}}}$ are constants. Collectively, they are known as the defect CFT data.
Similar expressions exist when either $\Om$ or $\hat \Om$ are spinning operators, and we follow the conventions of the appendix of \cite{Gimenez-Grau:2022ebb}.

The main observable we are interested in is the two-point function of bulk operators, which has the form \cite{Billo:2016cpy}
\begin{equation}
    \braket{\phi(x) \phi(y)}_{\mathcal{D}_j} = \frac{F_{\phi \phi}(z,\bar{z})}{|x_{\perp}|^{\Delta_{\phi}}|y_{\perp}|^{\Delta_{\phi}}},
\end{equation}
where $z$ and $\Bar{z}$ are lightcone coordinates that parametrize the hyperplane orthogonal to the defect. They are related to the external points in the following way
\begin{equation}\label{xi}
\xi = \frac{(1-z)(1-\bar{z})}{\sqrt{z\bar{z}}} =\frac{(x-y)^2}{|x_\perp||y_\perp|}, \qquad \frac{z+\bar{z}}{2\sqrt{z\bar{z}}} =\frac{x_\bot \cdot
 y_\bot}{|x_\perp||y_\perp|} \, .
\end{equation}
The combination we called $\xi$ appears often in equations, and it is convenient to give it a name.
One can also introduce radial coordinates $r$ and  $w$, defined by
\begin{equation}\label{coordinates}
    z=rw, \qquad \bar{z}=\frac{r}{w} \, .
\end{equation}
The two-point function can be expanded in the defect channel OPE as
\begin{equation}
     F_{\phi \phi}(r,w) = \sum_{\hat{\mathcal{O}}} b^2 _{\mathcal{O} \hat{\mathcal{O}}} \hat{f}_{\hat{\Delta}, s}(r,w) \, ,
\end{equation}
where the sum runs over defect operators labeled by their dimensions $\hat{\Delta}$ and (transverse) spins $s$.
The coefficients $b^2 _{\mathcal{O} \hat{\mathcal{O}}}$ are the bulk to defect couplings introduced above and $\hat{f}_{\hat{\Delta}, s}(r,w)$ are the defect conformal blocks, defined as\footnote{The normalization of our blocks differs by a factor of $2^{-s}$ from \cite{Gimenez-Grau:2022ebb}, which accounts for the fact that our block expansion is $\sum b_{\Om\hat\Om}^2 \hat f_{\hat\Delta,s}|_{\text{here}} = \sum 2^{-s} b_{\Om\hat\Om}^2 \hat f_{\hat\Delta,s}|_{\text{there}}$. Similarly, the bulk blocks in \eqref{eq:blkblocks} differ by a factor of $2^{-\ell}$ since $\sum \lambda_{\Om_1\Om_2\Om} a_{\Om} \hat f_{\Delta,\ell}|_{\text{here}} = \sum 2^{-\ell} \lambda_{\Om_1\Om_2\Om} a_{\Om} \hat f_{\Delta,\ell}|_{\text{there}}$. However, the meaning of the CFT data is always the same in both cases.}
\begin{equation}\label{defectblock}
        \hat{f}_{\hat\Delta,s}(r,w) =r^{\hat{\Delta}} \, {}_2 F_1 \left(\hat{\Delta},\frac{p}{2},\hat{\Delta}+1-\frac{p}{2},r^2\right) (2w)^{-s} {}_2 F_1 \left(-s,\frac{d-p}{2}-1,2-\frac{d-p}{2}-s,w^2\right)
\end{equation}
where $p$ is the dimension of the defect ($p=1$ in our case) and $d$ is the dimension of the bulk.
Alternatively, the correlator can be expanded as a combination of bulk blocks corresponding to the operators that appear in the bulk OPE. They are labeled by conformal dimensions $\Delta$ and spins $\ell$.
\begin{equation}
     F_{\phi \phi}(r,w) = \xi^{-\Delta_\phi} \sum_{\mathcal{O}}\lambda_{\phi \phi \mathcal{O}} a_{\mathcal{O}} f_{\Delta,\ell}(r,w)
\end{equation}
Here the coefficients are the product of one-point functions and three-point functions of bulk operators. The latter are unaffected by the defect and thus can be computed in the theory without the defect.
The bulk blocks $f_{\Delta,\ell}(r,w)$ are not known in a closed form, but can be expressed as a sum of Harish-Chandra functions \cite{Isachenkov:2018pef}
\begin{equation} \label{eq:blkblocks}
    f_{\Delta,\ell}(r,w)
    = 2^{-\ell} f^{HS}_{\Delta,\ell}(r,w)
    + \frac{\Gamma(\ell+d-2)\Gamma(-\ell-\frac{d-2}{2})}
           {2^\ell \Gamma(\ell+\frac{d-2}{2})\Gamma(-\ell)}
	  \frac{\Gamma(\frac{\ell+d-p-1}{2})\Gamma(\frac{1-\ell}{2})}
	       {\Gamma(\frac{\ell+d-1}{2})\Gamma(\frac{1-\ell-p}{2})}
	  f^{HS}_{\Delta,2-d-\ell}(r,w) \, ,
\end{equation}
where $f_{\Delta,\ell}^{HS}(r,w)$ can be expressed as a double infinite sum
\begin{equation}
\begin{split}\label{bulkblock}
        f_{\Delta,\ell}^{HS}(r,w) &=\sum_{m=0}^{\infty}\sum_{n=0}^{\infty} \left[(1-r w)(1-\frac{r}{w})\right]^{\frac{\Delta-\ell}{2}+m+n}
        h_n (\Delta,\ell)h_m (1-\ell,1-\Delta) \frac{4^{m-n}}{n! m!} \frac{(\frac{\Delta+\ell}{2})_{n-m}}{\left(\frac{\Delta+\ell}{2}-\frac{1}{2} \right)_{n-m}}  \\ &\textstyle \times {}_4 F_3 (-n,-m,\frac{1}{2},\frac{\Delta-\ell}{2}-\frac{d}{2}+1;-\frac{\Delta+\ell}{2}+1-n,\frac{\Delta+\ell}{2}-m,\frac{\Delta-\ell}{2}-\frac{d}{2}+\frac{3}{2};1) \\
        & (1-r^2)^{\ell-2m}  \textstyle {}_2 F_1 (\frac{\Delta+\ell}{2}-m+n,\frac{\Delta+\ell}{2}-m+n,\Delta+\ell-2(m-n),1-r^2),
\end{split}
\end{equation}
where
\begin{equation}
    h_n (\Delta,\ell)=\frac{\left(\frac{\Delta}{2}-\frac{1}{2}\right)_n \left(\frac{\Delta}{2}-\frac{p}{2}\right)_n \left(\frac{\Delta+\ell}{2} \right)_n }{\left( \Delta-\frac{d}{2}+1\right)_n \left(\frac{\Delta+\ell}{2}+\frac{1}{2}\right)_n}.
\end{equation}
An important feature of the bulk blocks is their analytic structure in the variable $w$. Since they are proportional to $\left(\frac{(1-r w)(w-r)}{r w}\right)^{\frac{\Delta-\ell}{2}} = \xi^{\frac{\Delta-\ell}{2}}$,  for generic $\Delta$ they have a branch cut between $w=0$ and $w=r$. This feature, together with the convergence of the bulk OPE around $w=r$, allows us to evaluate the discontinuity of the two-point function as
\begin{equation}
     \text{Disc} F(r,w)= \sum_{\mathcal{O}}\lambda_{\phi \phi \mathcal{O}} a_{\mathcal{O}} \text{Disc} \left[\xi^{-\Delta_\phi}f_{\Delta,\ell}(r,w)\right] \, .
\end{equation}

\section{Other perturbative computations}
In this appendix we explicitly evaluate some observables in perturbation theory via Feynman diagrams.

\subsection{Two-point function of defect spin operators}\label{twoptspinapp}
Here we compute the two loops contribution to the two-point function $\langle \hat{S}_a(\tau_1) \hat{S}_b(\tau_2) \rangle_{\Dm_j}$. 
Similarly to what has been done at one loop in \ref{corrdefectspin}, we need to compute connected diagrams. To get them, at two loops we will not only need to subtract the order zero connected contribution times pieces of two-loops bubble, but also the order one connected contributions times pieces of one-loop bubbles. 
We illustrate this with an example
\begin{equation}
\begin{split}
	  	\begin{tikzpicture}[scale=0.4]
	 \draw[double,thick,blue] (-4.5,0)--(1.25,0);
	  \draw[thick, black]    (-3,0) to[out=90,in=180] (-1.75,1) to[out=0,in=90]  (-0.5,0);
	  \draw[thick, black]   (-2.25,0)  to[out=90,in=0]  (-3,1) to[out=180,in=90] (-3.75,0);
	   \fill[blue] (-0.5,0) circle (5pt);
	   \fill[blue] (-3,0) circle (5pt);
	   	   \fill[blue] (-2.25,0) circle (5pt);
	   \fill[blue] (-3.75,0) circle (5pt);
	  	 \fill[blue] (0.5,0) circle (5pt);
	  	  \fill[blue] (-1.5,0) circle (5pt);
	\end{tikzpicture}_{\,c}
		\hspace{0.3 cm}
		&=
				\hspace{0.3 cm}
	  	\begin{tikzpicture}[scale=0.4]
	 \draw[double,thick,blue] (-4.5,0)--(1.25,0);
	  \draw[thick, black]    (-3,0) to[out=90,in=180] (-1.75,1) to[out=0,in=90]  (-0.5,0);
	  \draw[thick, black]   (-2.25,0)  to[out=90,in=0]  (-3,1) to[out=180,in=90] (-3.75,0);
	   \fill[blue] (-0.5,0) circle (5pt);
	   \fill[blue] (-3,0) circle (5pt);
	   	   \fill[blue] (-2.25,0) circle (5pt);
	   \fill[blue] (-3.75,0) circle (5pt);
	  	 \fill[blue] (0.5,0) circle (5pt);
	  	  \fill[blue] (-1.5,0) circle (5pt);
	\end{tikzpicture}
		\hspace{0.3 cm}
	 -
			\hspace{0.3 cm}
	\begin{tikzpicture}[scale=0.4]
	 \draw[double,thick,blue] (-2,0)--(2,0);
	  	 \fill[blue]   (-1,0) circle (5pt);
	  		 \fill[blue]    (1,0) circle (5pt);
	\end{tikzpicture}_{\,c}
	\hspace{0.1 cm}
	\times
		\text{bubbles}^{(2)} \ + \\
	&- \hspace{0.1 cm} 
		  	\begin{tikzpicture}[scale=0.4]
	 \draw[double,thick,blue] (-3.75,0)--(0.75,0);
	  \draw[thick, black]    (-3,0) to[out=90,in=180](-2,1)  to[out=00,in=90]  (-1,0);
	 \fill[blue] (-2,0) circle (5pt);
	   \fill[blue] (-1,0) circle (5pt);
	   \fill[blue] (-3,0) circle (5pt);
	  	 \fill[blue] (0,0) circle (5pt);
	\end{tikzpicture}_{\,c}
			\hspace{0.1 cm}
	\times \text{bubbles}^{(1)}
		\ =\hspace{0.1 cm} j(j+1) \hspace{0.1 cm}
			  	\begin{tikzpicture}[scale=0.4]
	 \draw[double,thick,blue] (-4.5,0)--(1.25,0);
	  \draw[thick, black]    (-3,0) to[out=90,in=180] (-1.75,1) to[out=0,in=90]  (-0.5,0);
	  \draw[thick, black]   (-2.25,0)  to[out=90,in=0]  (-3,1) to[out=180,in=90] (-3.75,0);
	  	 \draw[blue, fill=white]   (0.5,0) circle (5pt);
	  	  \draw[blue, fill=white]  (-1.5,0) circle (5pt);
	\end{tikzpicture}
		\hspace{0.3 cm},
	\end{split}
	\end{equation}	
	where the last diagram denotes just a kinematical integral stripped of the color factor.
	At the end of this procedure, what we are left is the following
	\begin{equation}
	\begin{split}
	I_c^{(2)}(\tau_1,\tau_2)&=\sum\, \Gamma^{(2)}_c = j(j+1) \, \bigg(
	\hspace{0.1 cm}
			  	\begin{tikzpicture}[scale=0.4]
	 \draw[double,thick,blue] (-4.5,0)--(1.25,0);
	  \draw[thick, black]    (-3,0) to[out=90,in=180] (-1.75,1) to[out=0,in=90]  (-0.5,0);
	  \draw[thick, black]   (-2.25,0)  to[out=90,in=0]  (-3,1) to[out=180,in=90] (-3.75,0);
	  	 \draw[blue, fill=white]   (0.5,0) circle (5pt);
	  	  \draw[blue, fill=white]  (-1.5,0) circle (5pt);
	\end{tikzpicture}
		\hspace{0.1 cm}
		+
		\hspace{0.1 cm}
			  	\begin{tikzpicture}[scale=0.4]
	 \draw[double,thick,blue] (0,0)--(-5.75,0);
	  \draw[thick, black]    (-1.5,0) to[out=90,in=0] (-2.75,1) to[out=180,in=90]  (-4,0);
	  \draw[thick, black]   (-2.25,0)  to[out=90,in=180]  (-1.5,1) to[out=0,in=90] (-0.75,0);
	  	 \draw[blue, fill=white]   (-5,0) circle (5pt);
	  	  \draw[blue, fill=white]  (-3,0) circle (5pt);
	\end{tikzpicture}
		\hspace{0.1 cm}
		+2
				\hspace{0.1 cm}
			  	\begin{tikzpicture}[scale=0.4]
	 \draw[double,thick,blue] (0,0)--(5.75,0);
	  \draw[thick, black]    (1.5,0) to[out=90,in=180] (2.75,1) to[out=0,in=90]  (4,0);
	  \draw[thick, black]   (3.25,0)  to[out=90,in=0]  (2,1) to[out=180,in=90] (0.75,0);
	  	 \draw[blue, fill=white]   (5,0) circle (5pt);
	  	  \draw[blue, fill=white]  (2.375,0) circle (5pt);
	\end{tikzpicture}
		\hspace{0.1 cm}+ \\
		\hspace{0.1 cm}
		&+2
				\hspace{0.1 cm}
			  	\begin{tikzpicture}[scale=0.4]
	 \draw[double,thick,blue] (0,0)--(-5.75,0);
	  \draw[thick, black]    (-1.5,0) to[out=90,in=0] (-2.75,1) to[out=180,in=90]  (-4,0);
	  \draw[thick, black]   (-3.25,0)  to[out=90,in=180]  (-2,1) to[out=0,in=90] (-0.75,0);
	  	 \draw[blue, fill=white]   (-5,0) circle (5pt);
	  	  \draw[blue, fill=white]  (-2.375,0) circle (5pt);
	\end{tikzpicture}
		\hspace{0.1 cm}
		+
		\hspace{0.1 cm}
			  	\begin{tikzpicture}[scale=0.4]
	 \draw[double,thick,blue] (0,0)--(5.75,0);
	  \draw[thick, black]    (1.5,0) to[out=90,in=180] (2.375,0.5) to[out=0,in=90]  (3.25,0);
	  \draw[thick, black]   (4,0)  to[out=90,in=0]  (2.375,1) to[out=180,in=90] (0.75,0);
	  	 \draw[blue, fill=white]   (5,0) circle (5pt);
	  	  \draw[blue, fill=white]  (2.375,0) circle (5pt);
	\end{tikzpicture}
		\hspace{0.1 cm}
		+
			\hspace{0.1 cm}
			  	\begin{tikzpicture}[scale=0.4]
	 \draw[double,thick,blue] (0,0)--(-5.75,0);
	  \draw[thick, black]    (-1.5,0) to[out=90,in=0] (-2.375,0.5) to[out=180,in=90]  (-3.25,0);
	  \draw[thick, black]   (-4,0)  to[out=90,in=180]  (-2.375,1) to[out=0,in=90] (-0.75,0);
	  	 \draw[blue, fill=white]   (-5,0) circle (5pt);
	  	  \draw[blue, fill=white]  (-2.375,0) circle (5pt);
	\end{tikzpicture}
		\hspace{0.1 cm}
		+
				\hspace{0.1 cm}
			  	\begin{tikzpicture}[scale=0.4]
	 \draw[double,thick,blue] (0,0)--(5.75,0);
	  \draw[thick, black]    (1.5,0) to[out=90,in=180] (3.25,1) to[out=0,in=90]  (5,0);
	  \draw[thick, black]   (3.25,0)  to[out=90,in=0]  (2,1) to[out=180,in=90] (0.75,0);
	  	 \draw[blue, fill=white]   (4,0) circle (5pt);
	  	  \draw[blue, fill=white]  (2.375,0) circle (5pt);
	\end{tikzpicture}
		\hspace{0.1 cm}+\\
		\hspace{0.1 cm}
		&+
				\hspace{0.1 cm}
			  	\begin{tikzpicture}[scale=0.4]
	 \draw[double,thick,blue] (0,0)--(-5.75,0);
	  \draw[thick, black]    (-1.5,0) to[out=90,in=0] (-3.25,1) to[out=180,in=90]  (-5,0);
	  \draw[thick, black]   (-3.25,0)  to[out=90,in=180]  (-2,1) to[out=0,in=90] (-0.75,0);
	  	 \draw[blue, fill=white]   (-4,0) circle (5pt);
	  	  \draw[blue, fill=white]  (-2.375,0) circle (5pt);
	\end{tikzpicture}
		\hspace{0.1 cm}
		+
			\hspace{0.1 cm}
			  	\begin{tikzpicture}[scale=0.4]
	 \draw[double,thick,blue] (0,0)--(5.75,0);
	  \draw[thick, black]    (2.625,0) to[out=90,in=180] (3.25,1) to[out=0,in=90]  (3.875,0);
	  \draw[thick, black]   (3.25,0)  to[out=90,in=0]  (2,1) to[out=180,in=90] (0.75,0);
	  	 \draw[blue, fill=white]   (4.75,0) circle (5pt);
	  	  \draw[blue, fill=white]  (2,0) circle (5pt);
	\end{tikzpicture}
		\hspace{0.1 cm}+
			\hspace{0.1 cm}
			  	\begin{tikzpicture}[scale=0.4]
	 \draw[double,thick,blue] (0,0)--(-5.75,0);
	  \draw[thick, black]    (-2.625,0) to[out=90,in=0] (-3.25,1) to[out=180,in=90]  (-3.875,0);
	  \draw[thick, black]   (-3.25,0)  to[out=90,in=180]  (-2,1) to[out=0,in=90] (-0.75,0);
	  	 \draw[blue, fill=white]   (-4.75,0) circle (5pt);
	  	  \draw[blue, fill=white]  (-2,0) circle (5pt);
	\end{tikzpicture}
		\hspace{0.1 cm}+ 
		  	\begin{tikzpicture}[scale=0.4]
	 \draw[double,thick,blue] (0,0)--(6,0);
	  \draw[thick, black]    (0.75,0) to[out=90,in=180](1.75,1)  to[out=00,in=90]  (2.75,0);
	   \draw[blue, fill=white]   (1.75,0) circle (5pt);
	   \draw[thick, black]    (3.25,0) to[out=90,in=180](4.25,1)  to[out=00,in=90]  (5.25,0);
	   \draw[blue, fill=white]   (4.25,0) circle (5pt);
	\end{tikzpicture}
			\hspace{0.1 cm}
			+\\
			\hspace{0.1 cm}
		&+2		\hspace{0.1 cm} 
		  	\begin{tikzpicture}[scale=0.4]
	 \draw[double,thick,blue] (0,0)--(6,0);
	  \draw[thick, black]    (0.75,0) to[out=90,in=180](2,1)  to[out=00,in=90]  (3.25,0);
	   \draw[blue, fill=white]   (1.75,0) circle (5pt);
	   \draw[thick, black]    (2.75,0) to[out=90,in=180](4,1)  to[out=00,in=90]  (5.25,0);
	   \draw[blue, fill=white]   (4.25,0) circle (5pt);
	\end{tikzpicture}
			\hspace{0.1 cm} \bigg) .
		\end{split}
	\end{equation}
	This sum of kinematical terms can be reorganized and further simplified since an exponentiation of the previous orders occurs. 
	Indeed we have
		\begin{equation}\label{twoptconn}
	\begin{split}
	I_c^{(2)}(\tau_1,\tau_2)&= \frac{j(j+1)}{2}\left( 
 \hspace{0.1 cm} 
		  	\begin{tikzpicture}[scale=0.4]
	 \draw[double,thick,blue] (-3.75,0)--(0.75,0);
	  \draw[thick, black]    (-3,0) to[out=90,in=180](-2,1)  to[out=00,in=90]  (-1,0);
	 \draw[blue, fill=white] (-2,0) circle (5pt);
	  \draw[blue, fill=white] (0,0) circle (5pt);
	\end{tikzpicture}
			\hspace{0.1 cm}
			+
			 \hspace{0.1 cm} 
		  	\begin{tikzpicture}[scale=0.4]
	 \draw[double,thick,blue] (0,0)--(-4.5,0);
	  \draw[thick, black]    (-0.75,0) to[out=90,in=0](-1.75,1)  to[out=180,in=90]  (-2.75,0);
	 \draw[blue, fill=white]  (-1.75,0) circle (5pt);
	 \draw[blue, fill=white]  (-3.75,0) circle (5pt);
	\end{tikzpicture}
			\hspace{0.1 cm}
	\right)^2+j(j+1) \, \bigg( 
	\hspace{0.1 cm}
			  	\begin{tikzpicture}[scale=0.4]
	 \draw[double,thick,blue] (-4.5,0)--(1.25,0);
	  \draw[thick, black]    (-3,0) to[out=90,in=180] (-1.75,1) to[out=0,in=90]  (-0.5,0);
	  \draw[thick, black]   (-2.25,0)  to[out=90,in=0]  (-3,1) to[out=180,in=90] (-3.75,0);
	  	 \draw[blue, fill=white]   (0.5,0) circle (5pt);
	  	  \draw[blue, fill=white]  (-1.5,0) circle (5pt);
	\end{tikzpicture}
		\hspace{0.1 cm}
		+ \\
		\hspace{0.1 cm} &+ \hspace{0.1 cm}
			  	\begin{tikzpicture}[scale=0.4]
	 \draw[double,thick,blue] (0,0)--(-5.75,0);
	  \draw[thick, black]    (-1.5,0) to[out=90,in=0] (-2.75,1) to[out=180,in=90]  (-4,0);
	  \draw[thick, black]   (-2.25,0)  to[out=90,in=180]  (-1.5,1) to[out=0,in=90] (-0.75,0);
	  	 \draw[blue, fill=white]   (-5,0) circle (5pt);
	  	  \draw[blue, fill=white]  (-3,0) circle (5pt);
	\end{tikzpicture}
		\hspace{0.1 cm}
		+ 
				\hspace{0.1 cm}
			  	\begin{tikzpicture}[scale=0.4]
	 \draw[double,thick,blue] (0,0)--(5.75,0);
	  \draw[thick, black]    (1.5,0) to[out=90,in=180] (2.75,1) to[out=0,in=90]  (4,0);
	  \draw[thick, black]   (3.25,0)  to[out=90,in=0]  (2,1) to[out=180,in=90] (0.75,0);
	  	 \draw[blue, fill=white]   (5,0) circle (5pt);
	  	  \draw[blue, fill=white]  (2.375,0) circle (5pt);
	\end{tikzpicture}
		\hspace{0.1 cm}+ 
				\hspace{0.1 cm}
			  	\begin{tikzpicture}[scale=0.4]
	 \draw[double,thick,blue] (0,0)--(-5.75,0);
	  \draw[thick, black]    (-1.5,0) to[out=90,in=0] (-2.75,1) to[out=180,in=90]  (-4,0);
	  \draw[thick, black]   (-3.25,0)  to[out=90,in=180]  (-2,1) to[out=0,in=90] (-0.75,0);
	  	 \draw[blue, fill=white]   (-5,0) circle (5pt);
	  	  \draw[blue, fill=white]  (-2.375,0) circle (5pt);
	\end{tikzpicture}
		\hspace{0.1 cm}
		+
				\hspace{0.1 cm}
			  	\begin{tikzpicture}[scale=0.4]
	 \draw[double,thick,blue] (0,0)--(5.75,0);
	  \draw[thick, black]    (1.5,0) to[out=90,in=180] (3.25,1) to[out=0,in=90]  (5,0);
	  \draw[thick, black]   (3.25,0)  to[out=90,in=0]  (2,1) to[out=180,in=90] (0.75,0);
	  	 \draw[blue, fill=white]   (4,0) circle (5pt);
	  	  \draw[blue, fill=white]  (2.375,0) circle (5pt);
	\end{tikzpicture}
		\hspace{0.1 cm}
		+ \\
		\hspace{0.1 cm} &+ \hspace{0.1 cm}
			  	\begin{tikzpicture}[scale=0.4]
	 \draw[double,thick,blue] (0,0)--(-5.75,0);
	  \draw[thick, black]    (-1.5,0) to[out=90,in=0] (-3.25,1) to[out=180,in=90]  (-5,0);
	  \draw[thick, black]   (-3.25,0)  to[out=90,in=180]  (-2,1) to[out=0,in=90] (-0.75,0);
	  	 \draw[blue, fill=white]   (-4,0) circle (5pt);
	  	  \draw[blue, fill=white]  (-2.375,0) circle (5pt);
	\end{tikzpicture}
		\hspace{0.1 cm}+
			\hspace{0.1 cm}
			  	\begin{tikzpicture}[scale=0.4]
	 \draw[double,thick,blue] (0,0)--(5.75,0);
	  \draw[thick, black]    (2.625,0) to[out=90,in=180] (3.25,1) to[out=0,in=90]  (3.875,0);
	  \draw[thick, black]   (3.25,0)  to[out=90,in=0]  (2,1) to[out=180,in=90] (0.75,0);
	  	 \draw[blue, fill=white]   (4.75,0) circle (5pt);
	  	  \draw[blue, fill=white]  (2,0) circle (5pt);
	\end{tikzpicture}
		\hspace{0.1 cm}+
			\hspace{0.1 cm}
			  	\begin{tikzpicture}[scale=0.4]
	 \draw[double,thick,blue] (0,0)--(-5.75,0);
	  \draw[thick, black]    (-2.625,0) to[out=90,in=0] (-3.25,1) to[out=180,in=90]  (-3.875,0);
	  \draw[thick, black]   (-3.25,0)  to[out=90,in=180]  (-2,1) to[out=0,in=90] (-0.75,0);
	  	 \draw[blue, fill=white]   (-4.75,0) circle (5pt);
	  	  \draw[blue, fill=white]  (-2,0) circle (5pt);
	\end{tikzpicture}
			+	\hspace{0.1 cm} 
		  	\begin{tikzpicture}[scale=0.4]
	 \draw[double,thick,blue] (0,0)--(6,0);
	  \draw[thick, black]    (0.75,0) to[out=90,in=180](2,1)  to[out=00,in=90]  (3.25,0);
	   \draw[blue, fill=white]   (1.75,0) circle (5pt);
	   \draw[thick, black]    (2.75,0) to[out=90,in=180](4,1)  to[out=00,in=90]  (5.25,0);
	   \draw[blue, fill=white]   (4.25,0) circle (5pt);
	\end{tikzpicture}
			\hspace{0.1 cm} \bigg) .
		\end{split}
	\end{equation}
	Moreover, one can note that specular diagrams yield the same contribution. Hence one is left with the evaluation of only five diagrams. The integrals are not difficult and the result is
	\begin{equation}
\begin{split}
	 			 & 	\begin{tikzpicture}[scale=0.4]
	 \draw[double,thick,blue] (-4.5,0)--(1.25,0);
	  \draw[thick, black]    (-3,0) to[out=90,in=180] (-1.75,1) to[out=0,in=90]  (-0.5,0);
	  \draw[thick, black]   (-2.25,0)  to[out=90,in=0]  (-3,1) to[out=180,in=90] (-3.75,0);
	  	 \draw[blue, fill=white]   (0.5,0) circle (5pt);
	  	  \draw[blue, fill=white]  (-1.5,0) circle (5pt);
	\end{tikzpicture}
		\hspace{0.1 cm}
		=
\zeta_0^4 \, |\tau_1-\tau_2|^{2\veps} \,\frac{\delta_{ab}}{3} \cdot \frac{2^{-1-2\veps} \, \Gamma \left(\frac{1}{2}-\veps\right)\Gamma(\veps-1)}{\sqrt{\pi}(\veps-1)\veps}, \\
		&	\begin{tikzpicture}[scale=0.4]
	 \draw[double,thick,blue] (0,0)--(5.75,0);
	  \draw[thick, black]    (1.5,0) to[out=90,in=180] (2.75,1) to[out=0,in=90]  (4,0);
	  \draw[thick, black]   (3.25,0)  to[out=90,in=0]  (2,1) to[out=180,in=90] (0.75,0);
	  	 \draw[blue, fill=white]   (5,0) circle (5pt);
	  	  \draw[blue, fill=white]  (2.375,0) circle (5pt);
	\end{tikzpicture}
		\hspace{0.1 cm}
		= \zeta_0^4 \, |\tau_1-\tau_2|^{2\veps} \,\frac{\delta_{ab}}{3} \cdot \left(\frac{1}{2(\veps-1)\veps^2(2\veps-1)}-\frac{\Gamma(\veps-1)\Gamma(-2\veps)}{\Gamma(2-\veps)} \right),\\
		& 	\begin{tikzpicture}[scale=0.4]
	 \draw[double,thick,blue] (0,0)--(5.75,0);
	  \draw[thick, black]    (1.5,0) to[out=90,in=180] (3.25,1) to[out=0,in=90]  (5,0);
	  \draw[thick, black]   (3.25,0)  to[out=90,in=0]  (2,1) to[out=180,in=90] (0.75,0);
	  	 \draw[blue, fill=white]   (4,0) circle (5pt);
	  	  \draw[blue, fill=white]  (2.375,0) circle (5pt);
	\end{tikzpicture}
		\hspace{0.1 cm}
		=\zeta_0^4 \, |\tau_1-\tau_2|^{2\veps} \,\frac{\delta_{ab}}{3}  \cdot \left(-\frac{1}{2(\veps-1)^2\veps^2}+\frac{\Gamma(\veps-1)\Gamma(-2\veps)}{\Gamma(2-\veps)} \right),\\
		& 			  	\begin{tikzpicture}[scale=0.4]
	 \draw[double,thick,blue] (0,0)--(5.75,0);
	  \draw[thick, black]    (2.625,0) to[out=90,in=180] (3.25,1) to[out=0,in=90]  (3.875,0);
	  \draw[thick, black]   (3.25,0)  to[out=90,in=0]  (2,1) to[out=180,in=90] (0.75,0);
	  	 \draw[blue, fill=white]   (4.75,0) circle (5pt);
	  	  \draw[blue, fill=white]  (2,0) circle (5pt);
	\end{tikzpicture}
		\hspace{0.1 cm}
		=\zeta_0^4 \, |\tau_1-\tau_2|^{2\veps} \,\frac{\delta_{ab}}{3}  \cdot \left( \frac{-1+\frac{2^{1-2\veps}\sqrt{\pi} \,\veps \Gamma(\veps)}{\Gamma \left( \frac{1}{2}+\veps\right)}}{2(\veps-1)^2\,\veps^2} \right), \\
		& 	\begin{tikzpicture}[scale=0.4]
	 \draw[double,thick,blue] (0,0)--(6,0);
	  \draw[thick, black]    (0.75,0) to[out=90,in=180](2,1)  to[out=00,in=90]  (3.25,0);
	   \draw[blue, fill=white]   (1.75,0) circle (5pt);
	   \draw[thick, black]    (2.75,0) to[out=90,in=180](4,1)  to[out=00,in=90]  (5.25,0);
	   \draw[blue, fill=white]   (4.25,0) circle (5pt);
	\end{tikzpicture}
			\hspace{0.1 cm} 
			=\zeta_0^4 \, |\tau_1-\tau_2|^{2\veps} \,\frac{\delta_{ab}}{3} \cdot \left( \frac{-\Gamma(1+\veps)^2+\Gamma(1+2\veps)}{(\veps-1)^2\,\veps^2 \,\Gamma (1+2\veps)}\right).
\end{split}
	\end{equation}
	Substituting these into \eqref{twoptconn} gives the two loops contribution to the bare two-point function, which can then be used to compute the renormalized one.

\newpage

\providecommand{\href}[2]{#2}\begingroup\raggedright\endgroup

\end{document}